\documentclass[11pt]{article}
\usepackage[letterpaper,margin=1in]{geometry}
\usepackage{times}
\usepackage{natbib}
\setcitestyle{authoryear,round,citesep={;},aysep={,},yysep={;}}

\usepackage{amsmath,amsfonts,bm}

\def\eqref#1{equation~\ref{#1}}

\def\1{\bm{1}}

\DeclareMathAlphabet{\mathsfit}{\encodingdefault}{\sfdefault}{m}{sl}
\SetMathAlphabet{\mathsfit}{bold}{\encodingdefault}{\sfdefault}{bx}{n}

\usepackage[pagebackref=true]{hyperref} 
\usepackage{url}            
\usepackage{booktabs}       
\usepackage{amsfonts}       
\usepackage{nicefrac}       
\usepackage{microtype}      
\usepackage[T1]{fontenc}
\usepackage[table,dvipsnames]{xcolor}
\usepackage{mathtools, amsmath, amssymb, amsthm}
\usepackage{makecell}
\usepackage{graphicx} 
\usepackage{array}    
\usepackage{caption}  
\usepackage{wrapfig}
\usepackage{sidecap}
\usepackage{multirow}
\usepackage{balance}
\usepackage{algorithm}
\usepackage{algorithmic}
\newcommand{\algcmt}[1]{\hfill{\color{black!60}$\triangleright$ #1}}
\usepackage{fancyvrb}       
\DefineVerbatimEnvironment{shellcmd}{Verbatim}{fontsize=\small, frame=single, framesep=1.5mm, framerule=0.4pt, rulecolor=\color{black!35}, samepage=true}

\usepackage{bm}

\usepackage[capitalize,noabbrev]{cleveref}
\usepackage{pifont}

\theoremstyle{plain}

\theoremstyle{definition}

\theoremstyle{remark}

\definecolor{darkblue}{RGB}{0, 51, 153}   
\definecolor{darkgreen}{RGB}{0, 102, 51}  
\definecolor{darkred}{RGB}{153, 0, 0}   
\hypersetup{colorlinks=true,
  citecolor=darkblue,  
  linkcolor=darkred,   
  urlcolor=darkgreen,  
  pdftitle={Fold'EM: Direct atomic structure inference from raw Cryo-EM particles}}
\renewcommand*{\backref}[1]{}
\renewcommand*{\backrefalt}[4]{%
  \ifcase #1 \or {\footnotesize\textcolor{black!60}{(Cited on page~#2.)}}%
  \else {\footnotesize\textcolor{black!60}{(Cited on pages~#2.)}}\fi}

\renewcommand{\eqref}[1]{Equation~\ref{#1}}

\usepackage[textsize=tiny]{todonotes}

\newcommand{\ens}[1]{\bm{\mathcal{#1}}}

\newcommand{\brecovar}{B-RECOVAR}

\newcommand{\hd}[1]{\multicolumn{1}{c}{\boldmath\thead{#1}}}

\makeatletter
\let\foldem@AA\AA
\DeclareRobustCommand{\AA}{\ifmmode\text{\foldem@AA}\else\foldem@AA\fi}
\makeatother

\title{Fold'EM: Direct atomic structure inference from Cryo-EM particles}

\author{%
  Advaith Maddipatla\textsuperscript{1,*} \quad
  M\"{a}rt-Erik M\"{a}eots\textsuperscript{2,*} \quad
  Marco Pegoraro\textsuperscript{1,*} \quad
  Nikolaus Dr\"{a}ger\textsuperscript{1}\\
  Roberto Covino\textsuperscript{3} \quad
  Sanketh Vedula\textsuperscript{4,5,\dag} \quad
  Martin Pacesa\textsuperscript{2,\dag} \quad
  Alex Bronstein\textsuperscript{1,\dag}\\[0.5em]
  {\normalsize \textsuperscript{1}Institute of Science and Technology Austria \qquad \textsuperscript{2}University of Zurich}\\
  {\normalsize \textsuperscript{3}Frankfurt Institute for Advanced Studies \qquad \textsuperscript{4}Princeton University}\\
  {\normalsize \textsuperscript{5}Broad Institute of MIT and Harvard}
}
\date{\vspace{-1em}}

\begin{document}
\maketitle
\begingroup
\renewcommand{\thefootnote}{}
\footnotetext{\textsuperscript{*}Equally contributing first authors. \quad \textsuperscript{\dag}Equally contributing senior authors.}
\endgroup
\begin{abstract}
Single-particle cryo-electron microscopy (cryo-EM) has become a widely adopted technique for biomolecular structure determination. The conventional cryo-EM computational pipeline first combines many particle images to reconstruct an electrostatic potential (ESP) map and then fits an atomic model to the recovered map. Density reconstruction has high sample complexity, requiring large numbers of particle images and making structure determination high-cost and low-throughput, particularly for structurally heterogeneous samples. Downstream atomic model building, in turn, becomes increasingly difficult as the resolution of the reconstructed map deteriorates. Protein structure prediction models provide strong sequence-derived priors on atomic structure, and experiment-guided approaches can use these priors to recover structures consistent with experimental measurements. Yet, in cryo-EM, such priors are typically integrated only after density reconstruction during atomic model fitting. We introduce Fold'EM, an inference-time framework that combines priors from protein structure generative models directly with cryo-EM particle images to determine atomic models from a small number of single particle images, bypassing both intermediate density reconstruction and downstream model building against the reconstructed map. Across synthetic benchmarks and experimental cryo-EM data, Fold'EM recovers accurate atomic structures both with known particle orientations and in an \textit{ab initio} setting where orientations are inferred jointly with structure. In heterogeneous datasets, Fold'EM further resolves distinct conformational states from mixed particle populations without separately reconstructing a density map and building an atomic model for each state. We believe these results open new avenues for structure determination in the low-sample regime and for characterizing low-population conformational states directly from cryo-EM particles.
\end{abstract}
\section{Introduction}
Single-particle cryo-electron microscopy (cryo-EM) has transformed structural
biology by enabling high-resolution structure determination of large
macromolecular assemblies without crystallization
\citep{fernandez-leiro2016unravelling}. Its ability to image individual
molecules in different conformations also provides a route to understanding
how structural changes drive biological function
\citep{ourmazd2019cryo,maeots2022structural}. Yet the fundamental measurement
is extremely noisy. Radiation damage limits the electron dose that can be
applied to each particle, leaving individual images unsuitable for direct
structural interpretation and requiring large numbers of observations to be
aligned and combined into a three-dimensional electrostatic potential (ESP)
map \citep{scheres2012relion,kimanius2024data}. The resulting density must
then be interpreted as an atomic model, a step that can remain difficult
despite advances in automated model building, particularly where local
resolution is poor \citep{jamali2024automated,fadini2026alphafold}.

The demand for coherent averaging creates a tension between spatial
resolution and biological completeness. Combining particles from different
conformations blurs structural features, whereas separating them into
increasingly homogeneous populations leaves fewer observations available for
each state \citep{frank2016continuous}. Low-population and transient states
are especially difficult because they may be merged with more abundant
conformations, excluded during particle selection, or remain too sparsely
represented for independent reconstruction. For scarce samples or proteins
that cannot be obtained in sufficient quantity, the observations required to
reconstruct each state may simply not be available.

Heterogeneity modeling can recover conformational variability beyond a small
set of discrete structural classes
\citep{frank2016continuous,zhong2021cryodrgn,gilles2025cryo}. However, these
approaches generally require prior particle orientations, and thus remain
dependent on upstream particle selection and alignment. Improving
heterogeneous reconstruction therefore addresses only part of the difficulty
of obtaining mechanistic information from sparse, noisy observations, while
leaving unresolved the fundamental requirement for large numbers of
particles.

Conventional single-particle cryo-EM reconstruction pipelines primarily
estimate densities from image measurements, making limited use of prior
knowledge about protein structures in general, or the sequence of the sample
under study. Blush regularization in RELION demonstrates the value of
incorporating such knowledge by using a denoiser trained on experimental
cryo-EM maps to provide richer regularization than a generic smoothness
prior, improving reconstruction when particle signals are weak
\citep{kimanius2024data}.

AlphaFold and related prediction models encode knowledge of
sequence-dependent folds and interactions learned from evolutionary and
structural data, providing strong sequence-conditioned structural priors
\citep{jumper2021highly,baek2021accurate}. Building on these priors, a recent
line of work has shown that experimental guidance can steer these models
toward conformations consistent with the experimental data but missed by
unconditioned predictions
\citep{maddipatla2025inverse,Maddipatla2026,fadini2026alphafold,raghu2026multiscale}. In cryo-EM,
however, these approaches have focused on reconstructed maps, applying
structural knowledge after particle selection, alignment, and averaging have
already taken place. They therefore inherit the need for sufficiently many
particles to support reliable reconstruction. In this work, we ask:
\textit{can structural priors be introduced directly at the particle level,
enabling atomic inference from sparse cryo-EM observations without requiring
a high-resolution intermediate reconstruction?}

\subsection{Contributions and main results}

\begin{figure*}[t]
    \centering
    \includegraphics[width=1.\linewidth]{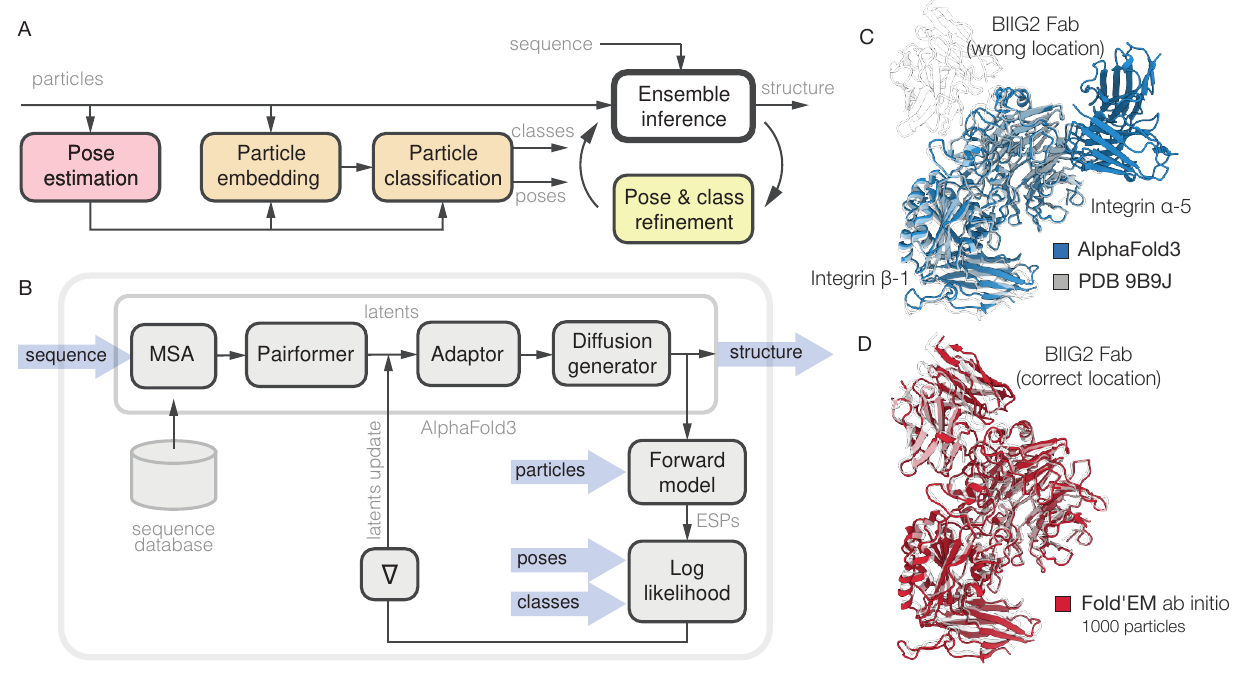}
    \caption{\textbf{Fold'EM infers atomic structures directly from sparse
    cryo-EM particle images.}
    (A) Fold'EM alternates structural inference with pose and class refinement.
    (B) Particle evidence is back-propagated into intermediate Pairformer states
    $\mathbf H$, filtering experimental updates through the remaining Pairformer
    blocks before AlphaFold3 diffusion.
    (C) For the integrin--BIIG2 Fab complex (PDB \texttt{9B9J}), AlphaFold3
    predicts the individual components but places the Fab at the wrong binding
    site.
    (D) \textit{Ab initio} Fold'EM recovers the bound complex and epitope from only
    $1{,}000$ particles---$0.26\%$ of those used for the deposited
    reconstruction---with RMSD $2.49\,\text{\AA}$, TM-score $0.956$, and FSC
    resolution $3.52\,\text{\AA}$ (deposited map: $2.6\,\text{\AA}$).}
    \label{fig:method}
\end{figure*}

We introduce \textbf{Fold'EM}, an inference-time framework that guides
protein structure generative models directly using cryo-EM particle images.
Rather than first reconstructing a three-dimensional density and subsequently
building an atomic model into it, Fold'EM uses sequence-conditioned
structural priors to infer atomic structures directly from sparse particle
observations. We formulate structure determination as joint inference over
the structural representation and particle-specific latent variables,
including pose and conformational assignment (Figure~\ref{fig:method}).

\paragraph{Particle vs.\ density guidance.}
We characterize when inference from individual particle images can be
replaced by inference from a reconstructed density. For a homogeneous sample
with known poses, we show that the two are equivalent only when density
mismatch is measured under the experiment-induced metric defined by the
imaging operators and noise model. The standard density-space $L_2$ objective
therefore does not, in general, reproduce the particle likelihood. When poses
or conformational assignments are uncertain, the distinction becomes more
fundamental, as collapsing particles into a consensus density can discard
particle-specific information required to infer these latent variables
(Figure~\ref{fig:coarse_map}). This provides the statistical motivation for
introducing structural priors directly at the particle level.

\paragraph{Improved experimental guidance.}
Existing approaches condition AlphaFold on experimental data either by
optimizing its terminal embedding before sampling
\citep{maddipatla2026inference,li2026robust}, or by directly guiding the
diffusion trajectory in coordinate space
\citep{maddipatla2025inverse,Maddipatla2026,raghu2026multiscale}. Fold'EM instead optimizes
intermediate representations within the Pairformer module. Passing
experimental updates through the remaining Pairformer blocks induces a
sequence- and state-dependent geometry in the terminal conditioning space,
constraining perturbations through transformations learned by the pretrained
model. We find that this reparameterization substantially strengthens
experimental guidance (Figure~\ref{fig:guidance_comp}) and is a key enabling
component of direct particle-level structure inference.

\paragraph{Low-sample structure inference.}
We first evaluate Fold'EM in the known-pose setting, where it recovers
accurate atomic structures on both experimental (Figure~\ref{fig:homo};
Tables~\ref{tab:real_res}--~\ref{tab:real_tm_rmsd_fraction}) and synthetic
(Tables~\ref{tab:syn_res_particles}--\ref{tab:syn_tm_rmsd_particles})
datasets using fewer than $1\%$ of the particles used for conventional
density reconstruction. Structures inferred directly from particle images
reach comparable structural resolution to those obtained through density
reconstruction, and on experimental homogeneous data from EMPIAR, as few as
$0.1\%$ of the particles are sufficient to reach comparable Fourier shell
correlation (FSC) and atomic RMSD. Heterogeneous results are shown in
Figures~\ref{fig:hetero} and~\ref{fig:synthetic_het},
Tables~\ref{tab:het_real_fsc0143}--\ref{tab:het_real_rmsd_tm}
and~\ref{tab:het_syn_fsc0143}--\ref{tab:het_syn_rmsd_tm}. In continuously
heterogeneous systems, we further find that the organization of
conformational states in particle space is reflected in the AlphaFold
embedding space, indicating that experimentally observed conformational
structure is preserved within the learned representation and allowing a
continuous trajectory to be represented by a small number of optimized anchor
states (Figure~\ref{fig:continuous-het}).

\paragraph{\textit{Ab initio} Fold'EM.}
We further propose \textit{ab initio} Fold'EM, a framework for structure
determination from cryo-EM particles with unknown poses and conformational
states. The problem is formulated as a joint inference of the atomic
structures, particle poses, and class assignments under the particle-level
likelihood and sequence-conditioned structural prior. We solve this as a
joint optimization problem by alternating between two stages:
(i) \textit{structure inference given the current poses and class
assignments}, which uses particle-level guidance to optimize the atomic
structures; (ii) \textit{pose and class inference given the inferred
structures}, which refines each particle's pose and class assignment under
the current structural ensemble. For low-sample heterogeneous data, we
further introduce a RECOVAR-inspired residual-subspace estimator,
\brecovar{}, that provides more robust particle assignments and confidence
estimates, which are used to prioritize reliable assignments during
heterogeneous structure inference (Figure~\ref{fig:brecovar}, and
Table~\ref{tab:confidence_tm_sep}). In experimental datasets with unknown
poses, \textit{ab initio} Fold'EM reaches comparable structural accuracy
using approximately $3\%$ of the particles used in the deposited
reconstruction (Figure~\ref{fig:homo},
Tables~\ref{tab:real_res_abinitio}--\ref{tab:abinitio_tm_rmsd_real}). In
heterogeneous data, it resolves distinct conformational states using
approximately $5\%$ of the particles (experimental: Figure~\ref{fig:hetero},
Tables~\ref{tab:het_real_fsc0143_abinitio}--\ref{tab:het_real_rmsd_tm_abinitio};
synthetic:
Tables~\ref{tab:het_syn_fsc0143_abinitio}--\ref{tab:het_syn_rmsd_tm_abinitio}).

Together, these results show that strong structural priors can substantially
reduce the particle counts needed for atomic and conformational inference and
suggest a route toward characterizing low-population states directly from
cryo-EM particle images.

\section{Problem formulation}
\label{sec:problem_formulation}

\paragraph{Notation.}
Let $\mathbf a$ denote the molecular sequence specification supplied to
AlphaFold3 and $\mathbf X=\{\mathbf x_1,\ldots,\mathbf x_m\}$ the resulting
3D structure, where $\mathbf x_i\in\mathbb R^3$ is the Cartesian coordinate
of atom $i$. We view AlphaFold3 as a sequence-conditioned latent model whose
structure prediction comprises two stages. The Pairformer first maps the
sequence information to a latent representation $\mathbf Z=f(\mathbf a)$,
which then conditions a diffusion generator producing atomic coordinates
$\mathbf X\sim p(\mathbf X\mid\mathbf Z)$.

The Pairformer consists of $N$ blocks, $f=f_N\circ\cdots\circ f_1$, acting on
single and pair representations
$\mathbf s\in\mathbb R^{|\mathbf a|\times c_s}$ and
$\mathbf z\in\mathbb R^{|\mathbf a|\times|\mathbf a|\times c_z}$ that
together form the latent state $\mathbf Z=(\mathbf s,\mathbf z)$. We split
the Pairformer trunk at depth $N-\ell$ as $f=g\circ h$, where
$h=f_{N-\ell}\circ\cdots\circ f_1$ produces an intermediate state
$\mathbf H=h(\mathbf a)$, and $g=f_N\circ\cdots\circ f_{N-\ell+1}$ denotes
the final $\ell$ blocks, so that $\mathbf Z=g(\mathbf H)$. Since every
Pairformer block preserves the tensor shapes of the single and pair
representations, $\mathbf H$ and $\mathbf Z$ belong to the same
representation space. We use $\ell=3$ throughout.

\paragraph{Inference objective.}
Let $\ens Y=\{\mathbf y_k\}_{k=1}^K$ denote a set of $K$ raw cryo-EM
particle images. Our goal is to infer an ensemble of $n$ structures
$\ens X=\{\mathbf X^1,\ldots,\mathbf X^n\}$ that explains the observed
particles while remaining faithful to the AlphaFold3 prior. We parameterize
these structures through a corresponding set of intermediate Pairformer
states $\ens H=\{\mathbf H^j\}_{j=1}^n$, $\ens Z=g(\ens H)$, and optimize
$\ens H$ directly at inference time. Each particle is associated with
nuisance variables $\boldsymbol\eta_k=(\boldsymbol\phi_k,c_k,\ldots)$, where
$\boldsymbol\phi_k$ denotes its pose and $c_k\in\{1,\ldots,n\}$ the
structural class from which it originates. The nuisance variables may also
include additional per-particle or shared acquisition parameters entering
the image-formation model, such as the contrast transfer function (CTF) or
other imaging parameters. We denote the complete set of nuisance variables by
$\boldsymbol\eta=\{\boldsymbol\eta_k\}_{k=1}^K$. In this work we assume all
imaging parameters other than pose and class to be known, suppress them from
the notation, and therefore infer only $\boldsymbol\phi_k$ and $c_k$.
Conditioned on the structures and nuisance variables, and assuming particles
are independent,
\[
p(\ens Y\mid\ens X,\boldsymbol\eta)
=
\prod_{k=1}^K
p(\mathbf y_k\mid\mathbf X^{c_k},\boldsymbol\phi_k),
\]
where $\mathbf X^{c_k}$ denotes the conformation corresponding to class label
$c_k$. This can be generalized to continuous heterogeneity. In this case,
$c_k$ becomes a vector of barycentric coordinates and the conformation is
decoded from the interpolated state
$\mathbf H(c_k)=\sum_j c_k^j\mathbf H^j$.
We relax the deterministic Pairformer state $\mathbf H_0=h(\mathbf a)$ into
a separable prior
$p(\ens H\mid\mathbf a)
=
\prod_{j=1}^n
\mathcal N(\mathbf H^j;\mathbf H_0,\sigma^2\mathbf I)$.
Treating the nuisance variables as unknowns, our inference problem can then
be written as joint MAP estimation of the marginal objective
\begin{equation}
\mathcal J(\ens H,\boldsymbol\eta)
=
\log
\mathbb E_{\ens X\sim p(\ens X\mid g(\ens H))}
p(\ens Y\mid\ens X,\boldsymbol\eta)
+
\log p(\ens H\mid\mathbf a)
+
\log p(\boldsymbol\eta),
\label{eq:exact_MAP}
\end{equation}
where
$p(\ens X\mid g(\ens H))=\prod_{j=1}^n p(\mathbf X^j\mid g(\mathbf H^j))$.
The term $p(\boldsymbol\eta)$ can encode priors over particle orientations,
translations, and class frequencies; in the absence of additional information
these may be taken as uninformative.

The stochastic AlphaFold generator makes direct optimization of
\eqref{eq:exact_MAP} inconvenient. Applying Jensen's inequality,
$\log\mathbb E p\geq\mathbb E\log p$, gives the tractable ELBO-like surrogate
\begin{equation}
(\ens H^\ast,\boldsymbol\eta^\ast)
=
\arg\max_{\ens H,\boldsymbol\eta}
\left[
\mathbb E_{\ens X\sim p(\ens X\mid g(\ens H))}
\log p(\ens Y\mid\ens X,\boldsymbol\eta)
+
\log p(\ens H\mid\mathbf a)
+
\log p(\boldsymbol\eta)
\right].
\label{eq:optim_problem}
\end{equation}

\paragraph{Optimizing $\ens H$ induces a learned geometry in $\ens Z$.}
Optimizing the intermediate Pairformer state $\ens H$ constrains the
optimization to representations of the form $\ens Z=g(\ens H)$, but more
importantly, it changes the local geometry of optimization in conditioning
space. Consider a single ensemble member $\mathbf Z=g(\mathbf H)$, let
$\mathbf J(\mathbf H)=\partial\mathbf Z/\partial\mathbf H$ be its Jacobian,
and let
$L(\ens Z)
=
\mathbb E_{\ens X\sim p(\ens X\mid\ens Z)}
\log p(\ens Y\mid\ens X,\boldsymbol\eta)$
denote the experimental data term in \eqref{eq:optim_problem}. By the chain
rule, $\nabla_{\mathbf H}L=\mathbf J^\top\nabla_{\mathbf Z}L$. Consequently,
a gradient-ascent step
$\Delta\mathbf H=\gamma\,\mathbf J^\top\nabla_{\mathbf Z}L$ induces, to first
order,
\[
\Delta\mathbf Z
=
g(\mathbf H+\Delta\mathbf H)-g(\mathbf H)
\approx
\gamma\,\mathbf J\mathbf J^\top\nabla_{\mathbf Z}L .
\]
Thus, Euclidean optimization in $\mathbf H$ corresponds locally to
preconditioned optimization in $\mathbf Z$, with positive-semidefinite
preconditioner $\mathbf J\mathbf J^\top$. Directions in conditioning space to
which the final Pairformer blocks are highly sensitive are amplified, whereas
directions to which they are insensitive are suppressed. The resulting local
geometry is learned by the Pairformer itself and depends on both the sequence
and the current representation.

The same interpretation applies to the prior. We place an isotropic Gaussian
anchor on the precursor representation,
$p(\mathbf H\mid\mathbf a)=\mathcal N(\mathbf H_0,\sigma^2\mathbf I)$, where
$\mathbf H_0=h(\mathbf a)$. Let $\mathbf J_0=\mathbf J(\mathbf H_0)$ and
$\mathbf Z_0=g(\mathbf H_0)$. Linearizing $g$ around $\mathbf H_0$ gives
approximately
$p(\mathbf Z\mid\mathbf a)
=
\mathcal N(\mathbf Z_0,\sigma^2\mathbf J_0\mathbf J_0^\top)$.
Hence, an isotropic relaxation of the precursor state becomes a
sequence-dependent anisotropic relaxation of the terminal Pairformer
representation. In contrast, treating $\mathbf Z$ as a free optimization
variable with an isotropic prior assumes the same local geometry in every
direction of the conditioning space.

Intuitively, perturbations of $\mathbf H$ are filtered through the final
Pairformer blocks before they affect the structure generator. They are
therefore expressed through transformations that the trained trunk itself has
learned to realize, rather than as arbitrary perturbations of its terminal
representation. Figure~\ref{fig:guidance_comp} and
Section~\ref{sec:guidance_comp} empirically demonstrate this.

\paragraph{Advantages of particle-domain fitting.}
Our formulation operates directly on particle-level likelihoods rather than
first averaging the particles into a reconstructed density. This distinction
has two consequences. First, when the imaging operators are known,
particle-level fitting is equivalent to density fitting only under an
experiment-specific metric determined by the imaging operators and noise
covariance. Second, when poses are uncertain, the particles generally cannot
be reduced to a single sufficient density statistic.

Consider the linearized image formation model
$\mathbf y_k=\mathbf A_k\rho+\boldsymbol\varepsilon_k$,
$\boldsymbol\varepsilon_k\sim\mathcal N(\mathbf0,\boldsymbol\Sigma_k)$, where
$\mathbf A_k$ are known operators encapsulating the pose, CTF, and the
sampling of a particle $k$, and $\rho$ is the underlying 3D density. When all
imaging operators $\mathbf A_k$ are known exactly, no information is lost by
first reconstructing a density---provided that the subsequent structure fit
uses the correct experiment-induced metric. Indeed, defining
$\displaystyle{\mathbf Q=\sum_k\mathbf A_k^\ast\boldsymbol\Sigma_k^{-1}\mathbf A_k}$,
$\displaystyle{\mathbf q=\sum_k\mathbf A_k^\ast\boldsymbol\Sigma_k^{-1}\mathbf y_k}$,
and letting $\tilde\rho=\mathbf Q^\dagger\mathbf q$ denote the reconstructed
density, the particle-level data term satisfies
\[
\frac12\sum_k
\|\mathbf y_k-\mathbf A_k\rho(\mathbf H)\|_{\boldsymbol\Sigma_k^{-1}}^2
=
\frac12
\|\rho(\mathbf H)-\tilde\rho\|_{\mathbf Q}^2+\mathrm{const},
\]
where $\rho(\mathbf H)$ is the density generated by the structural latent
$\mathbf H$ being solved for. Thus, with known poses, fitting a structural
model directly to all particle likelihoods is equivalent to first
reconstructing a density and fitting to that density, but only when the
latter uses the experiment-induced metric $\mathbf Q$.

Importantly, the correct metric is generally not the ordinary density-space
$L_2$ norm. Even for equal white noise and uniformly distributed viewing
directions, the idealized normal operator is isotropic but frequency
dependent, satisfying
$\widehat{\mathbf Q\rho}(\boldsymbol\xi)
\propto
\|\boldsymbol\xi\|^{-1}\hat\rho(\boldsymbol\xi)$ in the Fourier domain, with
$\|\boldsymbol\xi\|$ denoting radial spatial frequency, or equivalently
$\mathbf Q\propto(-\Delta)^{-1/2}$, with $\Delta$ denoting the Laplacian,
very far from $\mathbf Q=\mathbf I$. Consequently, guidance against a
reconstructed density using a Euclidean map loss, as done in recent works
\citep{Maddipatla2026,raghu2026multiscale}, \textit{does not} in general
reproduce the original particle likelihood, even when the poses are known;
equivalence requires the density-space loss to incorporate the
experiment-specific metric $\mathbf Q$.

\paragraph{Particle averaging under uncertainty.}
The distinction becomes more fundamental when poses or classes are
uncertain. In this case the imaging operators themselves depend on latent
variables, $\mathbf A_k=\mathbf A(\boldsymbol\phi_k)$, and inference depends
on the individual residuals
$\mathbf y_k-\mathbf A(\boldsymbol\phi_k)\rho(\mathbf H)$. Collapsing the
particles into a single density at nominal or estimated poses generally
destroys information required to update the $\boldsymbol\phi_k$. A simple
scalar analogue illustrates this. If $y_k=\phi_k\rho+\varepsilon_k$, with
$\phi_k\sim\mathcal N(1,\tau^2)$ and
$\varepsilon_k\sim\mathcal N(0,\sigma^2)$, then after marginalizing the
uncertain pose one obtains
$y_k\mid\rho\sim\mathcal N(\rho,\sigma^2+\tau^2\rho^2)$. The likelihood
therefore depends not only on $\sum_k y_k$ but also on $\sum_k y_k^2$; the
sample mean retains the former but not the latter and is therefore not a
sufficient statistic.

Particle-level inference therefore has two advantages in our setting. With
known poses, it preserves the statistically correct image-space metric
without requiring it to be approximated in density space. With uncertain
poses or class assignments, it additionally retains particle-specific
information that can be lost when the observations are first collapsed into
a reconstructed map. This motivates structure inference directly against the
individual particle likelihoods while refining nuisance variables, rather
than reconstructing an intermediate density and guiding the structural model
against that density.

\section{Methods}
\label{sec:methods}

Fold'EM jointly infers an ensemble of atomic structures and particle-specific
nuisance variables directly from raw cryo-EM particle images. Each ensemble
member is represented by an intermediate Pairformer state that conditions the
AlphaFold3 diffusion model, and the nuisance variables of a particle are its
pose and, in a heterogeneous sample, its discrete class. Fold'EM first places
the particles in a common reference frame and, for heterogeneous data,
assigns initial classes; it then alternates between updating the structural
states given the particle images and refining the nuisance variables under
the current ensemble. Reconstructed volumes, when used, serve only for
initialization and class-specific frame alignment; structural guidance itself
operates on individual particles. In what follows, we describe the
ingredients of this solver of \eqref{eq:optim_problem}, schematically
depicted in Figure~\ref{fig:method} and summarized in
Algorithm~\ref{alg:FoldEM_alg}.

\subsection{Inputs, initialization, and reference frame}
\label{sec:initialization}

\paragraph{Inputs.}
The inputs to Fold'EM are a collection of particle images and the molecular
sequence specification $\mathbf a$ of the target. We assume knowledge of the
pixel/box size and CTF parameters from preprocessing; additionally, the
particles are assumed to be centered. The sequence is the only structural
input. AlphaFold3 constructs its MSA and unguided prediction from it, and no
deposited atomic model or high-resolution experimental reconstruction is
supplied. For heterogeneous data the number of classes $n$ is specified; for
homogeneous data, $n=1$. Poses and class labels are provided only in the
\textit{known-pose} and \textit{known-class} evaluations, in which the poses
are those of the deposited consensus refinement.

\paragraph{Pose estimation.}
In the \textit{ab initio} setting, poses are estimated in three stages:
(i) VDAM, a variable-metric gradient descent algorithm with adaptive moments
estimation \citep{kimanius2021new}, jointly estimates a coarse volume and
particle poses from random initialization by marginalizing over poses;
(ii) a consensus density is reconstructed from these poses by Wiener
back-projection; and
(iii) an AlphaFold3 prediction is docked into this density by $L_1$ map
guidance to define a structural reference frame.
Since the cryo-EM likelihood is invariant to a global rigid transformation of
the structure and the corresponding inverse transformation of the poses, the
solution is defined only up to a global $\mathrm{SE}(3)$ transformation. We
fix this gauge by expressing all particle poses relative to the docked
structure before particle classification and structural inference.

\paragraph{Particle embedding and classification.}
When class labels are unknown, we infer them directly from the particle
images, without state-specific reference structures or atomic models. We
consider cryoDRGN \citep{zhong2021cryodrgn}, which embeds particles using a
learned latent representation, and RECOVAR \citep{gilles2025cryo}, which
embeds them by projection onto principal components of the estimated volume
covariance. In the low-particle regime, however, experimental nuisance
variation can be comparable to the structural differences of interest, and
we find that these embeddings do not reliably separate classes.

We therefore introduce \brecovar{}, a bootstrapped residual-subspace
estimator inspired by RECOVAR. For the two-state case, it
(i) fits class-deviation volumes to residuals from a consensus
reconstruction, pooled through the current poses and restricted by a solvent
mask;
(ii) refines assignments by cross-validated comparison against per-class
reconstructions; and
(iii) combines the assignments from randomized subsamples and frequency bands
using a label-free spectral meta-learner \citep{parisi2014ranking}.
For more than two states, successive modes are extracted by deflation; for
continuous heterogeneity, the same modes are interpreted as variability
directions. The ensemble vote margin defines a per-particle confidence score,
which is used to prioritize reliable assignments during structural
inference.

After classification, we reconstruct one auxiliary consensus volume per
class, used only to establish its reference frame. We dock the reference
structure into each volume as in the pose-estimation stage and re-synchronize
particle poses to the corresponding class reference. The resulting classified
particles, reference ensemble, and poses are then passed to structural
inference. The full procedure is given in Algorithm~\ref{alg:residual_mode}
and Appendix~\ref{app:class_assignments}, which also details its relation to
RECOVAR.

\subsection{Ensemble inference}

The inputs to structure inference are the particles, the reference
structure, and the poses expressed with respect to it; for a heterogeneous
sample, they are the classified particles, the ensemble of references, and
the poses re-synchronized to their classes.
Equation~\ref{eq:optim_problem} couples the structural states $\ens H$ with
the particle-specific nuisance variables $\boldsymbol\eta$, and we
approximately optimize the two blocks by alternating coordinate ascent. Given
the current nuisance estimates, we infer structures by an IT-Opt-style
update \citep{maddipatla2026inference}; given the resulting structures, we
refine the nuisance variables. Depending on the evaluation setting, known
poses or classes are held fixed rather than updated. The full procedure is
summarized in Algorithm~\ref{alg:FoldEM_alg}.

\paragraph{Step 1: structure inference given nuisance variables.}
At outer iteration $r$, we hold the current nuisance estimates
$\boldsymbol\eta^{(r)}
=\{(\boldsymbol\phi_k^{(r)},c_k^{(r)})\}_{k=1}^K$
fixed and update the structural latent variables as
\begin{equation}
\ens H^{(r+1)}
\approx
\arg\max_{\ens H}
\left[
\mathbb E_{\ens X\sim p(\ens X\mid g(\ens H))}
\log p(\ens Y\mid\ens X,\boldsymbol\eta^{(r)})
+
\log p(\ens H\mid\mathbf a)
\right].
\label{eq:structure_step}
\end{equation}
For fixed nuisance variables, the likelihood factorizes over particles as
\[
\log p(\ens Y\mid\ens X,\boldsymbol\eta^{(r)})
=
\sum_{k=1}^K
\log p(
\mathbf y_k\mid
\mathbf X^{c_k^{(r)}},
\boldsymbol\phi_k^{(r)}).
\]
Hence each ensemble member is guided by the particles currently assigned to
it, projected according to their current pose estimates. The implemented
particle likelihood and image formation model are described in
Appendix~\ref{sec:image-formation-model}.

We solve \eqref{eq:structure_step} by an IT-Opt-style inference-time update
\citep{maddipatla2026inference} that back-propagates likelihood gradients
into AlphaFold3 latents, but optimize the intermediate state $\ens H$ rather
than the terminal Pairformer representation. AlphaFold3
\citep{abramson2024accurate} is a diffusion-based generative model
\citep{ho2020denoising,karras2022elucidating}. Let $\mathcal R_\theta$ denote
one step of its pretrained reverse diffusion process,
\begin{equation}
\mathbf X_{t-1}
=
\mathcal R_\theta(
\mathbf X_t;g(\mathbf H),t,\boldsymbol n_t), \quad
\boldsymbol n_t\sim\mathcal N(\mathbf0,\mathbf I),
\label{eq:sde}
\end{equation}
and let $D_\theta$ denote its current denoised structure estimate.
Integrating the reverse process from
$\mathbf X_T\sim\mathcal N(\mathbf 0,\sigma_T^2\mathbf I)$ to $t=0$
generates a structure conditioned on $g(\mathbf H)$.

For each structural update we perform $Q$ independent reverse-diffusion
runs, retaining the optimized $\ens H$ between runs. At run $q$, we
initialize each ensemble member independently as
$\mathbf X_T^{q,j}\sim\mathcal N(\mathbf 0,\sigma_T^2\mathbf I)$ for
$j=1,\ldots,n$. Resampling the diffusion noise while retaining the optimized
Pairformer states encourages $\ens H$ to explain the experimental
observations across different stochastic realizations of the AlphaFold
generator.

At every reverse-diffusion step $t$, the current denoised structures are used
to update the Pairformer states by gradient ascent,
\begin{equation}
\ens H^{q}_{t-1}
=
\ens H^{q}_{t}
+
\gamma_h
\nabla_{\ens H^{q}_{t}}
\left[
\log p\!\left(
\ens Y\mid
D_\theta(\ens X^{q}_{t};g(\ens H^{q}_{t}),t),
\boldsymbol\eta^{(r)}
\right)
+
\lambda_p\log p(\ens H^{q}_{t}\mid\mathbf a)
\right],
\label{eq:emb_update}
\end{equation}
where $\gamma_h$ denotes the step size. The updated latent states are then
used to condition the next reverse-diffusion step. For ensemble member $j$,
this amounts to conditioning $\mathcal R_\theta$ in \eqref{eq:sde} on
$g(\mathbf H^{q,j}_{t-1})$. At the end of each reverse process,
$\ens H^{q}_{0}$ initializes the next run. Thus, $\ens H$ acts as persistent
memory across stochastic diffusion samples, progressively biasing the
AlphaFold generator toward structures supported by the experimental
particles. After the final run, $g(\ens H^{(r+1)})$ is decoded to obtain the
updated structural ensemble $\ens X^{(r+1)}$.

\paragraph{Step 2: nuisance inference given structure.}
We next hold the structural ensemble $\ens X^{(r+1)}$ fixed and update the
nuisance variables as
\begin{equation}
\boldsymbol\eta^{(r+1)}
=
\arg\max_{\boldsymbol\eta}
\left[
\log p(
\ens Y\mid\ens X^{(r+1)},\boldsymbol\eta)
+
\log p(\boldsymbol\eta)
\right].
\label{eq:nuisance_step}
\end{equation}
In this work, the unknown nuisance variables are particle pose and
structural class. Assuming their priors factorize across particles, the
update can be performed independently for each particle as
\[
(\boldsymbol\phi_k^{(r+1)},c_k^{(r+1)})
=
\arg\max_{\boldsymbol\phi,c}
\left[
\log p(
\mathbf y_k\mid
\mathbf X^{c,(r+1)},\boldsymbol\phi)
+
\log p(\boldsymbol\phi,c)
\right].
\]
Pose refinement therefore aligns each particle against the currently
inferred structures, while class refinement assigns it to the ensemble member
that best explains its image. Poses remain expressed in the docked reference
frame, fixing the otherwise arbitrary global $\mathrm{SE}(3)$ gauge.

The updated nuisance variables are then returned to
\eqref{eq:structure_step}, and the two coordinate steps are alternated. This
coupling is particularly important in the low-signal regime. Inaccurate poses
or class assignments can produce misleading structural gradients, while an
inaccurate structural model can in turn bias particle alignment and
classification.

If both coordinate steps were solved exactly for a deterministic objective,
each iteration would be non-decreasing in \eqref{eq:optim_problem} and
convergence would yield a coordinate-wise stationary solution. In practice,
the structural step uses stochastic diffusion samples, minibatches of
particles, and finite gradient updates, so monotonic improvement is not
guaranteed at every iteration. We therefore use the coordinate-ascent
interpretation primarily as the organizing principle for joint structure and
nuisance inference.

\section{Experiments}
\label{sec:experiments}
\suppressfloats[t] 

We evaluate Fold'EM on synthetic particles rendered from deposited structures
and on experimental particle stacks from EMPIAR (Tables~\ref{tab:synthetic}
and~\ref{tab:real}). Sections~\ref{sec:exp_particle}
and~\ref{sec:guidance_comp} test in isolation the two design choices of
Sections~\ref{sec:problem_formulation} and~\ref{sec:methods}, namely guidance
by the particles rather than by a reconstructed density, and optimization of
the intermediate Pairformer state rather than of the terminal embedding or
the atomic coordinates. Sections~\ref{sec:exp_homo}
and~\ref{sec:exp_discrete} evaluate structure determination for homogeneous
samples and for two-state mixtures, in each case first with known poses,
which tests structure inference alone, and then \textit{ab initio}, with
poses and classes inferred from the particles as described in
Section~\ref{sec:initialization}. In the known-pose setting the poses are
those used to render the synthetic particles or, for experimental data, those
of the deposited consensus refinement. Section~\ref{sec:exp_brecovar}
evaluates the label-free class assignment by \brecovar{} that initializes
the heterogeneous experiments, and Section~\ref{sec:exp_continuous} treats
continuous heterogeneity. Unless stated otherwise, we report the C$\alpha$ RMSD and
TM-score against the deposited model, and the resolution at which the Fourier
shell correlation (FSC) against the deposited map crosses $0.143$
(Appendix~\ref{app:metrics}); for synthetic data the deposited model is the
ground truth from which the particles were rendered. The density-based
baselines are RELION \citep{zivanov2018new} and cryoDRGN
\citep{zhong2021cryodrgn} with known poses, and RELION \textit{ab initio} and
cryoDRGN-AI \citep{levy2025cryodrgn} without them
(Appendix~\ref{app:baselines}). We also report the unguided AlphaFold3 prediction, which
represents the structure obtained without experimental conditioning and
serves as a control for its effect.

\subsection{Particle guidance vs.\ density guidance}
\label{sec:exp_particle}

\begin{figure*}[p]
    \centering
    \includegraphics[width=0.9\linewidth]{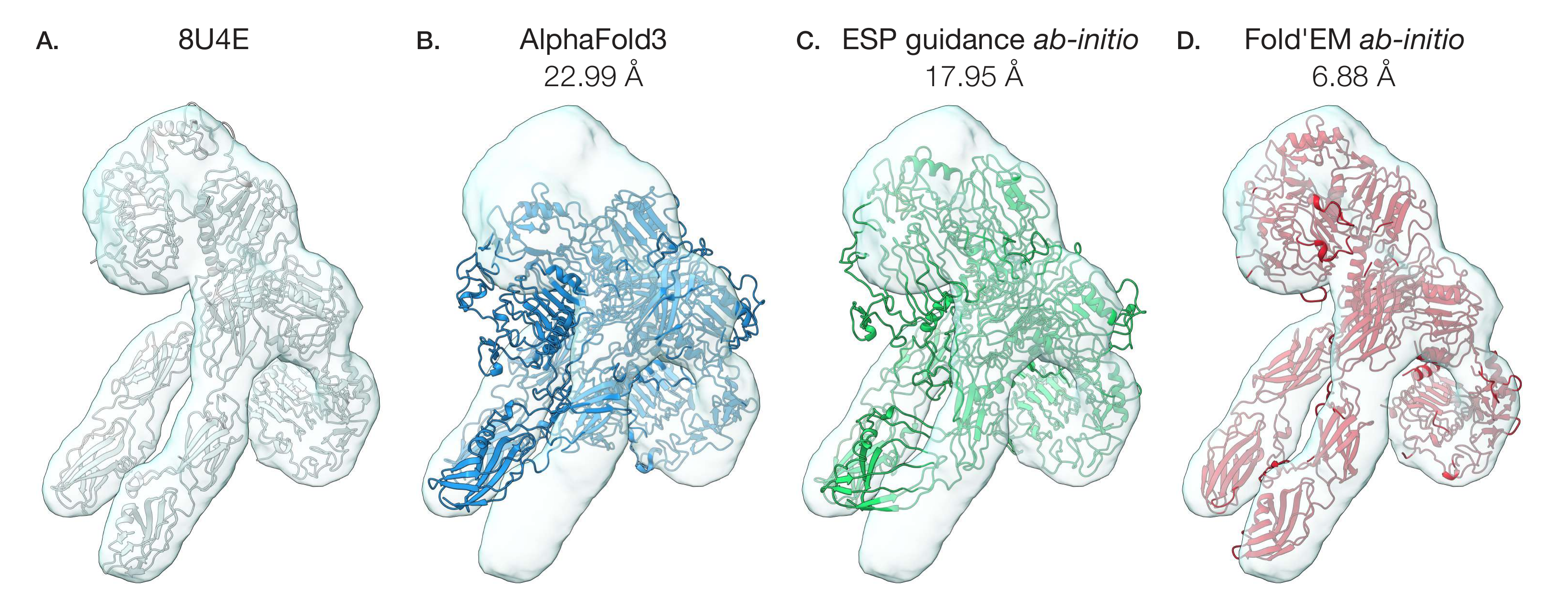}

\caption{\textbf{Guiding with the particles recovers the structure, whereas guiding with their reconstructed ESP fails.} \textit{Ab initio} inference on synthetic \texttt{8U4E} ($3{,}304$ residues) from $500$ particles at SNR $0.005$. The grey surface is the consensus density reconstructed from these particles at the estimated poses, and the numbers are C$\alpha$ RMSD to the deposited model. (A) The deposited model. (B) AlphaFold3 misplaces the domains of the assembly. (C) ESP guidance optimizes the same Pairformer state with the same particles, poses and optimizer, but replaces the particle likelihood by a loss against the consensus density, and improves the prediction without recovering the arrangement. (D) Fold'EM, guided by the particle likelihood, recovers it, with a TM-score of $0.81$ against $0.48$ for AlphaFold3. As argued in Section~\ref{sec:problem_formulation}, a density loss reproduces the particle likelihood only under the experiment-induced metric, and with estimated poses the consensus density is not a sufficient statistic for the particles, so the information needed to refine the poses together with the istructure is lost.}

    \label{fig:coarse_map}

    \vspace{1.5em}

    \centering
    \includegraphics[width=0.88\linewidth]{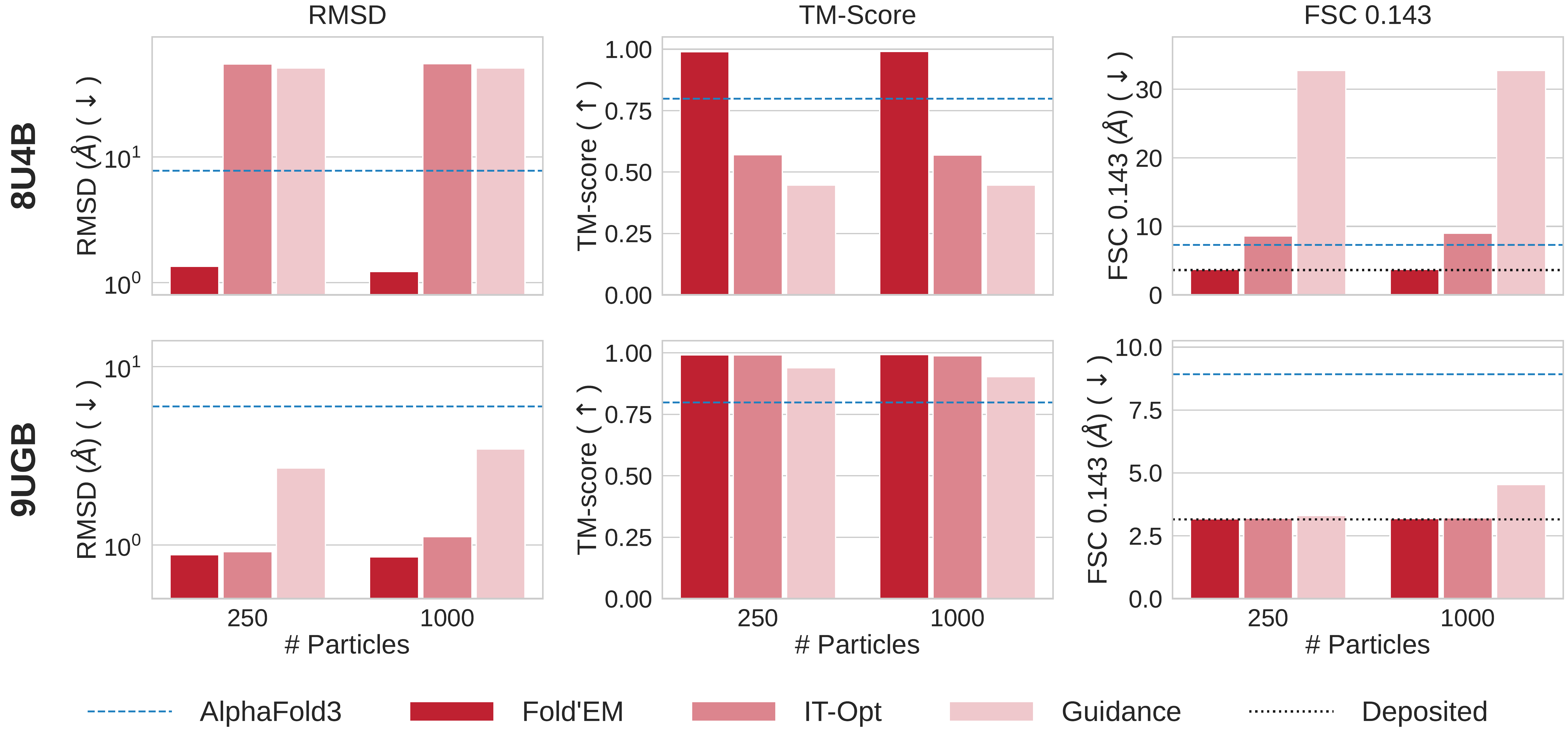}
    \vspace{-10pt}
    \caption{\textbf{Optimizing the intermediate Pairformer state improves particle-level guidance.} We compare three ways of steering AlphaFold3 with the same particle likelihood, poses and particle subsets: Fold'EM, which optimizes the intermediate Pairformer state $\ens{H}$ and passes the updates through the final $\ell$ Pairformer blocks (Section~\ref{sec:problem_formulation}); IT-Opt \citep{maddipatla2026inference}, which optimizes the terminal embedding $\ens{Z}$ directly under an isotropic prior; and coordinate-space guidance of the reverse diffusion process \citep{Maddipatla2026}. \textit{Rows:} synthetic \texttt{8U4B} and \texttt{9UGB} from Table~\ref{tab:synthetic} at SNR $0.005$ with known poses, using $250$ and $1000$ particles. \textit{Columns:} C$\alpha$-RMSD ($\downarrow$, log scale) and TM-score ($\uparrow$) against the deposited PDB model; the dashed line marks the AlphaFold3 prediction. On both targets Fold'EM improves substantially over both alternatives and reaches near-atomic agreement with the deposited model; the learned geometry induced by optimizing $\ens{H}$ thus keeps particle-level guidance stable at low SNR and on large assemblies.}
    \label{fig:guidance_comp}
\end{figure*}

\paragraph{Experimental setup.}
We first ask whether guidance by the individual particles recovers more of
the structure than guidance by the density reconstructed from them, at equal
particle count. To evaluate this, we take the synthetic \texttt{8U4E}
assembly, the largest target in Table~\ref{tab:synthetic} ($3{,}304$
residues), and perform \textit{ab initio} inference from $500$ particles at
SNR $0.005$. A consensus density is reconstructed from these particles at the
estimated poses. Density guidance, the baseline, optimizes the same Pairformer
state with the same particles, poses and optimizer as Fold'EM, but replaces
the particle likelihood by a loss against this consensus density. The two
settings therefore differ only in the data term.

\paragraph{Results.}
At equal particle count, particle guidance recovers the assembly and density
guidance does not. The unguided AlphaFold3 prediction misplaces the domains
of \texttt{8U4E} (C$\alpha$ RMSD $22.99\,\text{\AA}$). Density guidance
reduces this only to $17.95\,\text{\AA}$, whereas Fold'EM, guided by the same
particles while refining their poses, reaches $6.88\,\text{\AA}$ with
TM-score $0.809$ (Table~\ref{tab:abinitio_tm_rmsd_syn}) and is the only one
whose domains follow the envelope of the density
(Figure~\ref{fig:coarse_map}). As argued in
Section~\ref{sec:problem_formulation}, a density loss reproduces the particle
likelihood only for known poses and under the experiment-induced metric
$\mathbf Q$, and with estimated poses the consensus density is not a
sufficient statistic for the particles. At $500$ particles the density is
coarse and formed at uncertain poses; it serves to initialize the reference
frame (Section~\ref{sec:initialization}) but cannot replace the particles as
the data term.

\subsection{Optimizing \texorpdfstring{$\ens H$}{H} vs.\ \texorpdfstring{$\ens Z$}{Z} vs.\ coordinate-space guidance}
\label{sec:guidance_comp}

\paragraph{Experimental setup.}
We next ask whether optimizing the intermediate Pairformer state $\ens H$
gives stronger particle guidance than optimizing the terminal embedding
$\ens Z$ or guiding the atomic coordinates directly. To evaluate this, we
take synthetic \texttt{9UGB} and \texttt{8U4B} at SNR $0.005$ with $250$ or
$1{,}000$ particles and known poses, so that differences reflect
the guidance mechanism rather than pose estimation. We compare Fold'EM
with IT-Opt \citep{maddipatla2026inference}, which optimizes $\ens Z$ under
an isotropic prior (see also \citealp{li2026robust}), and with
coordinate-space guidance of the reverse diffusion process
\citep{Maddipatla2026}. The three schemes use the same particle likelihood,
poses and particle subsets, and differ only in the variable that the
likelihood gradient updates.

\paragraph{Results.}
Optimizing $\ens H$ gives the strongest guidance of the three. All three
schemes correct the AlphaFold3 prediction of \texttt{9UGB}, and Fold'EM
attains the lowest C$\alpha$ RMSD, highest TM-score and finest FSC
resolution. On the larger \texttt{8U4B} assembly, Fold'EM recovers
the deposited structure, whereas IT-Opt and coordinate-space guidance degrade
the initial prediction (Figure~\ref{fig:guidance_comp}). As argued in
Section~\ref{sec:problem_formulation}, the update of \eqref{eq:emb_update} is
locally a gradient ascent in $\mathbf Z$ preconditioned by
$\mathbf J\mathbf J^\top$, under a sequence-dependent anisotropic prior on
$\mathbf Z$. The separation is largest on the larger assembly, where
unfiltered updates of $\ens Z$ or of the coordinates carry the prediction
away from the deposited model.

\subsection{Guidance on homogeneous samples}
\label{sec:exp_homo}

\begin{figure*}[p]
    \centering
    \includegraphics[width=\linewidth]{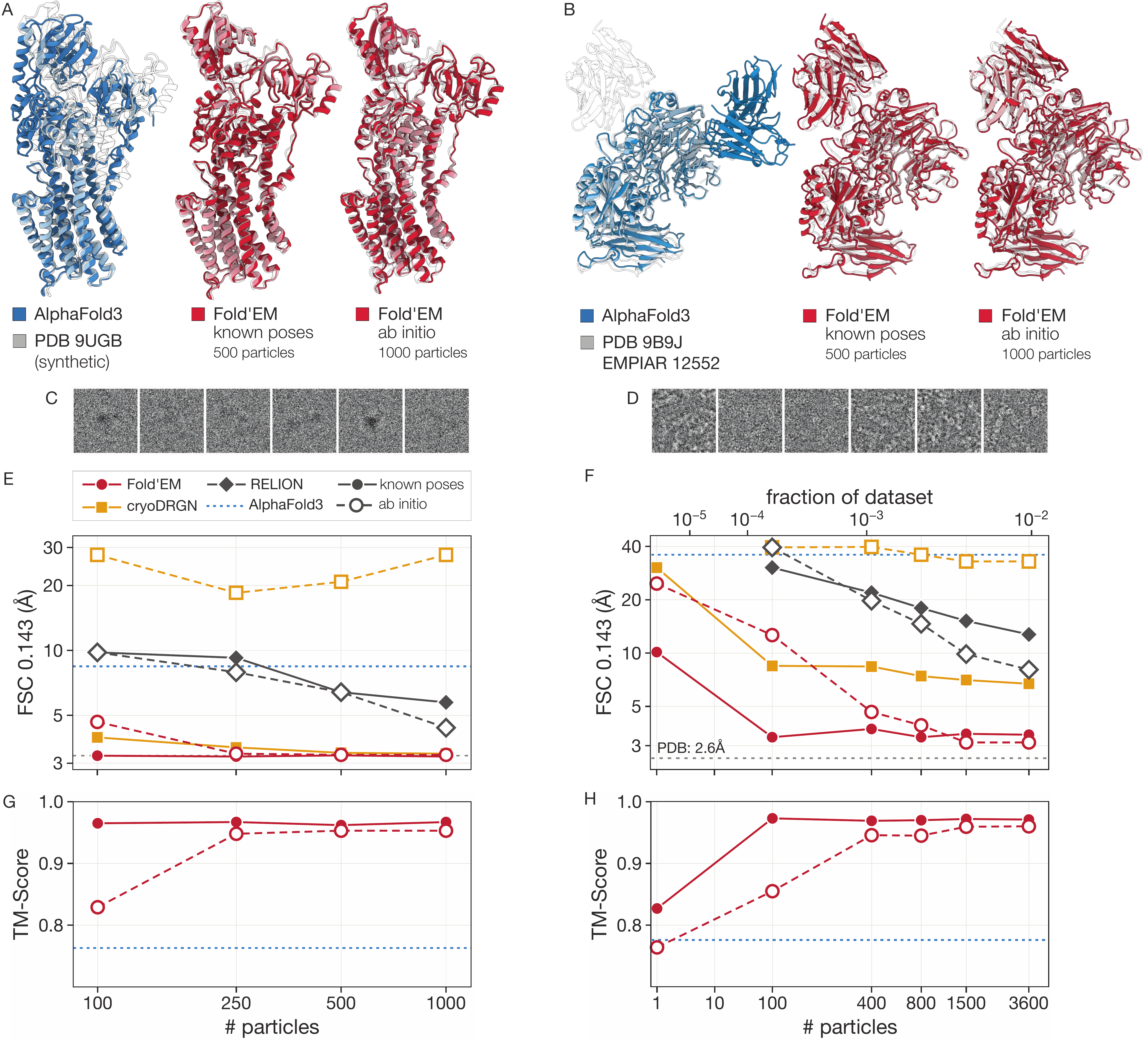}
    \caption{\textbf{Fold'EM recovers atomic structures from $0.03\%$ of the deposited particles with known poses and from $0.27\%$ \textit{ab initio}.}
Each column is one target: a synthetic dataset on the left (A, C, E, G) and an experimental one on the right (B, D, F, H). Rows show the structures (A, B; AlphaFold3 in blue, Fold'EM in red, deposited model in grey), representative particles (C, D), FSC resolution (E, F) and TM-score against the deposited model (G, H) as a function of the number of particles; filled markers denote known poses, open markers \textit{ab initio}, and the dotted line the AlphaFold3 prediction.
\textit{Left:} Pma1, the yeast plasma-membrane H$^+$-ATPase, in its activated (E2P) state (\texttt{9UGB}); synthetic particles at SNR $0.005$. AlphaFold3 predicts the transmembrane domain but misplaces the cytosolic A-, N- and P-domains, whose arrangement changes along the catalytic cycle (TM-score $0.70$). Fold'EM recovers the deposited state, reaching $3.3\,\text{\AA}$ and a TM-score of $0.97$ from $100$ particles with known poses, and $3.3\,\text{\AA}$ from $500$ particles \textit{ab initio}.
\textit{Right:} Integrin $\alpha_5 \beta_1$--BIIG2 (\texttt{9B9J}, EMPIAR-12552), where AlphaFold3 predicts both component folds but places the Fab at the wrong site (TM-score $0.78$); antibody--antigen pairs provide little paired coevolutionary information. Fold'EM recovers the bound complex and epitope, although not the interface variable loops, reaching $3.5\,\text{\AA}$ and a TM-score of $0.97$ from $100$ particles with known poses and $3.5\,\text{\AA}$ from $1{,}000$ particles \textit{ab initio}; no baseline reaches $6\,\text{\AA}$, and Fold'EM improves over AlphaFold3 even at the smallest particle counts.
Sparse particle evidence therefore corrects the global arrangement largely supplied by the structural prior. Resolution ($\text{\AA}$; FSC $=0.143$).}
    \label{fig:homo}
\end{figure*}

\paragraph{Experimental setup.}
Having fixed the guidance mechanism, we ask how few particles suffice to
determine the structure of a single conformational state, first with known
poses and then \textit{ab initio}. To evaluate this, we take the synthetic
targets \texttt{9UGB}, \texttt{9UGC}, \texttt{8U4B} and \texttt{8U4E}
(Table~\ref{tab:synthetic}), rendered at SNR $0.01$ and $0.005$, with $100$
to $1{,}000$ particles and known poses, or $500$ particles \textit{ab
initio}. We further take the five experimental datasets of
Table~\ref{tab:real}, namely integrin $\alpha_5\beta_1$ bound to the BIIG2 Fab
(\texttt{9B9J}), \texttt{9B77}, \texttt{8PWH}, and the two UvrA states
\texttt{9GXQ} and \texttt{9GXR}, each with its own deposited particle set.
With known poses we use between $1/1000$ and $1/50$ of the deposited
particles; \textit{ab initio} we use $1{,}000$ particles for \texttt{9B9J}
and \texttt{9B77} and $5{,}000$ for the remaining three. The baselines
reconstruct a density from the same particles. For the
baselines the FSC is computed from the reconstructed map, and for Fold'EM
from the map rendered from its atomic model; the deposited model is reported
for reference.

\paragraph{Results.}
On the synthetic targets, $100$ particles with known poses suffice.
AlphaFold3 mispredicts the Pma1 conformation \texttt{9UGB} (TM-score
$0.699$, C$\alpha$ RMSD $8.96\,\text{\AA}$; Figure~\ref{fig:homo}A). From
$100$ particles at SNR $0.005$, Fold'EM reaches $3.25\,\text{\AA}$ FSC
resolution, TM-score $0.965$ and C$\alpha$ RMSD $1.01\,\text{\AA}$,
comparable to the deposited model ($3.39\,\text{\AA}$;
Tables~\ref{tab:syn_res_particles} and~\ref{tab:syn_tm_rmsd_particles}), and
$3.45$, $3.80$ and
$4.18\,\text{\AA}$ on \texttt{9UGC}, \texttt{8U4B} and \texttt{8U4E}, with
TM-scores $0.983$, $0.979$ and $0.973$. cryoDRGN remains coarser than
$5.8\,\text{\AA}$ on the two larger targets at every particle count, and
RELION coarser than $5\,\text{\AA}$ on every target. Increasing the particle
count to $1{,}000$ alters the Fold'EM results little, consistent with the
prior supplying most of the molecular detail while the particles select the
realized conformation. \textit{Ab initio}, $500$ particles give comparable
accuracy on \texttt{9UGB}, $3.25\,\text{\AA}$ at SNR $0.005$ with TM-score
$0.976$ and C$\alpha$ RMSD $1.57\,\text{\AA}$, against $8.93$ and
$27.70\,\text{\AA}$ for RELION \textit{ab initio} and cryoDRGN-AI
(Tables~\ref{tab:syn_res_abinitio} and~\ref{tab:abinitio_tm_rmsd_syn}).
\texttt{8U4B} and \texttt{8U4E} are resolved less well \textit{ab initio}
($5.87$ and $11.38\,\text{\AA}$ at SNR $0.01$), although Fold'EM still raises
their TM-scores from $0.798$ and $0.477$ to $0.917$ and $0.810$.

\pagebreak 
On experimental data, a few hundred particles suffice when the component
folds are already present in the unguided prediction. AlphaFold3 predicts the
integrin and Fab components of \texttt{9B9J} individually but places the Fab
at the wrong site, illustrating the difficulty of antibody--antigen
interfaces, for which paired coevolutionary information is limited
(Figure~\ref{fig:homo}B). Fold'EM identifies the correct epitope from $400$
particles with poses estimated \textit{ab initio}, compared with $374{,}814$
particles used in the deposited reconstruction. With fewer than $1{,}500$
particles it reaches a resolution finer than $3.5\,\text{\AA}$ in both the
known-pose and \textit{ab initio} settings (Figure~\ref{fig:homo}F), and it
exceeds a TM-score of $0.9$ from $400$ particles (Figure~\ref{fig:homo}H),
while no baseline reaches $6\,\text{\AA}$ on this target. The particles thus
correct the relative arrangement of the predicted components and reveal the
epitope, although the interface variable loops of BIIG2 are not recovered
with atomic accuracy. With $1/1000$ of the deposited particles and known
poses, Fold'EM reaches $2.88$, $3.08$ and $3.18\,\text{\AA}$ on \texttt{9B77},
\texttt{9GXQ} and \texttt{9GXR} (Table~\ref{tab:real_res}). \texttt{8PWH} is
the one exception. Fold'EM raises the TM-score of the unguided prediction
from $0.397$ to $0.900$ (Table~\ref{tab:real_tm_rmsd_fraction}), but from
$1/100$ of the particles onward cryoDRGN attains a finer FSC resolution
($5.16$ and $5.01\,\text{\AA}$ against $7.06$ and $6.13\,\text{\AA}$), as does
RELION \textit{ab initio} from $5{,}000$ particles ($9.84$ against
$10.83\,\text{\AA}$; Table~\ref{tab:real_res_abinitio}). When the unguided
prediction departs this far from the deposited model, density reconstruction
attains the finer resolution once roughly $7{,}000$ particles are available.

\subsection{Guidance under two-state heterogeneity}
\label{sec:exp_discrete}

\begin{figure*}[p]
    \centering
    \includegraphics[width=1.\linewidth]{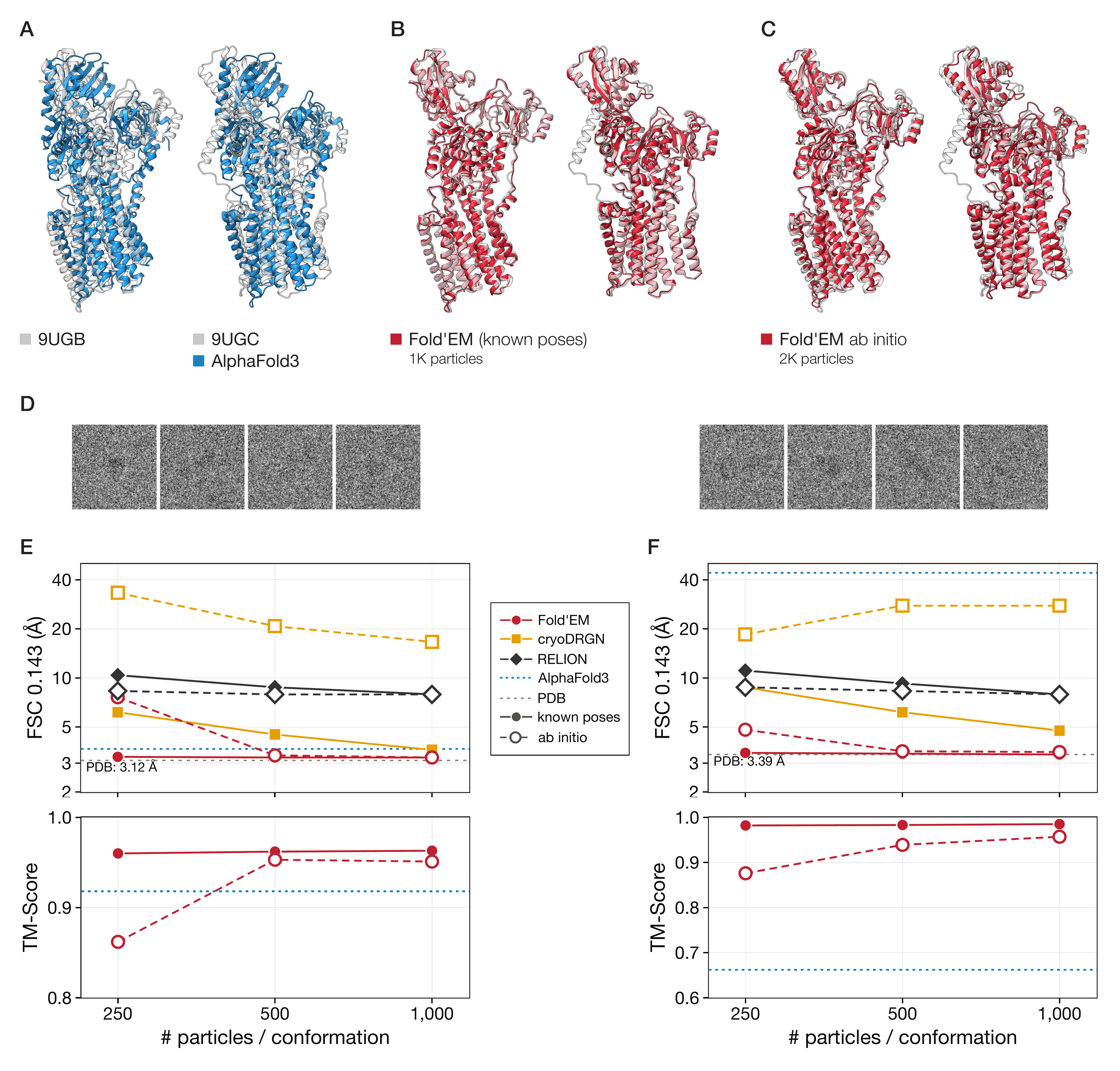}
\caption{\textbf{Fold'EM resolves both states of a synthetic \texttt{9UGB}/\texttt{9UGC} mixture from $250$ particles per state with known poses, and from $500$ \textit{ab initio}.} Pma1 is the plasma-membrane proton pump of yeast, a P-type ATPase whose cytosolic A-, N- and P-domains rearrange along the catalytic cycle; \texttt{9UGB} is the activated (E2P) state and \texttt{9UGC} the autoinhibited state. Particles are rendered from the two deposited conformations, approximately $9$\,\AA{} apart in C$\alpha$ RMSD, and mixed in equal proportion at SNR $0.005$. In every panel, \texttt{9UGB} is on the left and \texttt{9UGC} on the right. (A) AlphaFold3 (blue) over the deposited models (grey) recovers neither state (TM-score $0.70$ and $0.66$): the transmembrane domain is predicted correctly and the error is confined to the cytosolic domains, whose arrangement depends on the state of the cycle. (B) Fold'EM (red) with known poses resolves both from the mixed stack of $1{,}000$ particles ($500$ per state). (C) Fold'EM \textit{ab initio} resolves both from $2{,}000$ particles ($1{,}000$ per state). (D) Representative particles of each state. (E, F) FSC resolution (top) and TM-score against the deposited model (bottom) as a function of the number of particles per state, for \texttt{9UGB} (E) and \texttt{9UGC} (F); filled markers denote known poses and open markers \textit{ab initio}, the dashed grey line is the resolution of the deposited model and the dotted blue line the AlphaFold3 prediction.
}

    \label{fig:synthetic_het}
\end{figure*}

\begin{figure*}[p]
    \centering
    \includegraphics[width=\linewidth]{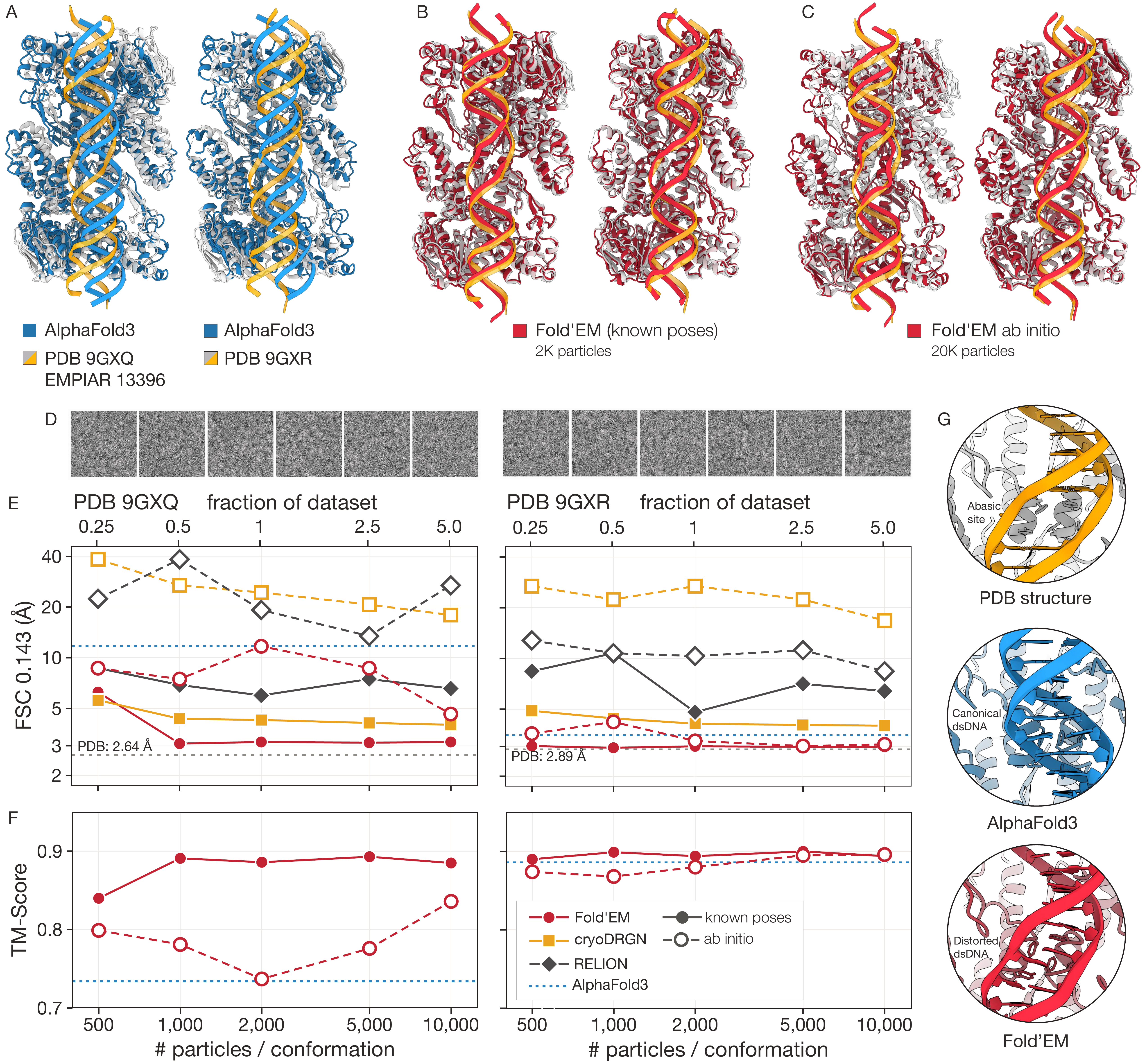}
    \caption{\textbf{Fold'EM resolves both UvrA--DNA conformations from $0.5\%$
    of the deposited particles with known poses and from $5\%$ \textit{ab initio}.}
    EMPIAR-13396 contains two states of UvrA bound to DNA carrying an abasic
    site (PDB \texttt{9GXQ}/\texttt{9GXR}); damage recognition involves insertion
    of the \texttt{Y62}/\texttt{Q65}/\texttt{F66} wedge into the duplex.
    (A) AlphaFold3 recovers \texttt{9GXR} but not \texttt{9GXQ}
    (TM-score $0.734$).
    (B,C) Fold'EM resolves both states from $1{,}000$ particles per state with
    known poses and from $10{,}000$ particles per state \textit{ab initio} (left: \texttt{9GXQ}, right: \texttt{9GXR}).
    (D) Representative particles.
    (E) FSC versus particles per state (filled: known poses; open: \textit{ab initio}).
    With known poses Fold'EM reaches $3.1/3.0\,\text{\AA}$ from $1{,}000$
    particles per state; \textit{ab initio} it reaches $3.0\,\text{\AA}$ on
    \texttt{9GXR} and $4.6\,\text{\AA}$ on \texttt{9GXQ}, while RELION and
    cryoDRGN-AI remain above $8\,\text{\AA}$.
    (F) TM-score against the deposited models.
    (G) AlphaFold3 retains a fully paired duplex, whereas Fold'EM recovers the
    wedge insertion and disrupted DNA, but not the flipped-out nucleotide, as
    expected because the lesion was supplied to both models as intact DNA. Resolution ($\text{\AA}$; FSC $=0.143$).}
    \label{fig:hetero}
\end{figure*}

\paragraph{Experimental setup.}
We then ask whether two conformational states can be resolved from a single
mixed particle stack, first with known poses and then with poses and class
labels inferred from the particles. To evaluate this, we take a synthetic
mixture of the activated (\texttt{9UGB}) and autoinhibited (\texttt{9UGC})
states of Pma1, about $9\,\text{\AA}$ apart in C$\alpha$ RMSD, rendered in
equal proportion at SNR $0.01$ and $0.005$ with $250$ to $1{,}000$ particles
per state. We further take the experimental mixture EMPIAR-13396, which
contains two states of UvrA bound to DNA carrying an abasic site
(\texttt{9GXQ} and \texttt{9GXR}), and draw balanced stacks of $500$ to
$10{,}000$ particles per state. In the \textit{ab initio} setting,
\brecovar{} initializes the class labels (Section~\ref{sec:exp_brecovar}),
and Fold'EM then alternates structure, pose and class refinement, guiding
each ensemble member with the more confident half of the particles assigned
to it (Section~\ref{sec:methods}). The baselines first cluster the particles
and then reconstruct one density per class from its assigned subset
(Appendix~\ref{app:baselines}), using RELION and cryoDRGN with known poses and
RELION \textit{ab initio} and cryoDRGN-AI otherwise.

\paragraph{Results.}
On the synthetic mixture, a few hundred particles per state suffice to
resolve both states. AlphaFold3 recovers neither (TM-score $0.699$ and
$0.661$; Figure~\ref{fig:synthetic_het}). With known poses, Fold'EM resolves
both from $250$ particles per state at SNR $0.005$, reaching $3.3$ and
$3.5\,\text{\AA}$ with TM-scores $0.96$ and $0.98$
(Tables~\ref{tab:het_syn_fsc0143} and~\ref{tab:het_syn_rmsd_tm}), whereas at
the same noise level RELION remains coarser than $7.9\,\text{\AA}$ and
cryoDRGN requires $1{,}000$ particles per state to reach $3.6$ and
$4.8\,\text{\AA}$. \textit{Ab initio}, $500$ particles per state give $3.4$
and $3.6\,\text{\AA}$ with TM-scores $0.95$ and $0.94$ at SNR $0.005$, within
about $0.1\,\text{\AA}$ of the known-pose results
(Tables~\ref{tab:het_syn_fsc0143_abinitio}
and~\ref{tab:het_syn_rmsd_tm_abinitio}), whereas RELION \textit{ab initio}
and cryoDRGN-AI remain coarser than $6.9$ and $16.6\,\text{\AA}$,
respectively, on either state at every sample size. At $250$ particles per
state and SNR $0.005$, \texttt{9UGB} degrades to $7.6\,\text{\AA}$, which
marks the sample size at which the signal no longer suffices for the joint
inference of pose, class and structure.

On UvrA, the particle count required to resolve a state depends on how
closely the unguided prediction already reproduces it. AlphaFold3 predicts
\texttt{9GXR} ($3.54\,\text{\AA}$, TM-score $0.886$) but not \texttt{9GXQ}
($11.69\,\text{\AA}$, TM-score $0.734$). With known poses, Fold'EM resolves
both from $1{,}000$ particles per state, reaching $3.09$ and
$2.95\,\text{\AA}$ on \texttt{9GXQ} and \texttt{9GXR}, against $4.34$ and
$4.41\,\text{\AA}$ for cryoDRGN and $6.89$ and $10.75\,\text{\AA}$ for RELION
(Table~\ref{tab:het_real_fsc0143}). \textit{Ab initio}, Fold'EM reaches
$3.58\,\text{\AA}$ on \texttt{9GXR} from $500$ particles per state and
$3.02\,\text{\AA}$ from $5{,}000$, whereas \texttt{9GXQ} is recovered from
$10{,}000$ particles per state, about $5\%$ of its deposited stack, at
$4.63\,\text{\AA}$ and TM-score $0.836$ (Figure~\ref{fig:hetero};
Tables~\ref{tab:het_real_fsc0143_abinitio}
and~\ref{tab:het_real_rmsd_tm_abinitio}). RELION \textit{ab initio} and
cryoDRGN-AI remain coarser than $8\,\text{\AA}$ on both states at every
sample size. The larger particle count required \textit{ab initio} is
therefore attributable to \texttt{9GXQ}, the state absent from the unguided
prediction.

\pagebreak 
The recovered difference is functionally meaningful. UvrA recognizes damaged DNA by inserting the \texttt{Y62}, \texttt{Q65}
and \texttt{F66} wedge into the duplex. AlphaFold3 predicts a largely
canonical duplex, whereas Fold'EM recovers the wedge insertion and the
associated distortion of the DNA (Figure~\ref{fig:hetero}G). It does not
recover the flipped-out abasic nucleotide, which was not represented as an
abasic site in the input molecular specification.

\subsection{Label-free class assignment with \texorpdfstring{\brecovar{}}{B-RECOVAR}}
\label{sec:exp_brecovar}

\begin{figure*}[t]
    \centering
    \includegraphics[width=1.\linewidth]{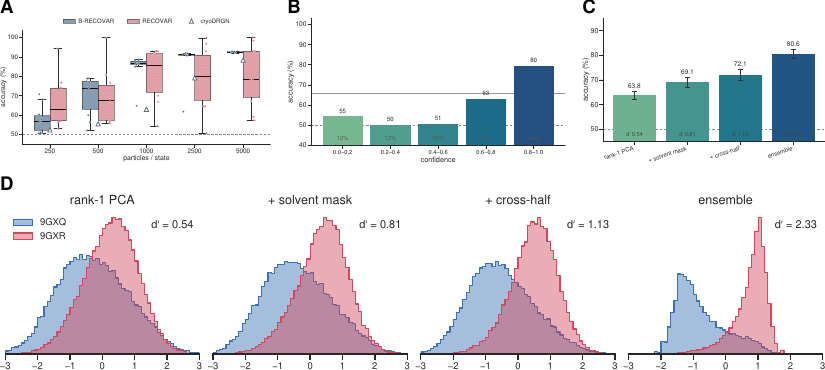}
    \caption{\textbf{\brecovar{} assigns particles to conformational states more accurately and more reliably than RECOVAR and cryoDRGN at low particle counts, and its confidence score orders the assignments by their reliability.} Label-free class assignment on the UvrA protein--DNA complex (EMPIAR-13396, \texttt{9GXQ} vs.\ \texttt{9GXR}), scored against the deposited labels. \textbf{(A)} Accuracy as a function of the number of particles per state, from $250$ to $5{,}000$, that is, $0.13$--$2.6\%$ of the particles deposited for each state. Boxes summarize ten random draws per size, shared by \brecovar{} and RECOVAR; triangles mark cryoDRGN. The RECOVAR estimator is ill-conditioned at these sample sizes and its accuracy ranges from chance to near-perfect between draws of the same size; \brecovar{} trails it at $250$ particles but is both more accurate and markedly more stable from $500$ particles onward, and cryoDRGN remains below \brecovar{} throughout. \textbf{(B)} Accuracy of \brecovar{} stratified by its confidence score $\kappa_k$, the normalized margin of the ensemble vote (Appendix~\ref{app:class_assignments}). The percentage in each bar is the fraction of particles in that bin; the solid line is the accuracy over all particles and the dashed line chance. Accuracy increases monotonically with confidence, so the score ranks the particles by the reliability of their label; Fold'EM uses this ordering to guide each ensemble member with the more confident half of its particles (Table~\ref{tab:confidence_tm_sep}). \textbf{(C, D)} Ablation of the estimator at $1000$ particles per state. The base estimator is a single rank-one fit to the consensus-subtracted residuals, a one-dimensional RECOVAR fit without noise correction, and is improved in turn by restricting the mode to a solvent mask, by the cross-validated refinement of the labels, and by ensembling replicates over particle subsamples, frequency bands and initializations. \textbf{(C)} Accuracy of each stage, with error bars over repeated draws. \textbf{(D)} Distribution of the per-particle class score at each stage for the particles of either state, with their separation $d'$, the difference of the two means in units of the pooled standard deviation. Every step widens the separation between the states, and the ensemble accounts for the largest gain.}
    \label{fig:brecovar}
\end{figure*}

\paragraph{Experimental setup.}
The heterogeneous experiments depend on the class labels that initialize
them. We therefore ask whether particles can be assigned to their
states without reference structures, at particle counts at which existing
embeddings are unreliable. To evaluate this, we take the UvrA mixture of
EMPIAR-13396 at $250$ to $5{,}000$ particles per state, between $0.13\%$ and
$2.6\%$ of the particles deposited for each state, with ten random draws per
size and poses from the consensus refinement, and score the assignments
against the deposited labels. We compare \brecovar{}
(Section~\ref{sec:initialization}; Appendix~\ref{app:class_assignments}) with
RECOVAR \citep{gilles2025cryo}, applied to the same draws with the same
label-free protocol, and with cryoDRGN, reported as the median over
ten seeds. The ablation of \brecovar{} is carried out at $1{,}000$ particles
per state.

\paragraph{Results.}
From $500$ particles per state onward, \brecovar{} assigns the particles more
accurately and far more consistently than either baseline
(Figure~\ref{fig:brecovar}A). At these sample sizes the leading direction of
the RECOVAR covariance depends as much on the particles drawn as on the
structural difference between the states, and across ten draws of the same size the accuracy of RECOVAR
ranges from chance to near-perfect, so that the median over draws does not
represent any single draw. \brecovar{} is less accurate at $250$
particles per state, but at $5{,}000$ it reaches $92.5\%$ with a range across
draws of under one percentage point, and cryoDRGN remains below it at every
sample size.

The normalized margin $\kappa_k$ of the ensemble vote serves as a confidence
score. Accuracy increases monotonically with it, from near chance in the
lowest bins to $80\%$ in the highest, which contains $44\%$ of the particles
(Figure~\ref{fig:brecovar}B), so the score orders the labels by reliability
although it is not a calibrated probability. Fold'EM uses this ordering in
the heterogeneous experiments of Section~\ref{sec:exp_discrete}, and
discarding the less confident half of the particles assigned to each member
increases the separation between the two modelled conformations, measured by
their C$\alpha$ RMSD, and raises their TM-score against the deposited models
(Table~\ref{tab:confidence_tm_sep}). This is consistent with the least
confident particles being the most likely to be mislabelled, since
mislabelled particles draw both members toward a common average rather than
their respective states.

The ablation of Figures~\ref{fig:brecovar}C and~D isolates the contribution
of each component of the estimator. The base estimator, a rank-one fit to the
consensus-subtracted residuals equivalent to a one-dimensional RECOVAR fit
without noise correction and without a solvent mask, reaches $63.8\%$
accuracy with strongly overlapping class scores ($d'=0.54$).
Restricting the fitted mode to the interior of a solvent mask raises the
accuracy to $69.1\%$, and refining the labels by cross-validated comparison
against per-class reconstructions raises it to $72.1\%$. The largest gain
comes from the ensemble over particle subsets, frequency bands and
initializations, which yields $80.6\%$ and doubles the separation between
the states, from $d'=1.13$ to $d'=2.33$. The class direction recurs across
replicates, whereas the nuisance directions that compete with it at these
particle counts do not, and aggregation over replicates therefore separates
the two states.

\subsection{Guidance under continuous heterogeneity}
\label{sec:exp_continuous}

\begin{figure*}[p]
    \centering
    \includegraphics[width=1.\linewidth]{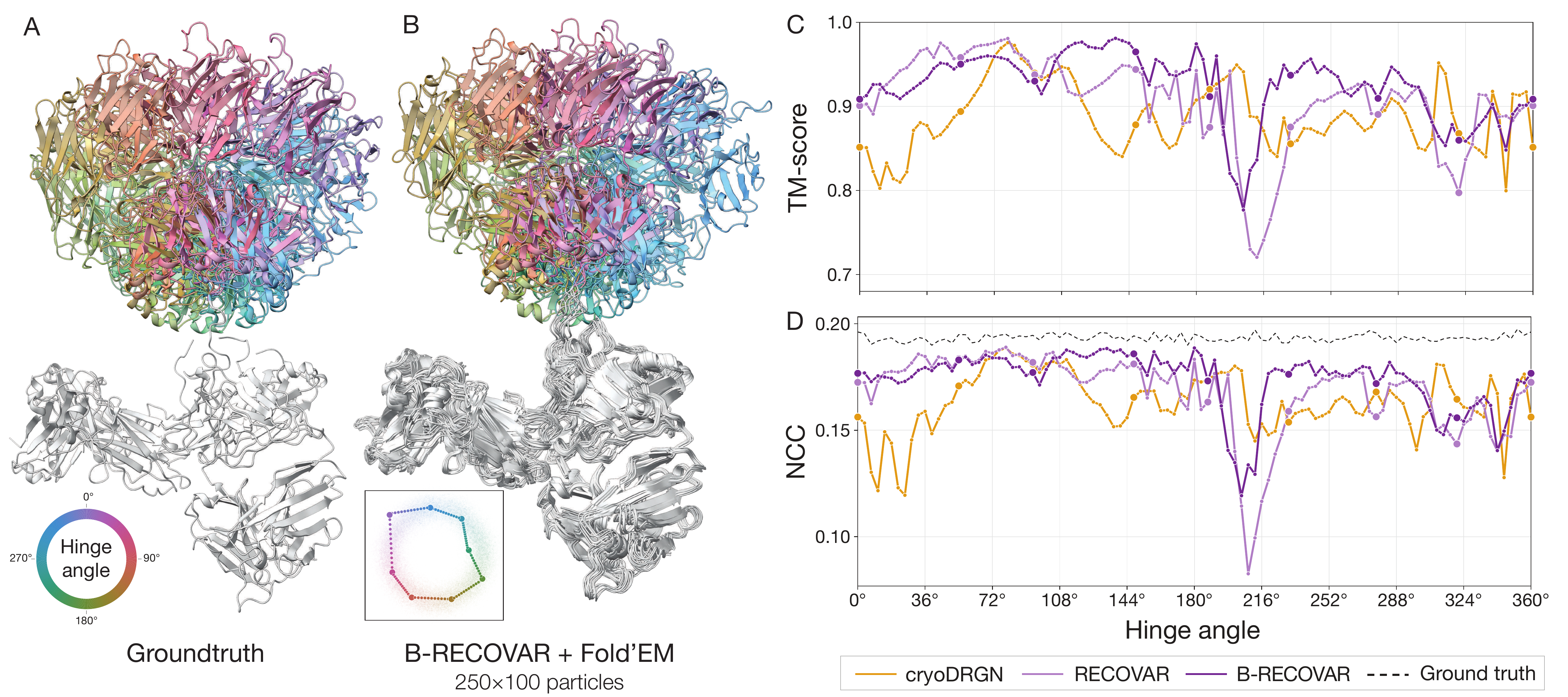}
    \caption{\textbf{Fold'EM resolves continuous heterogeneity of an IgG antibody complex.} (A) Shown as ensembles are $10$ of the $100$ ground-truth CryoBench IgG-1D synthetic models, colored by hinge angle; and (B) the corresponding structures recovered by Fold'EM with B-RECOVAR particle classification. The inset depicts the two leading principal components of the optimized Pairformer states, overlaid on the B-RECOVAR embedding of the particles. (C) TM-score and (D) Fourier NCC plotted as a function of hinge angle, for the three conformational coordinates.}
    \label{fig:continuous-het}

    \vspace{2.5em}

    \centering
    \includegraphics[width=1.\linewidth]{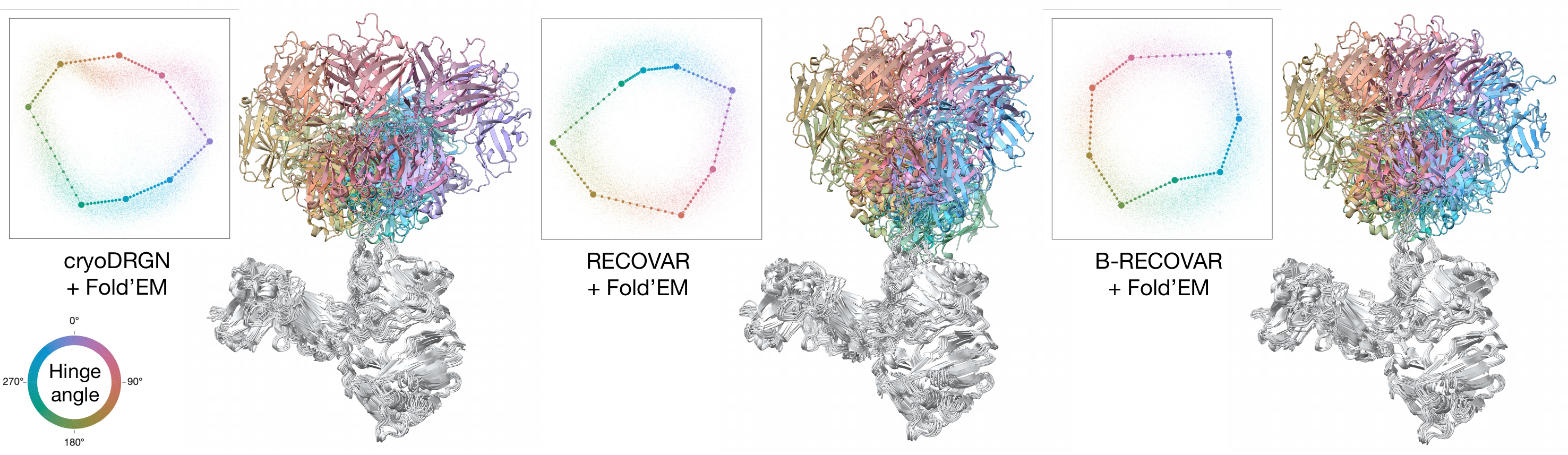}

    \caption{\textbf{Optimized Pairformer states preserve the hinge-angle ordering for all three conformational coordinates.} We show the geometry and the generated structures for each of the three particle classifiers: cryoDRGN, RECOVAR and \brecovar{}. Left of each panel: the two leading principal components of the optimized Pairformer states, with the eight anchors as large dots and the interpolated states as small dots, overlaid on the per-particle coordinates of the corresponding method mapped into the same plane; all are coloured by the hinge angle of the state they represent, as indicated by the colour wheel. Right of each panel: a subset of $10$ structures decoded from the interpolated states, superimposed on the fixed body of the antibody and coloured by hinge angle (top), and in white to show the sweep of the moving Fab about the hinge (bottom). In all three cases the anchors, ordered by hinge angle, form a simple closed ring in the representation and the decoded structures trace the full rotation of the Fab. }

    \label{fig:igg_latents}
\end{figure*}

\paragraph{Experimental setup.}
Finally, we ask whether a continuous conformational trajectory can be
represented by a small number of optimized Pairformer states and their
interpolation. To evaluate this, we take the IgG-1D dataset of CryoBench
\citep{jeon2024cryobench}, which consists of $100$ ground-truth atomic
models, generated from PDB entry \texttt{1HZH} by rotating one Fab domain
about the hinge through a closed $360^\circ$ trajectory, and $25{,}000$
synthetic particles at SNR $0.01$, $250$ per state, with ground-truth poses
and per-particle CTFs. We embed each particle in three label-free
conformational coordinates, namely the latent of a cryoDRGN VAE trained on
the stack, the leading principal components of RECOVAR, and the
residual-subspace embedding of \brecovar{}
(Appendix~\ref{app:class_assignments}). Each coordinate defines one arm of the
experiment, and the arms differ only in the coordinate used to select the
particles of each state.

In each arm, we fix eight \emph{anchor} conformations at equispaced positions
along the ring and, for each anchor, optimize one structure against the $250$
particles nearest to it in the conformational coordinate of that arm. This
yields eight Pairformer states $\mathbf H_1,\dots,\mathbf H_8$, optimized
independently against disjoint particle sets, with no term coupling them. An
intermediate conformation at position $\alpha$ along the ring is obtained as
the convex combination $\mathbf H(\alpha)=\sum_k c_k(\alpha)\,\mathbf H_k$ of
the two neighbouring anchors, with piecewise-linear coefficients $c_k$. For
each of the $100$ ground-truth states we decode $\mathbf H(\alpha)$ through a
full reverse-diffusion trajectory, with a brief coordinate-space refinement
against the $250$ particles that the same coordinate places nearest to that
state, and relax the result (Appendix~\ref{app:relaxation}).

\pagebreak 
The comparison is among the three coordinates, and between the interpolated
structures and the anchors, rather than against a density-based baseline.
Each structure is scored by the TM-score against the ground-truth model of
its state and by the Fourier NCC between the rendered structure and the
particles of that state, expressed as a fraction of the NCC of the
ground-truth model against the same particles.

\paragraph{Results.}
Eight anchors suffice to represent the trajectory. Over the full ring, the
$100$ structures of the \brecovar{} arm, all decoded from interpolated
states, attain a median NCC of $92\%$ of the ground-truth value and a median
TM-score of $0.93$, comparable to the eight anchors themselves ($89\%$ and
$0.91$). The cryoDRGN coordinate reaches $85\%$ and $0.89$, and the RECOVAR
coordinate $90\%$ and $0.92$ (Figure~\ref{fig:continuous-het}). The leading
principal components of the optimized Pairformer states preserve the ordering
in hinge angle for all three coordinates (Figure~\ref{fig:igg_latents}),
although the anchors are optimized independently and nothing in the
objective couples them. Table~\ref{tab:igg_molprobity} reports the
stereochemical quality of the anchors and of the interpolated structures.
The conformational organization of the particles is therefore mirrored in
the AlphaFold representation, and a continuous trajectory can be represented
by a small number of experimentally optimized states.

\section{Related work}
\label{sec:related}

\paragraph{Map reconstruction.} Reconstructing an electrostatic potential (ESP) map at a resolution sufficient for atomic model building typically requires tens to hundreds of thousands of particles. RELION \citep{zivanov2018new} and cryoSPARC \citep{punjani2017cryosparc} estimate the particle poses and the map jointly within a Bayesian framework. For heterogeneous data, these pipelines partition the particles into $K$ discrete classes and reconstruct one map per class. cryoDRGN \citep{zhong2021cryodrgn} instead learns a continuous latent space over the particles with a variational autoencoder and decodes each latent into a three-dimensional density, and RECOVAR \citep{gilles2025cryo} estimates the covariance of the underlying volume and embeds the particles in its leading principal components. Both take the poses from a prior consensus reconstruction. CryoDRGN-AI \citep{levy2025cryodrgn} extends this approach to the \textit{ab initio} setting by estimating the poses and the heterogeneous structural distribution jointly from the images. More recent approaches use a protein structure prediction model \citep{abramson2024accurate} as a structural prior and guide it with the experimental density. In X-ray crystallography, \citet{maddipatla2024generative} fit ensembles to electron densities by jointly guiding a diffusion model of protein structure, recovering alternate backbone conformations, and \citet{maddipatla2025inverse} extend this ensemble guidance to AlphaFold3, treated as a sequence-conditioned structural prior, to infer ensembles consistent with crystallographic densities and NMR restraints. In cryo-EM, CryoBoltz \citep{raghu2026multiscale} steers Boltz \citep{passaro2025boltz} toward conformations consistent with an ESP map, but requires an input structure roughly aligned to the map to establish the correspondence used during guidance. \citet{Maddipatla2026} instead guide AlphaFold3 \textit{ab initio}, from sequence alone, toward structures consistent with experimental ESP maps. Embed-Opt \citep{li2026robust} and IT-Opt \citep{maddipatla2026inference} take a complementary inference-time approach and optimize the latent representations of AlphaFold to produce structures consistent with synthetic ESP maps.

\citep{riley2021qfit}

The closest prior work is CoCoFold \citep{liao2026cocofold}, which constrains AlphaFold directly with individual cryo-EM particle images. CoCoFold performs target-specific fine-tuning of the AlphaFold-Multimer structure module, with the Evoformer frozen, using an FRC loss between experimental particles and differentiable projections; it assumes upstream pose/CTF estimates and uses a coarse reconstruction to establish a fixed coordinate frame. 

\section{Conclusion}
\label{sec:conclusion}
We introduced Fold'EM, an inference-time framework that infers atomic structures directly from cryo-EM particle images by guiding AlphaFold3 with particle-level evidence, bypassing density reconstruction and model building. We showed that density fitting reproduces the particle likelihood only under the experiment-induced metric and that, under pose or class uncertainty, a consensus density need not be a sufficient statistic; and that optimizing intermediate rather than terminal Pairformer states induces a learned geometry that substantially strengthens guidance. Fold'EM recovers accurate structures from fewer than $1\%$ of the particles used by conventional reconstruction with known poses, roughly $3\%$ \textit{ab initio}, and roughly $5\%$ for heterogeneous samples, recovering features missed by the unguided prediction, such as the Integrin $\alpha_5\beta_1$--BIIG2 binding site and the distorted DNA of UvrA. Conformational organization in particle space is also reflected in the AlphaFold representation, so that continuous motions can be represented by a few anchor states. Together, these results suggest a shorter route to structure determination in which the prior supplies what is already known and the experiment determines what it does not: assembly, conformation, and interaction geometry.

\paragraph{Limitations.}
Fold'EM is not yet an end-to-end replacement for the cryo-EM pipeline: inputs are curated particle collections with known CTF parameters, and \textit{ab initio} inference relies on accurate pose estimates for initialization. Components absent from the model, such as glycans, ligands or flexible regions, are treated as unexplained signal, and fine local details, such as interface loops, remain harder to recover than global assemblies and major conformational changes. Uncurated particles, unmodelled density, better pose initialization, and finer local refinement are natural directions for future work.

\section*{AI use statement}
Generative AI tools, primarily OpenAI ChatGPT and Anthropic Claude, were used during manuscript preparation for language editing, improving clarity and organization, checking mathematical exposition, assisting with code development and debugging, and literature and reference formatting. All scientific claims, derivations, code, experimental results, interpretations, and citations were reviewed and verified by the authors. The authors take full responsibility for the content of the paper and full credit for its scientific claims.

\section*{Ethics statement}
This work does not involve human subjects or sensitive personal data, and poses no identifiable risks to individuals or their security. All datasets used in this work were either synthetically generated from protein structures available in the Protein Data Bank (PDB) \citep{burley2017protein} or derived from publicly available cryo-EM particle datasets in EMPIAR \citep{iudin2016empiar}.
\section*{Reproducibility statement}
All hyperparameters and implementation details are provided in Appendix~\ref{sec:impl_details}, and the full Fold'EM procedure is given as pseudocode in Algorithm~\ref{alg:FoldEM_alg}. The hardware resources, model weights and software versions are listed in Appendix~\ref{sec:hardware}. The datasets, their pre-processing and the baseline protocols are described in Appendices~\ref{app:tab_datasets}, \ref{app:preprocessing} and~\ref{app:baselines}. The image formation model and likelihood are derived in Appendix~\ref{sec:image-formation-model}, the stereochemical losses and relaxation in Appendices~\ref{app:losses} and~\ref{app:relaxation}, and the evaluation metrics in Appendix~\ref{app:metrics}. The code will be publicly released.
\bibliography{preprint}
\bibliographystyle{plainnat}
\clearpage
\appendix
\setcounter{figure}{0}
\renewcommand{\thefigure}{A\arabic{figure}}
\renewcommand{\theHfigure}{A\arabic{figure}}
\setcounter{table}{0}
\renewcommand{\thetable}{A\arabic{table}}
\renewcommand{\theHtable}{A\arabic{table}}

\section{Overview of the appendix}
\label{app:overview}
The appendix is organized as follows.
\begin{itemize}
    \item \textbf{Appendix~\ref{app:tables}: Additional tables.} The experimental and synthetic datasets (Appendix~\ref{app:tab_datasets}), and the full quantitative results for homogeneous reconstruction with known poses (Appendix~\ref{app:tab_known}), homogeneous \textit{ab initio} reconstruction (Appendix~\ref{app:tab_abinitio}), heterogeneous reconstruction (Appendix~\ref{app:tab_hetero}), and continuous heterogeneity on IgG-1D (Appendix~\ref{app:tab_igg}).
    \item \textbf{Appendix~\ref{app:algorithms}: Algorithms.} Pseudocode for Fold'EM (Algorithm~\ref{alg:FoldEM_alg}) and for the residual-subspace estimator of \brecovar{} (Algorithm~\ref{alg:residual_mode}).
    \item \textbf{Appendix~\ref{app:details}: Additional details.} Data pre-processing and baselines (Appendices~\ref{app:preprocessing} and~\ref{app:baselines}), software and hardware (Appendix~\ref{sec:hardware}), the stereochemical losses and relaxation (Appendices~\ref{app:losses} and~\ref{app:relaxation}), the evaluation metrics (Appendix~\ref{app:metrics}), the image formation model used to render particles from atomic structures, the contrast transfer function, and likelihood computation (Appendix~\ref{sec:image-formation-model}), the \brecovar{} class-assignment method (Appendix~\ref{app:class_assignments}) and the hyperparameters and implementation details (Appendix~\ref{sec:impl_details}).
\end{itemize}

\clearpage

\clearpage
\section{Additional tables}
\label{app:tables}
\subsection{Datasets}
\label{app:tab_datasets}
\begin{table}[H]
\centering
\caption{\textbf{Experimental cryo-EM datasets from EMPIAR \citep{iudin2016empiar}.} For each PDB entry, we report the corresponding EMPIAR and EMDB entries \citep{wwpdb2024emdb}, reported resolution, molecular weight (kDa), and number of particles in original dataset.}
\label{tab:real}
\small
\begin{tabular}{@{}cccccc@{}}
\toprule
\textbf{PDB ID} & \textbf{EMPIAR ID} & \textbf{EMD ID} & \textbf{Resolution(\AA)} & \textbf{\# Particles} & \makecell[l]{\textbf{Protein size (kDa)}} \\
\midrule
\texttt{9B9J} & $12552$ & $44386$ & $2.60$ & $374814$ & $115.11$ \\
\texttt{9B77} & $12002$ & $44302$ & $2.74$ & $490807$ & $141.72$ \\
\texttt{8PWH} & $11665$ & $17993$ & $3.17$ & $708609$ & $162.75$ \\
\texttt{9GXR} & $13396$ & $51671$ & $3.04$ & $197058$ & $162.70$ \\
\texttt{9GXQ} & $13396$ & $51672$ & $2.74$ & $195233$ & $162.70$ \\
\bottomrule
\end{tabular}
\end{table}

\begin{table}[H]
\centering
\caption{\textbf{Synthetic datasets constructed using PDB-deposited structures \citep{burley2017protein}.} For each entry, we report the corresponding EMDB entry \citep{wwpdb2024emdb}, reported resolution, molecular weight (kDa), and amino acid sequence length.}
\label{tab:synthetic}
\small
\begin{tabular}{@{}ccccc@{}}
\toprule
\textbf{PDB ID} & \textbf{EMD ID} & \textbf{Resolution(\AA)} & \textbf{Seq Length} & \textbf{Protein size (kDa)} \\
\midrule
\texttt{9UGB} & $64135$ & $3.25$ & $818$ & $88.97$ \\
\texttt{9UGC} & $64136$ & $3.52$ & $818$ & $99.62$ \\
\texttt{8U4B} & $41877$ & $3.80$ & $2764$ & $312.70$ \\
\texttt{8U4E} & $41880$ & $4.20$ & $3304$ & $373.10$ \\
\texttt{1HZH} & CryoBench (IgG-$1$D) & $2.70$ & $1331$ & $146.96$ \\
\bottomrule
\end{tabular}
\end{table}

\subsection{Homogeneous reconstruction with known poses}
\label{app:tab_known}

\begin{table}[H]
\centering
\caption{\textbf{Resolution at \mbox{Fourier shell correlation (FSC) $=0.143$} ($\downarrow, \AA$) on the experimental cryo-EM datasets from Table~\ref{tab:real}.} For Fold'EM, AlphaFold3 and PDB structures, FSC is computed using \texttt{phenix.model\_map}. All baselines at different fraction of the original dataset $500$ particles with known poses and are evaluated against the deposited EMD map. Entries highlighted in \textbf{bold} indicate the best result (excluding PDB).}
\label{tab:real_res}
\small
\begin{tabular}{@{}cc|ccccc@{}}
\toprule
\hd{Fraction of dataset} & \hd{Method} & \hd{\texttt{9B9J}} & \hd{\texttt{9B77}} & \hd{\texttt{8PWH}} & \hd{\texttt{9GXQ}} & \hd{\texttt{9GXR}} \\
\midrule
\multirow{5}{*}{$1/1000$} & Fold'EM & $\mathbf{3.73}$ & $\mathbf{2.88}$ & $\mathbf{6.21}$ & $\mathbf{3.08}$ & $\mathbf{3.18}$ \\
 & cryoDRGN & $8.40$ & $5.70$ & $7.56$ & $9.87$ & $4.33$ \\
 & RELION & $21.94$ & $11.48$ & $14.12$ & $14.15$ & $17.92$ \\
 & AlphaFold3 & $21.90$ & $5.96$ & $64.96$ & $11.69$ & $3.54$ \\
 & PDB & $2.38$ & $2.49$ & $3.06$ & $2.64$ & $2.89$ \\
\midrule
\multirow{5}{*}{$1/500$} & Fold'EM & $\mathbf{3.73}$ & $\mathbf{2.85}$ & $\mathbf{6.63}$ & $\mathbf{3.02}$ & $\mathbf{3.20}$ \\
 & cryoDRGN & $7.41$ & $4.87$ & $7.55$ & $9.60$ & $3.90$ \\
 & RELION & $17.95$ & $11.48$ & $15.47$ & $14.15$ & $14.93$ \\
 & AlphaFold3 & $21.90$ & $5.96$ & $64.96$ & $11.69$ & $3.54$ \\
 & PDB & $2.38$ & $2.49$ & $3.06$ & $2.64$ & $2.89$ \\
\midrule
\multirow{5}{*}{$1/250$} & Fold'EM & $\mathbf{3.35}$ & $\mathbf{2.73}$ & $\mathbf{6.63}$ & $\mathbf{3.95}$ & $\mathbf{3.09}$ \\
 & cryoDRGN & $7.04$ & $4.87$ & $\mathbf{6.63}$ & $9.27$ & $3.73$ \\
 & RELION & $15.19$ & $11.07$ & $12.99$ & $12.22$ & $12.22$ \\
 & AlphaFold3 & $21.90$ & $5.96$ & $64.96$ & $11.69$ & $3.54$ \\
 & PDB & $2.38$ & $2.49$ & $3.06$ & $2.64$ & $2.89$ \\
\midrule
\multirow{5}{*}{$1/100$} & Fold'EM & $\mathbf{3.50}$ & $\mathbf{2.73}$ & $7.06$ & $\mathbf{3.09}$ & $\mathbf{3.26}$ \\
 & cryoDRGN & $6.71$ & $3.92$ & $\mathbf{5.16}$ & $9.27$ & $3.73$ \\
 & RELION & $12.74$ & $7.21$ & $10.15$ & $10.75$ & $9.60$ \\
 & AlphaFold3 & $21.90$ & $5.96$ & $64.96$ & $11.69$ & $3.54$ \\
 & PDB & $2.38$ & $2.49$ & $3.06$ & $2.64$ & $2.89$ \\
\midrule
\multirow{5}{*}{$1/50$} & Fold'EM & $\mathbf{3.46}$ & $\mathbf{2.75}$ & $6.13$ & $\mathbf{3.36}$ & $\mathbf{3.45}$ \\
 & cryoDRGN & $6.46$ & $3.87$ & $\mathbf{5.01}$ & $8.67$ & $3.58$ \\
 & RELION & $12.34$ & $6.74$ & $9.02$ & $7.07$ & $7.07$ \\
 & AlphaFold3 & $21.90$ & $5.96$ & $64.96$ & $11.69$ & $3.54$ \\
 & PDB & $2.38$ & $2.49$ & $3.06$ & $2.64$ & $2.89$ \\
\bottomrule
\end{tabular}
\end{table}

\begin{table}[H]
\centering
\caption{\textbf{Quantitative evaluation of C$\alpha$ RMSD ($\downarrow$, \AA) and TM-score ($\uparrow$) on the experimental cryo-EM datasets from Table~\ref{tab:real}.} Fold'EM structures are obtained using different fractions of the original dataset. Entries highlighted in \textbf{bold} indicate the best result.}
\label{tab:real_tm_rmsd_fraction}
\small
\begin{tabular}{@{}c|ccccc|ccccc@{}}
\toprule
\multicolumn{1}{c}{} & \multicolumn{5}{c}{\textbf{C$\alpha$ RMSD ($\downarrow$, \AA)}} & \multicolumn{5}{c}{\textbf{TM-score ($\uparrow$)}} \\
\cmidrule(lr){2-6} \cmidrule(lr){7-11}
\hd{Fraction\\of dataset} & \hd{\texttt{9B9J}} & \hd{\texttt{9B77}} & \hd{\texttt{8PWH}} & \hd{\texttt{9GXQ}} & \hd{\texttt{9GXR}} & \hd{\texttt{9B9J}} & \hd{\texttt{9B77}} & \hd{\texttt{8PWH}} & \hd{\texttt{9GXQ}} & \hd{\texttt{9GXR}} \\
\midrule
$1/1000$ & $1.96$          & $1.90$          & $4.340$          & $2.20$          & $1.87$          & $0.969$          & $0.972$          & $0.900$          & $0.877$          & $0.899$          \\
$1/500$  & $1.97$          & $1.77$          & $4.989$          & $\mathbf{1.99}$ & $1.39$          & $0.970$          & $0.979$          & $0.897$          & $\mathbf{0.893}$ & $0.902$          \\
$1/250$  & $\mathbf{1.88}$ & $\mathbf{1.71}$ & $4.597$          & $2.48$          & $\mathbf{1.23}$ & $\mathbf{0.972}$ & $\mathbf{0.981}$ & $0.885$          & $0.880$          & $\mathbf{0.904}$ \\
$1/100$  & $1.92$          & $1.74$          & $5.349$          & $2.02$          & $1.35$          & $0.971$          & $0.980$          & $0.884$          & $\mathbf{0.893}$ & $0.903$          \\
$1/50$   & $\mathbf{1.88}$ & $\mathbf{1.71}$ & $\mathbf{4.174}$ & $2.09$          & $1.47$          & $\mathbf{0.972}$ & $\mathbf{0.981}$ & $\mathbf{0.909}$ & $0.885$          & $0.897$          \\
\midrule
AlphaFold3 & $21.90$       & $3.87$          & $33.16$         & $5.96$          & $2.01$          & $0.776$          & $0.916$          & $0.397$          & $0.734$          & $0.886$          \\
\bottomrule
\end{tabular}
\end{table}

\begin{table}[H]
\centering
\caption{\textbf{Resolution at \mbox{Fourier shell correlation (FSC) $=0.143$} ($\downarrow$) on the synthetic cryo-EM datasets from Table~\ref{tab:synthetic}.} For Fold'EM, AlphaFold3 and PDB structures, FSC is computed using \texttt{phenix.model\_map}. Fold'EM structures are generated using $100, 250, 500, \&\, 1000$ particles at SNR $0.01$ and $0.005$ with known poses and are evaluated against the deposited EMD map. Entries highlighted in \textbf{bold} indicate the best result (excluding PDB).}
\label{tab:syn_res_particles}
\small
\begin{tabular}{@{}cc|cc|cc|cc|cc@{}}
\toprule
\multicolumn{2}{c}{} & \multicolumn{2}{c}{\textbf{\texttt{9UGB}}} & \multicolumn{2}{c}{\textbf{\texttt{9UGC}}} & \multicolumn{2}{c}{\textbf{\texttt{8U4B}}} & \multicolumn{2}{c}{\textbf{\texttt{8U4E}}} \\
\cmidrule(lr){3-4} \cmidrule(lr){5-6} \cmidrule(lr){7-8} \cmidrule(lr){9-10}
\hd{\#\,particles} & \hd{Method} & \hd{SNR\\$0.01$} & \hd{SNR\\$0.005$} & \hd{SNR\\$0.01$} & \hd{SNR\\$0.005$} & \hd{SNR\\$0.01$} & \hd{SNR\\$0.005$} & \hd{SNR\\$0.01$} & \hd{SNR\\$0.005$} \\
\midrule
\multirow{5}{*}{$100$} & Fold'EM & $\mathbf{3.24}$ & $\mathbf{3.25}$ & $\mathbf{3.59}$ & $\mathbf{3.45}$ & $\mathbf{3.76}$ & $\mathbf{3.80}$ & $\mathbf{4.04}$ & $\mathbf{4.18}$ \\
 & cryoDRGN & $3.63$ & $3.95$ & $3.81$ & $3.96$ & $7.85$ & $9.80$ & $9.47$ & $10.03$ \\
 & RELION & $9.79$ & $9.79$ & $9.79$ & $9.79$ & $17.78$ & $16.84$ & $19.20$ & $18.29$ \\
 & AlphaFold3 & \multicolumn{2}{c|}{$8.21$} & \multicolumn{2}{c|}{$19.29$} & \multicolumn{2}{c|}{$7.29$} & \multicolumn{2}{c}{$8.84$} \\
 & PDB & \multicolumn{2}{c|}{$3.39$} & \multicolumn{2}{c|}{$3.47$} & \multicolumn{2}{c|}{$3.50$} & \multicolumn{2}{c}{$3.67$} \\
\midrule
\multirow{5}{*}{$250$} & Fold'EM & $\mathbf{3.22}$ & $\mathbf{3.22}$ & $\mathbf{3.39}$ & $\mathbf{3.43}$ & $\mathbf{3.75}$ & $\mathbf{3.73}$ & $\mathbf{3.99}$ & $\mathbf{3.89}$ \\
 & cryoDRGN & $3.39$ & $3.55$ & $3.59$ & $3.63$ & $7.00$ & $7.16$ & $7.23$ & $7.56$ \\
 & RELION & $7.92$ & $9.24$ & $6.66$ & $8.76$ & $10.32$ & $11.03$ & $13.71$ & $16.70$ \\
 & AlphaFold3 & \multicolumn{2}{c|}{$8.21$} & \multicolumn{2}{c|}{$19.29$} & \multicolumn{2}{c|}{$7.29$} & \multicolumn{2}{c}{$8.84$} \\
 & PDB & \multicolumn{2}{c|}{$3.39$} & \multicolumn{2}{c|}{$3.47$} & \multicolumn{2}{c|}{$3.50$} & \multicolumn{2}{c}{$3.67$} \\
\midrule
\multirow{5}{*}{$500$} & Fold'EM & $\mathbf{3.23}$ & $\mathbf{3.26}$ & $\mathbf{3.39}$ & $\mathbf{3.39}$ & $\mathbf{3.70}$ & $\mathbf{3.70}$ & $\mathbf{3.96}$ & $\mathbf{3.93}$ \\
 & cryoDRGN & $3.32$ & $3.35$ & $3.55$ & $3.59$ & $5.99$ & $6.88$ & $6.77$ & $7.40$ \\
 & RELION & $5.94$ & $6.40$ & $5.94$ & $6.40$ & $10.00$ & $10.32$ & $12.00$ & $12.39$ \\
 & AlphaFold3 & \multicolumn{2}{c|}{$8.21$} & \multicolumn{2}{c|}{$19.29$} & \multicolumn{2}{c|}{$7.29$} & \multicolumn{2}{c}{$8.84$} \\
 & PDB & \multicolumn{2}{c|}{$3.39$} & \multicolumn{2}{c|}{$3.47$} & \multicolumn{2}{c|}{$3.50$} & \multicolumn{2}{c}{$3.67$} \\
\midrule
\multirow{5}{*}{$1000$} & Fold'EM & $\mathbf{3.22}$ & $\mathbf{3.22}$ & $\mathbf{3.38}$ & $\mathbf{3.43}$ & $\mathbf{3.69}$ & $\mathbf{3.72}$ & $\mathbf{3.91}$ & $\mathbf{4.00}$ \\
 & cryoDRGN & $3.25$ & $3.32$ & $3.55$ & $3.55$ & $5.81$ & $5.91$ & $6.06$ & $6.95$ \\
 & RELION & $5.04$ & $5.74$ & $5.04$ & $6.16$ & $8.65$ & $10.00$ & $10.00$ & $10.53$ \\
 & AlphaFold3 & \multicolumn{2}{c|}{$8.21$} & \multicolumn{2}{c|}{$19.29$} & \multicolumn{2}{c|}{$7.29$} & \multicolumn{2}{c}{$8.84$} \\
 & PDB & \multicolumn{2}{c|}{$3.39$} & \multicolumn{2}{c|}{$3.47$} & \multicolumn{2}{c|}{$3.50$} & \multicolumn{2}{c}{$3.67$} \\
\bottomrule
\end{tabular}
\end{table}

\begin{table}[H]
\centering
\caption{\textbf{Quantitative evaluation of C$\alpha$ RMSD ($\downarrow$, \AA) and TM-score ($\uparrow$) on the synthetic cryo-EM datasets from Table~\ref{tab:synthetic}.} Fold'EM structures are obtained using $100, 250, 500, \&\, 1000$ particles with known poses at SNR $0.01$ and $0.005$; AlphaFold3 is the unguided baseline. Scores are computed with respect to the experimentally resolved structures deposited in the PDB. Entries highlighted in \textbf{bold} indicate the best result.}
\label{tab:syn_tm_rmsd_particles}
\small
\begin{tabular}{@{}cc|cc|cc|cc|cc@{}}
\toprule
\multicolumn{2}{c}{} & \multicolumn{2}{c}{\textbf{\texttt{9UGB}}} & \multicolumn{2}{c}{\textbf{\texttt{9UGC}}} & \multicolumn{2}{c}{\textbf{\texttt{8U4B}}} & \multicolumn{2}{c}{\textbf{\texttt{8U4E}}} \\
\cmidrule(lr){3-4} \cmidrule(lr){5-6} \cmidrule(lr){7-8} \cmidrule(lr){9-10}
\hd{Metric} & \hd{\#\,particles} & \hd{SNR\\$0.01$} & \hd{SNR\\$0.005$} & \hd{SNR\\$0.01$} & \hd{SNR\\$0.005$} & \hd{SNR\\$0.01$} & \hd{SNR\\$0.005$} & \hd{SNR\\$0.01$} & \hd{SNR\\$0.005$} \\
\midrule
\multirow{5}{*}{\makecell{C$\alpha$ RMSD\\($\downarrow$, \AA)}}
 & $100$      & $1.05$          & $1.01$          & $2.78$          & $1.45$          & $1.681$          & $1.907$          & $1.769$          & $2.116$          \\
 & $250$      & $0.88$          & $0.88$          & $1.34$          & $\mathbf{1.37}$ & $1.362$          & $1.350$          & $1.798$          & $1.512$          \\
 & $500$      & $\mathbf{0.81}$ & $1.12$          & $1.44$          & $1.41$          & $1.248$          & $\mathbf{1.227}$ & $\mathbf{1.503}$ & $\mathbf{1.479}$ \\
 & $1000$     & $0.84$          & $\mathbf{0.86}$ & $\mathbf{1.33}$ & $1.44$          & $\mathbf{1.132}$ & $1.228$          & $1.686$          & $1.711$          \\
\cmidrule(lr){2-10}
 & AlphaFold3 & \multicolumn{2}{c|}{$8.96$} & \multicolumn{2}{c|}{$11.22$} & \multicolumn{2}{c|}{$7.780$} & \multicolumn{2}{c}{$26.670$} \\
\midrule
\multirow{5}{*}{\makecell{TM-score\\($\uparrow$)}}
 & $100$      & $0.964$          & $0.965$          & $0.939$          & $0.983$          & $0.9832$          & $0.9785$          & $0.9810$          & $0.9731$          \\
 & $250$      & $0.967$          & $\mathbf{0.967}$ & $\mathbf{0.987}$ & $\mathbf{0.984}$ & $0.9887$          & $0.9890$          & $0.9807$          & $0.9861$          \\
 & $500$      & $\mathbf{0.968}$ & $0.962$          & $0.985$          & $\mathbf{0.984}$ & $0.9906$          & $0.9908$          & $\mathbf{0.9862}$ & $\mathbf{0.9866}$ \\
 & $1000$     & $0.967$          & $\mathbf{0.967}$ & $\mathbf{0.987}$ & $0.983$          & $\mathbf{0.9923}$ & $\mathbf{0.9909}$ & $0.9828$          & $0.9822$          \\
\cmidrule(lr){2-10}
 & AlphaFold3 & \multicolumn{2}{c|}{$0.699$} & \multicolumn{2}{c|}{$0.661$} & \multicolumn{2}{c|}{$0.7981$} & \multicolumn{2}{c}{$0.4772$} \\
\bottomrule
\end{tabular}
\end{table}

\begin{table}[H]
\centering
\caption{\textbf{MolProbity evaluation of Fold'EM structures over synthetic targets with known poses}. We report median MolProbity score $(\downarrow)$, clash score ($\downarrow$), and Ramachandran outlier percentage ($\downarrow$), computed using \texttt{phenix.molprobity} \citep{williams2018molprobity} over all relaxed structures in Table \ref{tab:syn_tm_rmsd_particles}.}
\label{tab:syn_molprobity_known}
\small
\begin{tabular}{@{}ccccccccc@{}}
\toprule
\multicolumn{1}{c}{} & \multicolumn{2}{c}{\textbf{\texttt{9UGB}}} & \multicolumn{2}{c}{\textbf{\texttt{9UGC}}} & \multicolumn{2}{c}{\textbf{\texttt{8U4B}}} & \multicolumn{2}{c}{\textbf{\texttt{8U4E}}} \\
\cmidrule(lr){2-3} \cmidrule(lr){4-5} \cmidrule(lr){6-7} \cmidrule(lr){8-9}
\hd{MolProbity statistic} & \hd{SNR\\$0.01$} & \hd{SNR\\$0.005$} & \hd{SNR\\$0.01$} & \hd{SNR\\$0.005$} & \hd{SNR\\$0.01$} & \hd{SNR\\$0.005$} & \hd{SNR\\$0.01$} & \hd{SNR\\$0.005$} \\
\midrule
MolProbity score      & $1.01$ & $1.05$    & $1.36$ & $1.44$ & $1.35$ & $1.35$ & $1.56$ & $1.54$ \\
Clashscore            & $2.29$ & $2.61$ & $3.38$ & $2.85$ & $1.59$ & $1.49$ & $3.01$ & $3.02$ \\
Ramachandran outliers & $0.00$    & $0.00$    & $0.28$ & $0.55$ & $1.49$ & $1.40$ & $1.72$ & $1.26$ \\
\bottomrule
\end{tabular}
\end{table}

\begin{table}[H]
\centering
\caption{\textbf{MolProbity evaluation of Fold'EM structures over real targets with known poses}. We report median MolProbity score $(\downarrow)$, clash score ($\downarrow$), and Ramachandran outlier percentage ($\downarrow$), computed using \texttt{phenix.molprobity} \citep{williams2018molprobity} over all structures in Table \ref{tab:real_tm_rmsd_fraction}.}
\label{tab:real_molprobity_known}
\small
\begin{tabular}{@{}cccccc@{}}
\toprule
\hd{MolProbity statistic} & \hd{\texttt{9B9J}} & \hd{\texttt{9B77}} & \hd{\texttt{8PWH}} & \hd{\texttt{9GXQ}} & \hd{\texttt{9GXR}} \\
\midrule
MolProbity score      & $0.84$ & $0.81$    & $1.71$ & $1.08$ & $1.07$ \\
Clashscore            & $0.38$ & $0.60$ & $5.42$ & $1.21$ & $0.99$ \\
Ramachandran outliers & $0.10$ & $0.24$ & $0.43$  & $0.16$ & $0.40$ \\
\bottomrule
\end{tabular}
\end{table}

\subsection{Homogeneous \textit{ab initio} reconstruction}
\label{app:tab_abinitio}
\begin{table}[H]
\centering
\caption{\textbf{Resolution at \mbox{Fourier shell correlation (FSC) $=0.143$} ($\downarrow$) on the experimental cryo-EM datasets from Table~\ref{tab:real}, with poses inferred \textit{ab initio}.} For Fold'EM \textit{ab initio}, AlphaFold3 and PDB structures, FSC is computed using \texttt{phenix.model\_map}. The number of particles used for the experiment is given below each identifier. Entries highlighted in \textbf{bold} indicate the best result.}
\label{tab:real_res_abinitio}
\small
\begin{tabular}{@{}cccccc@{}}
\toprule
\hd{Method\\\#\,particles} & \hd{\texttt{9B9J}\\$1000$} & \hd{\texttt{9B77}\\$1000$} & \hd{\texttt{8PWH}\\$5000$} & \hd{\texttt{9GXQ}\\$5000$} & \hd{\texttt{9GXR}\\$5000$} \\
\midrule
Fold'EM     & $\mathbf{3.52}$ & $\mathbf{2.85}$ & $10.83$              & $\mathbf{4.48}$ & $\mathbf{3.22}$ \\
cryoDRGN-AI & $35.90$         & $31.01$         & $32.80$         & $24.44$         & $26.88$         \\
RELION (\textit{ab initio})     & $11.62$         & $7.95$          & $\mathbf{9.84}$ & $6.25$          & $4.98$          \\
AlphaFold3  & $21.90$         & $4.40$          & $64.96$         & $11.70$         & $3.50$          \\
\midrule
PDB         & $2.38$          & $2.49$          & $3.06$          & $2.64$          & $2.89$          \\
\bottomrule
\end{tabular}
\end{table}

\begin{table}[H]
\centering
\caption{\textbf{Quantitative evaluation of TM-score ($\uparrow$) and $C\alpha$-RMSD ($\downarrow$) on the experimental cryo-EM datasets from Table~\ref{tab:real}, with poses inferred \textit{ab initio}.} The number of particles used is given below each identifier, and scores are computed with respect to the experimentally resolved structures deposited in the PDB. Entries highlighted in \textbf{bold} indicate the best result.}
\label{tab:abinitio_tm_rmsd_real}
\small
\setlength{\tabcolsep}{4pt}
\begin{tabular}{@{}c|ccccc|ccccc@{}}
\toprule
\multicolumn{1}{c}{} & \multicolumn{5}{c}{\textbf{C$\alpha$ RMSD ($\downarrow$, \AA)}} & \multicolumn{5}{c}{\textbf{TM-score ($\uparrow$)}} \\
\cmidrule(lr){2-6} \cmidrule(lr){7-11}
\hd{Method\\\#\,particles} & \hd{\texttt{9B9J}\\$1000$} & \hd{\texttt{9B77}\\$1000$} & \hd{\texttt{8PWH}\\$5000$} & \hd{\texttt{9GXQ}\\$5000$} & \hd{\texttt{9GXR}\\$5000$} & \hd{\texttt{9B9J}\\$1000$} & \hd{\texttt{9B77}\\$1000$} & \hd{\texttt{8PWH}\\$5000$} & \hd{\texttt{9GXQ}\\$5000$} & \hd{\texttt{9GXR}\\$5000$} \\
\midrule
Fold'EM    & $\mathbf{2.49}$ & $\mathbf{2.25}$ & $\mathbf{18.91}$ & $\mathbf{2.87}$ & $\mathbf{1.54}$ & $\mathbf{0.956}$ & $\mathbf{0.967}$ & $\mathbf{0.602}$ & $\mathbf{0.857}$ & $\mathbf{0.898}$ \\
AlphaFold3 & $21.90$         & $3.87$          & $33.16$          & $5.96$          & $2.01$          & $0.776$          & $0.916$          & $0.397$          & $0.734$          & $0.886$          \\
\bottomrule
\end{tabular}
\end{table}

\begin{table}[H]
\centering
\caption{\textbf{Resolution at \mbox{Fourier shell correlation (FSC) $=0.143$} ($\downarrow$) on the synthetic cryo-EM datasets from Table~\ref{tab:synthetic}, with poses inferred \textit{ab initio}.} For Fold'EM \textit{ab initio}, AlphaFold3 and PDB structures, FSC is computed using \texttt{phenix.model\_map}. All methods use $500$ particles at SNR $0.01$ and $0.005$. Entries highlighted in \textbf{bold} indicate the best result (excluding PDB).}
\label{tab:syn_res_abinitio}
\small
\setlength{\tabcolsep}{4pt}
\begin{tabular}{@{}cc|cc|cc|cc|cc@{}}
\toprule
\multicolumn{2}{c}{} & \multicolumn{2}{c}{\textbf{\texttt{9UGB}}} & \multicolumn{2}{c}{\textbf{\texttt{9UGC}}} & \multicolumn{2}{c}{\textbf{\texttt{8U4B}}} & \multicolumn{2}{c}{\textbf{\texttt{8U4E}}} \\
\cmidrule(lr){3-4} \cmidrule(lr){5-6} \cmidrule(lr){7-8} \cmidrule(lr){9-10}
\hd{\#\,particles} & \hd{Method} & \hd{SNR\\$0.01$} & \hd{SNR\\$0.005$} & \hd{SNR\\$0.01$} & \hd{SNR\\$0.005$} & \hd{SNR\\$0.01$} & \hd{SNR\\$0.005$} & \hd{SNR\\$0.01$} & \hd{SNR\\$0.005$} \\
\midrule
\multirow{5}{*}{$500$} & Fold'EM & $\mathbf{3.47}$ & $\mathbf{3.25}$ & $\mathbf{3.59}$ & $\mathbf{3.91}$ & $\mathbf{5.87}$ & $\mathbf{7.23}$ & $11.38$ & $11.53$ \\
 & cryoDRGN-AI & $12.80$ & $27.70$ & $33.30$ & $33.20$ & $22.59$ & $38.40$ & $38.40$ & $38.40$ \\
 & RELION (\textit{ab initio}) & $7.23$ & $8.93$ & $6.93$ & $7.56$ & $10.97$ & $16.00$ & $11.64$ & $15.36$ \\
 & AlphaFold3 & \multicolumn{2}{c|}{$8.21$} & \multicolumn{2}{c|}{$19.29$} & \multicolumn{2}{c|}{$7.29$} & \multicolumn{2}{c}{$\mathbf{8.84}$} \\
 & PDB & \multicolumn{2}{c|}{$3.39$} & \multicolumn{2}{c|}{$3.47$} & \multicolumn{2}{c|}{$3.5$} & \multicolumn{2}{c}{$3.67$} \\
\bottomrule
\end{tabular}
\end{table}

\begin{table}[H]
\centering
\caption{\textbf{Quantitative evaluation of TM-score ($\uparrow$) and $C\alpha$-RMSD ($\downarrow, \AA$) on the synthetic cryo-EM datasets from Table~\ref{tab:synthetic}, with poses inferred \textit{ab initio}.} Fold'EM \textit{ab initio} results are obtained using $500$ particles at SNR $0.01$ and $0.005$, and scores are computed with respect to the experimentally resolved structures deposited in the PDB. Entries highlighted in \textbf{bold} indicate the best result.}
\label{tab:abinitio_tm_rmsd_syn}
\small
\setlength{\tabcolsep}{4pt}
\begin{tabular}{@{}cc|cc|cc|cc|cc@{}}
\toprule
\multicolumn{2}{c}{} & \multicolumn{2}{c}{\textbf{\texttt{9UGB}}} & \multicolumn{2}{c}{\textbf{\texttt{9UGC}}} & \multicolumn{2}{c}{\textbf{\texttt{8U4B}}} & \multicolumn{2}{c}{\textbf{\texttt{8U4E}}} \\
\cmidrule(lr){3-4} \cmidrule(lr){5-6} \cmidrule(lr){7-8} \cmidrule(lr){9-10}
\hd{Metric} & \hd{Method} & \hd{SNR\\$0.01$} & \hd{SNR\\$0.005$} & \hd{SNR\\$0.01$} & \hd{SNR\\$0.005$} & \hd{SNR\\$0.01$} & \hd{SNR\\$0.005$} & \hd{SNR\\$0.01$} & \hd{SNR\\$0.005$} \\
\midrule
\multirow{2}{*}{\makecell{C$\alpha$ RMSD\\($\downarrow$, \AA)}}
 & Fold'EM    & $\mathbf{1.78}$ & $\mathbf{1.57}$ & $\mathbf{2.35}$ & $\mathbf{3.07}$ & $\mathbf{3.948}$ & $\mathbf{3.941}$ & $\mathbf{7.684}$ & $\mathbf{6.885}$ \\
\cmidrule(lr){2-10}
 & AlphaFold3 & \multicolumn{2}{c|}{$8.96$} & \multicolumn{2}{c|}{$11.22$} & \multicolumn{2}{c|}{$7.780$} & \multicolumn{2}{c}{$26.670$} \\
\midrule
\multirow{2}{*}{\makecell{TM-score\\($\uparrow$)}}
 & Fold'EM    & $\mathbf{0.969}$ & $\mathbf{0.976}$ & $\mathbf{0.948}$ & $\mathbf{0.917}$ & $\mathbf{0.9174}$ & $\mathbf{0.9182}$ & $\mathbf{0.810}$ & $\mathbf{0.809}$ \\
\cmidrule(lr){2-10}
 & AlphaFold3 & \multicolumn{2}{c|}{$0.699$} & \multicolumn{2}{c|}{$0.661$} & \multicolumn{2}{c|}{$0.7981$} & \multicolumn{2}{c}{$0.4772$} \\
\bottomrule
\end{tabular}
\end{table}

\begin{table}[H]
\centering
\caption{\textbf{MolProbity evaluation of the \textit{ab initio} Fold'EM structures in Table~\ref{tab:abinitio_tm_rmsd_syn}}. For each relaxed structure, we report MolProbity score $(\downarrow)$, clash score ($\downarrow$), and Ramachandran outlier percentage ($\downarrow$), computed using \texttt{phenix.molprobity} \citep{williams2018molprobity}.}
\label{tab:syn_molprobity_abinitio}
\small
\begin{tabular}{@{}c|cc|cc|cc|cc@{}}
\toprule
\multicolumn{1}{c}{} & \multicolumn{2}{c}{\textbf{\texttt{9UGB}}} & \multicolumn{2}{c}{\textbf{\texttt{9UGC}}} & \multicolumn{2}{c}{\textbf{\texttt{8U4B}}} & \multicolumn{2}{c}{\textbf{\texttt{8U4E}}} \\
\cmidrule(lr){2-3} \cmidrule(lr){4-5} \cmidrule(lr){6-7} \cmidrule(lr){8-9}
\hd{Statistic} & \hd{SNR\\$0.01$} & \hd{SNR\\$0.005$} & \hd{SNR\\$0.01$} & \hd{SNR\\$0.005$} & \hd{SNR\\$0.01$} & \hd{SNR\\$0.005$} & \hd{SNR\\$0.01$} & \hd{SNR\\$0.005$} \\
\midrule
MolProbity score      & $1.18$ & $1.15$ & $1.25$ & $1.15$ & $1.50$ & $1.53$ & $1.58$ & $1.79$ \\
Clashscore            & $3.72$ & $3.64$ & $4.83$ & $3.56$ & $1.21$ & $2.35$ & $1.56$ & $2.21$ \\
Ramachandran outliers & $0.00$ & $0.25$ & $0.12$ & $0.00$ & $1.26$ & $0.71$ & $0.98$ & $0.66$ \\
\bottomrule
\end{tabular}
\end{table}

\begin{table}[H]
\centering
\caption{\textbf{MolProbity evaluation of the \textit{ab initio} Fold'EM structures in Table \ref{tab:abinitio_tm_rmsd_real}}. For each relaxed structure, we report MolProbity score $(\downarrow)$, clash score ($\downarrow$), and Ramachandran outlier percentage ($\downarrow$), computed using \texttt{phenix.molprobity} \citep{williams2018molprobity}. Particle counts are as in Table~\ref{tab:real_res_abinitio}.}
\label{tab:real_molprobity_abinitio}
\small
\begin{tabular}{@{}cccccc@{}}
\toprule
\hd{Statistic} & \hd{\texttt{9B9J}} & \hd{\texttt{9B77}} & \hd{\texttt{8PWH}} & \hd{\texttt{9GXQ}} & \hd{\texttt{9GXR}} \\
\midrule
MolProbity score      & $0.93$ & $0.80$ & $1.48$ & $1.22$ & $1.11$ \\
Clashscore            & $0.82$ & $1.04$ & $4.01$ & $2.52$ & $1.57$ \\
Ramachandran outliers & $0.19$ & $0.00$ & $0.27$ & $0.16$ & $0.24$ \\
\bottomrule
\end{tabular}
\end{table}

\subsection{Heterogeneous reconstruction}
\label{app:tab_hetero}
\begin{table}[H]
\centering
\caption{\textbf{Resolution at \mbox{Fourier shell correlation (FSC) $=0.143$} ($\downarrow$) on the heterogeneous experimental dataset EMPIAR-13396, with known poses.} The two conformational states \texttt{9GXQ} and \texttt{9GXR} are resolved from the same mixed particle stack, and the sweep is over the number of particles per state. For Fold'EM, AlphaFold3 and PDB structures, FSC is computed using \texttt{phenix.model\_map}. Entries highlighted in \textbf{bold} indicate the best result.}
\label{tab:het_real_fsc0143}
\small
\begin{tabular}{@{}cc|cc@{}}
\toprule
\hd{\#\,particles\\per conf.} & \hd{Method} & \hd{\texttt{9GXQ}} & \hd{\texttt{9GXR}} \\
\midrule
\multirow{3}{*}{$500$} & Fold'EM & $6.25$ & $\mathbf{3.02}$ \\
 & RELION & $8.67$ & $8.40$ \\
 & cryoDRGN & $\mathbf{5.60}$ & $4.89$ \\
\midrule
\multirow{3}{*}{$1000$} & Fold'EM & $\mathbf{3.09}$ & $\mathbf{2.95}$ \\
 & RELION & $6.89$ & $10.75$ \\
 & cryoDRGN & $4.34$ & $4.41$ \\
\midrule
\multirow{3}{*}{$2000$} & Fold'EM & $\mathbf{3.16}$ & $\mathbf{3.01}$ \\
 & RELION & $5.97$ & $4.80$ \\
 & cryoDRGN & $4.27$ & $4.10$ \\
\midrule
\multirow{3}{*}{$5000$} & Fold'EM & $\mathbf{3.13}$ & $\mathbf{2.99}$ \\
 & RELION & $7.47$ & $7.07$ \\
 & cryoDRGN & $4.10$ & $4.03$ \\
\midrule
\multirow{3}{*}{$10000$} & Fold'EM & $\mathbf{3.16}$ & $\mathbf{2.99}$ \\
 & RELION & $6.56$ & $6.40$ \\
 & cryoDRGN & $4.00$ & $3.99$ \\
\midrule
\multicolumn{2}{c|}{AlphaFold3} & $11.69$ & $3.54$ \\
\multicolumn{2}{c|}{PDB}        & $2.64$  & $2.89$ \\
\bottomrule
\end{tabular}
\end{table}

\begin{table}[H]
\centering
\caption{\textbf{Quantitative evaluation of C$\alpha$ RMSD ($\downarrow$, \AA) and TM-score ($\uparrow$) on the heterogeneous experimental dataset EMPIAR-13396, with known poses.} Fold'EM resolves the two conformational states \texttt{9GXQ} and \texttt{9GXR} from the same mixed particle stack, and the sweep is over the number of particles per state. Entries highlighted in \textbf{bold} indicate the best result.}
\label{tab:het_real_rmsd_tm}
\small
\begin{tabular}{@{}c|cc|cc@{}}
\toprule
\multicolumn{1}{c}{} & \multicolumn{2}{c}{\textbf{C$\alpha$ RMSD ($\downarrow$, \AA)}} & \multicolumn{2}{c}{\textbf{TM-score ($\uparrow$)}} \\
\cmidrule(lr){2-3} \cmidrule(lr){4-5}
\hd{\#\,particles\\per conf.} & \hd{\texttt{9GXQ}} & \hd{\texttt{9GXR}} & \hd{\texttt{9GXQ}} & \hd{\texttt{9GXR}} \\
\midrule
$500$    & $3.30$          & $1.85$          & $0.840$          & $0.890$          \\
$1000$   & $1.63$          & $\mathbf{1.45}$ & $0.891$          & $0.899$          \\
$2000$   & $1.76$          & $1.71$          & $0.886$          & $0.894$          \\
$5000$   & $\mathbf{1.48}$ & $\mathbf{1.45}$ & $\mathbf{0.893}$ & $\mathbf{0.900}$ \\
$10000$  & $1.82$          & $1.73$          & $0.885$          & $0.894$          \\
\midrule
AlphaFold3 & $5.96$        & $2.01$          & $0.734$          & $0.886$          \\
\bottomrule
\end{tabular}
\end{table}

\begin{table}[H]
\centering
\caption{\textbf{Resolution at \mbox{Fourier shell correlation (FSC) $=0.143$} ($\downarrow$) on the heterogeneous experimental dataset EMPIAR-13396, with poses inferred \textit{ab initio}.} The two conformational states \texttt{9GXQ} and \texttt{9GXR} are resolved from the same mixed particle stack, and the sweep is over the number of particles per state. For Fold'EM \textit{ab initio}, AlphaFold3 and PDB structures, FSC is computed using \texttt{phenix.model\_map}. Entries highlighted in \textbf{bold} indicate the best result  (excluding PDB).}
\label{tab:het_real_fsc0143_abinitio}
\small
\begin{tabular}{@{}cc|cc@{}}
\toprule
\hd{\#\,particles\\per conf.} & \hd{Method} & \hd{\texttt{9GXQ}} & \hd{\texttt{9GXR}} \\
\midrule
\multirow{3}{*}{$500$} & Fold'EM & $\mathbf{8.67}$ & $3.58$ \\
 & RELION (\textit{ab initio}) & $22.40$ & $12.80$ \\
 & cryoDRGN-AI & $38.40$ & $26.90$ \\
\midrule
\multirow{3}{*}{$1000$} & Fold'EM & $\mathbf{7.47}$ & $4.20$ \\
 & RELION (\textit{ab initio}) & $38.40$ & $10.75$ \\
 & cryoDRGN-AI & $26.90$ & $22.40$ \\
\midrule
\multirow{3}{*}{$2000$} & Fold'EM & $\mathbf{11.69}$ & $\mathbf{3.24}$ \\
 & RELION (\textit{ab initio}) & $19.20$ & $10.34$ \\
 & cryoDRGN-AI & $24.40$ & $26.90$ \\
\midrule
\multirow{3}{*}{$5000$} & Fold'EM & $\mathbf{8.67}$ & $\mathbf{3.02}$ \\
 & RELION (\textit{ab initio}) & $13.44$ & $11.20$ \\
 & cryoDRGN-AI & $20.68$ & $22.40$ \\
\midrule
\multirow{3}{*}{$10000$} & Fold'EM & $\mathbf{4.63}$ & $\mathbf{3.09}$ \\
 & RELION (\textit{ab initio}) & $26.88$ & $8.40$ \\
 & cryoDRGN-AI & $17.92$ & $16.80$ \\
\midrule
\multicolumn{2}{c|}{AlphaFold3} & $11.69$ & $\mathbf{3.54}$ \\
\multicolumn{2}{c|}{PDB}        & $2.64$  & $2.89$ \\
\bottomrule
\end{tabular}
\end{table}

\begin{table}[H]
\centering
\caption{\textbf{Quantitative evaluation of C$\alpha$ RMSD ($\downarrow$, \AA) and TM-score ($\uparrow$) on the heterogeneous experimental dataset EMPIAR-13396, with poses inferred \textit{ab initio}.} Fold'EM \textit{ab initio} resolves the two conformational states \texttt{9GXQ} and \texttt{9GXR} from the same mixed particle stack, and the sweep is over the number of particles per state. Entries highlighted in \textbf{bold} indicate the best result.}
\label{tab:het_real_rmsd_tm_abinitio}
\small
\begin{tabular}{@{}c|cc|cc@{}}
\toprule
\multicolumn{1}{c}{} & \multicolumn{2}{c}{\textbf{C$\alpha$ RMSD ($\downarrow$, \AA)}} & \multicolumn{2}{c}{\textbf{TM-score ($\uparrow$)}} \\
\cmidrule(lr){2-3} \cmidrule(lr){4-5}
\hd{\#\,particles\\per conf.} & \hd{\texttt{9GXQ}} & \hd{\texttt{9GXR}} & \hd{\texttt{9GXQ}} & \hd{\texttt{9GXR}} \\
\midrule
$500$    & $4.54$          & $2.61$          & $0.799$          & $0.874$          \\
$1000$   & $5.48$          & $2.80$          & $0.781$          & $0.868$          \\
$2000$   & $5.84$          & $2.39$          & $0.737$          & $0.880$          \\
$5000$   & $5.06$          & $\mathbf{1.63}$ & $0.776$          & $0.895$          \\
$10000$  & $\mathbf{3.72}$ & $1.66$          & $\mathbf{0.836}$ & $\mathbf{0.896}$ \\
\midrule
AlphaFold3 & $5.96$        & $2.01$          & $0.734$          & $0.886$          \\
\bottomrule
\end{tabular}
\end{table}

\begin{table}[H]
\centering
\caption{\textbf{Resolution at \mbox{Fourier shell correlation (FSC) $=0.143$} ($\downarrow$) on the heterogeneous synthetic \texttt{9UGB}/\texttt{9UGC} mixture, with known poses.} The two conformational states are resolved from the same mixed particle stack at SNR $0.01$ and $0.005$, and the sweep is over the number of particles per state. Entries highlighted in \textbf{bold} indicate the best result.}
\label{tab:het_syn_fsc0143}
\small
\begin{tabular}{@{}cc|cc|cc@{}}
\toprule
\multicolumn{2}{c}{} & \multicolumn{2}{c}{\textbf{\texttt{9UGB}}} & \multicolumn{2}{c}{\textbf{\texttt{9UGC}}} \\
\cmidrule(lr){3-4} \cmidrule(lr){5-6}
\hd{\#\,particles\\per conf.} & \hd{Method} & \hd{SNR\\$0.01$} & \hd{SNR\\$0.005$} & \hd{SNR\\$0.01$} & \hd{SNR\\$0.005$} \\
\midrule
\multirow{3}{*}{$250$} & Fold'EM & $\mathbf{3.25}$ & $\mathbf{3.28}$ & $\mathbf{3.43}$ & $\mathbf{3.47}$ \\
 & RELION & $9.24$ & $10.40$ & $9.24$ & $11.09$ \\
 & cryoDRGN & $4.38$ & $6.16$ & $6.40$ & $8.76$ \\
\midrule
\multirow{3}{*}{$500$} & Fold'EM & $\mathbf{3.24}$ & $\mathbf{3.25}$ & $\mathbf{3.39}$ & $\mathbf{3.43}$ \\
 & RELION & $7.92$ & $8.76$ & $8.32$ & $9.24$ \\
 & cryoDRGN & $3.62$ & $4.50$ & $4.75$ & $6.16$ \\
\midrule
\multirow{3}{*}{$1000$} & Fold'EM & $\mathbf{3.22}$ & $\mathbf{3.25}$ & $\mathbf{3.43}$ & $\mathbf{3.39}$ \\
 & RELION & $6.93$ & $7.93$ & $7.23$ & $7.92$ \\
 & cryoDRGN & $3.70$ & $3.62$ & $4.38$ & $4.75$ \\
\midrule
\multicolumn{2}{c|}{AlphaFold3} & \multicolumn{2}{c|}{$8.21$} & \multicolumn{2}{c}{$19.29$} \\
\multicolumn{2}{c|}{PDB}        & \multicolumn{2}{c|}{$3.39$} & \multicolumn{2}{c}{$3.47$} \\
\bottomrule
\end{tabular}
\end{table}

\begin{table}[H]
\centering
\caption{\textbf{Quantitative evaluation of C$\alpha$ RMSD ($\downarrow$, \AA) and TM-score ($\uparrow$) on the heterogeneous synthetic \texttt{9UGB}/\texttt{9UGC} mixture, with known poses.} Fold'EM resolves the two conformational states from the same mixed particle stack at SNR $0.01$ and $0.005$, and the sweep is over the number of particles per state. Entries highlighted in \textbf{bold} indicate the best result.}
\label{tab:het_syn_rmsd_tm}
\small
\begin{tabular}{@{}ccccc|cccc@{}}
\toprule
 & \multicolumn{4}{c}{\textbf{C$\alpha$ RMSD ($\downarrow$, \AA)}} & \multicolumn{4}{c}{\textbf{TM-score ($\uparrow$)}} \\
\cmidrule(lr){2-5} \cmidrule(lr){6-9}
 & \multicolumn{2}{c}{\textbf{\texttt{9UGB}}} & \multicolumn{2}{c}{\textbf{\texttt{9UGC}}} & \multicolumn{2}{c}{\textbf{\texttt{9UGB}}} & \multicolumn{2}{c}{\textbf{\texttt{9UGC}}} \\
\cmidrule(lr){2-3} \cmidrule(lr){4-5} \cmidrule(lr){6-7} \cmidrule(lr){8-9}
\hd{\#\,particles\\per conf.} & \hd{SNR\\$0.01$} & \hd{SNR\\$0.005$} & \hd{SNR\\$0.01$} & \hd{SNR\\$0.005$} & \hd{SNR\\$0.01$} & \hd{SNR\\$0.005$} & \hd{SNR\\$0.01$} & \hd{SNR\\$0.005$} \\
\midrule
$250$  & $\mathbf{0.76}$ & $0.99$          & $1.55$          & $1.38$          & $\mathbf{0.964}$ & $0.960$          & $\mathbf{0.987}$ & $0.982$          \\
$500$  & $0.99$          & $0.89$          & $1.36$          & $\mathbf{1.28}$ & $0.960$          & $0.962$          & $0.982$          & $0.983$          \\
$1000$ & $0.93$          & $\mathbf{0.81}$ & $\mathbf{1.18}$ & $1.59$          & $0.961$          & $\mathbf{0.963}$ & $0.986$          & $\mathbf{0.985}$ \\
\midrule
AlphaFold3 & \multicolumn{2}{c}{$8.96$} & \multicolumn{2}{c|}{$11.22$} & \multicolumn{2}{c}{$0.699$} & \multicolumn{2}{c}{$0.661$} \\
\bottomrule
\end{tabular}
\end{table}

\begin{table}[H]
\centering
\caption{\textbf{Resolution at \mbox{Fourier shell correlation (FSC) $=0.143$} ($\downarrow$) on the heterogeneous synthetic \texttt{9UGB}/\texttt{9UGC} mixture, with poses inferred \textit{ab initio}.} The two conformational states are resolved from the same mixed particle stack at SNR $0.01$ and $0.005$, and the sweep is over the number of particles per state. Entries highlighted in \textbf{bold} indicate the best result.}
\label{tab:het_syn_fsc0143_abinitio}
\small
\begin{tabular}{@{}cc|cc|cc@{}}
\toprule
\multicolumn{2}{c}{} & \multicolumn{2}{c}{\textbf{\texttt{9UGB}}} & \multicolumn{2}{c}{\textbf{\texttt{9UGC}}} \\
\cmidrule(lr){3-4} \cmidrule(lr){5-6}
\hd{\#\,particles\\per conf.} & \hd{Method} & \hd{SNR\\$0.01$} & \hd{SNR\\$0.005$} & \hd{SNR\\$0.01$} & \hd{SNR\\$0.005$} \\
\midrule
\multirow{3}{*}{$250$} & Fold'EM & $\mathbf{3.22}$ & $\mathbf{7.61}$ & $\mathbf{3.63}$ & $\mathbf{4.81}$ \\
 & RELION (\textit{ab initio}) & $8.32$ & $8.32$ & $23.77$ & $8.76$ \\
 & cryoDRGN-AI & $33.28$ & $33.28$ & $41.60$ & $18.49$ \\
\midrule
\multirow{3}{*}{$500$} & Fold'EM & $\mathbf{3.25}$ & $\mathbf{3.35}$ & $\mathbf{3.51}$ & $\mathbf{3.55}$ \\
 & RELION (\textit{ab initio}) & $6.93$ & $7.92$ & $7.23$ & $8.32$ \\
 & cryoDRGN-AI & $27.73$ & $20.80$ & $33.28$ & $27.73$ \\
\midrule
\multirow{3}{*}{$1000$} & Fold'EM & $\mathbf{3.22}$ & $\mathbf{3.25}$ & $\mathbf{3.51}$ & $\mathbf{3.51}$ \\
 & RELION (\textit{ab initio}) & $6.93$ & $7.92$ & $7.23$ & $7.92$ \\
 & cryoDRGN-AI & $28.81$ & $16.64$ & $27.73$ & $27.73$ \\
\midrule
\multicolumn{2}{c|}{AlphaFold3} & \multicolumn{2}{c|}{$8.21$} & \multicolumn{2}{c}{$19.29$} \\
\multicolumn{2}{c|}{PDB}        & \multicolumn{2}{c|}{$3.39$} & \multicolumn{2}{c}{$3.47$} \\
\bottomrule
\end{tabular}
\end{table}

\begin{table}[H]
\centering
\caption{\textbf{Quantitative evaluation of C$\alpha$ RMSD ($\downarrow$, \AA) and TM-score ($\uparrow$) on the heterogeneous synthetic \texttt{9UGB}/\texttt{9UGC} mixture, with poses inferred \textit{ab initio}.} Fold'EM resolves the two conformational states from the same mixed particle stack at SNR $0.01$ and $0.005$, and the sweep is over the number of particles per state. Entries highlighted in \textbf{bold} indicate the best result.}
\label{tab:het_syn_rmsd_tm_abinitio}
\small
\begin{tabular}{@{}ccccc|cccc@{}}
\toprule
 & \multicolumn{4}{c}{\textbf{C$\alpha$ RMSD ($\downarrow$, \AA)}} & \multicolumn{4}{c}{\textbf{TM-score ($\uparrow$)}} \\
\cmidrule(lr){2-5} \cmidrule(lr){6-9}
 & \multicolumn{2}{c}{\textbf{\texttt{9UGB}}} & \multicolumn{2}{c}{\textbf{\texttt{9UGC}}} & \multicolumn{2}{c}{\textbf{\texttt{9UGB}}} & \multicolumn{2}{c}{\textbf{\texttt{9UGC}}} \\
\cmidrule(lr){2-3} \cmidrule(lr){4-5} \cmidrule(lr){6-7} \cmidrule(lr){8-9}
\hd{\#\,particles\\per conf.} & \hd{SNR\\$0.01$} & \hd{SNR\\$0.005$} & \hd{SNR\\$0.01$} & \hd{SNR\\$0.005$} & \hd{SNR\\$0.01$} & \hd{SNR\\$0.005$} & \hd{SNR\\$0.01$} & \hd{SNR\\$0.005$} \\
\midrule
$250$  & $1.52$          & $4.13$          & $2.84$          & $5.09$          & $0.967$          & $0.862$          & $0.929$          & $0.876$          \\
$500$  & $\mathbf{1.50}$ & $1.95$          & $2.71$          & $2.61$          & $\mathbf{0.968}$ & $\mathbf{0.953}$ & $0.936$          & $0.939$          \\
$1000$ & $1.97$          & $\mathbf{1.71}$ & $\mathbf{2.42}$ & $\mathbf{2.17}$ & $0.952$          & $0.951$          & $\mathbf{0.948}$ & $\mathbf{0.957}$ \\
\midrule
AlphaFold3 & \multicolumn{2}{c}{$8.96$} & \multicolumn{2}{c|}{$11.22$} & \multicolumn{2}{c}{$0.699$} & \multicolumn{2}{c}{$0.661$} \\
\bottomrule
\end{tabular}
\end{table}

\begin{table}[H]
\centering
\caption{\textbf{MolProbity evaluation of Fold'EM structures on the heterogeneous synthetic \texttt{9UGB}/\texttt{9UGC} mixture}. We report median MolProbity score $(\downarrow)$, clash score ($\downarrow$), and Ramachandran outlier percentage ($\downarrow$), computed using \texttt{phenix.molprobity} \citep{williams2018molprobity} over all structures in Tables \ref{tab:het_syn_rmsd_tm} and \ref{tab:het_syn_rmsd_tm_abinitio}.}
\label{tab:het_syn_molprobity}
\small
\begin{tabular}{@{}cc|cc|cc@{}}
\toprule
\multicolumn{2}{c}{} & \multicolumn{2}{c}{\textbf{\texttt{9UGB}}} & \multicolumn{2}{c}{\textbf{\texttt{9UGC}}} \\
\cmidrule(lr){3-4} \cmidrule(lr){5-6}
\hd{Poses} & \hd{MolProbity statistic} & \hd{SNR\\$0.01$} & \hd{SNR\\$0.005$} & \hd{SNR\\$0.01$} & \hd{SNR\\$0.005$} \\
\midrule
\multirow{3}{*}{Known}
 & MolProbity score      & $1.16$ & $1.13$ & $1.09$ & $1.11$ \\
 & Clashscore            & $3.72$ & $3.40$ & $3.01$ & $3.16$ \\
 & Ramachandran outliers & $0.00$ & $0.00$ & $0.12$ & $0.00$ \\
\midrule
\multirow{3}{*}{\textit{Ab initio}}
 & MolProbity score      & $1.39$ & $1.48$ & $1.44$ & $1.46$ \\
 & Clashscore            & $5.85$ & $7.28$ & $6.65$ & $6.72$ \\
 & Ramachandran outliers & $0.12$ & $0.12$ & $0.12$ & $0.25$ \\
\bottomrule
\end{tabular}
\end{table}

\begin{table}[H]
\centering
\caption{\textbf{MolProbity evaluation of Fold'EM structures on the heterogeneous experimental dataset EMPIAR-13396}. We report median MolProbity score $(\downarrow)$, clash score ($\downarrow$), and Ramachandran outlier percentage ($\downarrow$), computed using \texttt{phenix.molprobity} \citep{williams2018molprobity} over all structures in Tables \ref{tab:het_real_rmsd_tm} and \ref{tab:het_real_rmsd_tm_abinitio}.}
\label{tab:het_real_molprobity}
\small
\begin{tabular}{@{}cc|cc@{}}
\toprule
\hd{Poses} & \hd{MolProbity statistic} & \hd{\texttt{9GXQ}} & \hd{\texttt{9GXR}} \\
\midrule
\multirow{3}{*}{Known}
 & MolProbity score      & $1.16$ & $1.20$ \\
 & Clashscore            & $1.48$ & $1.66$ \\
 & Ramachandran outliers & $0.16$ & $0.16$ \\
\midrule
\multirow{3}{*}{\textit{Ab initio}}
 & MolProbity score      & $1.32$ & $1.38$ \\
 & Clashscore            & $2.74$ & $2.61$ \\
 & Ramachandran outliers & $0.16$ & $0.08$ \\
\bottomrule
\end{tabular}
\end{table}

\subsection{Continuous heterogeneity on IgG-1D}
\label{app:tab_igg}
\begin{table}[H]
\centering
\caption{\textbf{MolProbity evaluation of Fold'EM structures on IgG-1D}. For each arm of the experiment, we report MolProbity score $(\downarrow)$, clash score ($\downarrow$), and Ramachandran outlier percentage ($\downarrow$) of the relaxed structures, computed using \texttt{phenix.molprobity} \citep{williams2018molprobity}, averaged over the eight anchors and over the $100$ interpolated states.}
\label{tab:igg_molprobity}
\small
\begin{tabular}{@{}ccccccc@{}}
\toprule
\multicolumn{1}{c}{} & \multicolumn{2}{c}{\textbf{cryoDRGN}} & \multicolumn{2}{c}{\textbf{RECOVAR}} & \multicolumn{2}{c}{\textbf{\brecovar{}}} \\
\cmidrule(lr){2-3} \cmidrule(lr){4-5} \cmidrule(lr){6-7}
\hd{MolProbity statistic} & \hd{Anchors\\($n=8$)} & \hd{Interpolated\\($n=100$)} & \hd{Anchors\\($n=8$)} & \hd{Interpolated\\($n=100$)} & \hd{Anchors\\($n=8$)} & \hd{Interpolated\\($n=100$)} \\
\midrule
MolProbity score      & $0.72$ & $0.71$ & $0.70$ & $0.72$ & $0.74$ & $0.72$ \\
Clashscore            & $0.59$ & $0.59$ & $0.59$ & $0.66$ & $0.66$ & $0.54$ \\
Ramachandran outliers & $0.15$ & $0.15$ & $0.15$ & $0.15$ & $0.15$ & $0.15$ \\
\bottomrule
\end{tabular}
\end{table}

\subsection{Confidence-based particle selection}

\begin{table}[H]
\centering
\caption{\textbf{Confidence-based particle selection described in Section \ref{app:class_assignments} prevents mode collapse in the ensemble.} On the heterogeneous experimental dataset EMPIAR-13396, embeddings optimized using high-confidence particles typically yield better-separated conformations, accompanied by higher TM-scores to the corresponding conformations. The ``separation'' column reports the difference between the RMSD of the true conformations \texttt{9GXQ} and \texttt{9GXR} and that of the predicted conformations; smaller values indicate closer agreement with the true conformational separation. \textbf{Bold} entries indicate the best result.}
\label{tab:confidence_tm_sep}
\small
\begin{tabular}{@{}ccccc@{}}
\toprule
& & \multicolumn{2}{c}{\textbf{TM-score ($\uparrow$)}} & \\
\cmidrule(lr){3-4}
\hd{\#\,particles\\per conf.} & \hd{Confidence\\method} & \hd{\texttt{9GXQ}} & \hd{\texttt{9GXR}} & \hd{Separation ($\downarrow$)} \\
\midrule
\multirow{2}{*}{$7.5$k} & Cut by confidence & $\mathbf{0.835}$ & $\mathbf{0.894}$ & $\mathbf{0.58}$ \\
                        & No confidence     & $0.822$          & $0.888$          & $1.35$          \\
\midrule
\multirow{2}{*}{$10$k}  & Cut by confidence & $\mathbf{0.832}$ & $\mathbf{0.894}$ & $\mathbf{0.47}$ \\
                        & No confidence     & $0.823$          & $0.888$          & $1.48$          \\
\bottomrule
\end{tabular}
\end{table}

\section{Algorithms}
\label{app:algorithms}
Algorithm~\ref{alg:FoldEM_alg} gives the full Fold'EM procedure described in Section~\ref{sec:methods}: the coordinate ascent that alternates structure inference by guided reverse diffusion with the refinement of the nuisance variables. Algorithm~\ref{alg:residual_mode} gives a single replicate of the residual-subspace estimator of \brecovar{} (Appendix~\ref{app:class_assignments}), which provides the initial class assignment in the \textit{ab initio} heterogeneous setting.

\renewcommand{\algorithmiccomment}[1]{\algcmt{#1}} 
\begin{algorithm}[H]
      \caption{Fold'EM}
      \label{alg:FoldEM_alg}
      \begin{algorithmic}[1]
    \STATE \textbf{Input:} Initial trunk embeddings $\ens{H}$; particle images $\ens{Y}$; initial nuisance variables $\boldsymbol{\eta}^{(0)}$; noise schedule $[\sigma_{T}, \dots, \sigma_0]$; learning rate $\gamma_{h}$; ensemble size $n$; coordinate iterations $R$; diffusion runs $Q$; prior weight $\lambda_p$
    \STATE \textbf{Output:} Structural ensemble $\ens{X}=\{\mathbf{X}^{1},\ldots,\mathbf{X}^{n}\}$ decoded from the optimized Pairformer states $g(\ens{H})$
    \FOR[Coordinate ascent]{$r = 0$ to $R-1$} 
        \FOR[Step 1: structure inference, Equation~\ref{eq:structure_step}]{$q = 1$ to $Q$ }
            \STATE $\ens{X} \sim \sigma_T \cdot [\mathbf{N}^1, \dots, \mathbf{N}^{n}]^T$ \algcmt{$\mathbf{N}^{i} \sim \mathcal{N}(\mathbf{0}, \mathbf{I})$}
            \FOR{$t \in [T-1, \dots, 0]$}
              \STATE $\ens{H} \gets  \ens{H} + \gamma_{h}\nabla_{\ens{H}}\Big(  \log p\!\left(\ens{Y}\mid D_{\theta}(\ens{X}; g(\ens{H}), t), \boldsymbol{\eta}^{(r)}\right) + \lambda_{\mathrm{p}} \log p(\ens{H}|\mathbf{a}) \Big)$ \algcmt{Equation~\ref{eq:emb_update}}
              \STATE $\ens{X} \gets \mathrm{ReverseSDE}(\ens{X}; g(\ens{H}),  t, \sigma_t)$ \algcmt{Equation~\ref{eq:sde}}
            \ENDFOR
        \ENDFOR
        \STATE $\boldsymbol{\eta}^{(r+1)} \gets \arg\max_{\boldsymbol{\eta}} \Big( \log p(\ens{Y} \mid \ens{X}, \boldsymbol{\eta}) + \log p(\boldsymbol{\eta}) \Big)$ \algcmt{Step 2: pose and class refinement, Equation~\ref{eq:nuisance_step}}
    \ENDFOR
    \STATE $\ens{Z} \gets g(\ens{H})$ \algcmt{Optimized conditioning variables}
    \STATE $\ens{X} \sim p(\ens{X} \mid \ens{Z})$ \algcmt{Decode ensemble: unguided reverse diffusion, Equation~\ref{eq:sde}}
    \STATE \textbf{return} $\ens{X}$ \algcmt{Ensemble from the final reverse-diffusion run, conditioned on $g(\ens{H})$}
    \end{algorithmic}
\end{algorithm}

\begin{algorithm}[h]
\caption{\brecovar{} residual-subspace estimator for one replicate}
\label{alg:residual_mode}
\begin{algorithmic}[1]
\STATE \textbf{Input:} particles with poses and CTFs; band \(\Omega_{\mathrm a}\); number of classes \(n\) and of modes \(q\); fitting subset \(\mathcal I\); mask fraction \(f\); \(T\) alternating updates; \(L\) refinement iterations
\STATE Reconstruct the consensus \(\hat\rho\) from all particles and derive the solvent mask \(\mathcal M\) from it
\STATE Scale every particle to the consensus and subtract its projection \algcmt{residuals \(\mathbf r_k\)}
\FOR{\(m=1,\ldots,q\)}
  \STATE From a random start, alternate \(T\) times between back-projecting the residuals of \(\mathcal I\) into a masked mode \(W_m\) and projecting them onto it \algcmt{alternating least squares}
  \STATE Remove the fitted mode from the residuals \algcmt{deflation}
\ENDFOR
\STATE Fit the coefficients \(\hat{\mathbf b}_k\) of every particle on all modes jointly \algcmt{joint coefficients}
\STATE Cluster \(\{\hat{\mathbf b}_k\}\) into \(n\) classes by \(k\)-means \algcmt{initial labels}
\STATE Refine the labels by \(L\) cross-validated updates of \eqref{eq:lf_centred} \algcmt{two folds, centered rule}
\STATE \textbf{return} labels \(\{\hat c_k\}_{k=1}^K\)
\end{algorithmic}
\end{algorithm}

\clearpage
\section{Additional details}
\label{app:details}
\subsection{Pre-processing}
\label{app:preprocessing}
\paragraph{Synthetic datasets.} All synthetic datasets were prepared using the PDB deposited structure \citep{burley2017protein}. First, noiseless particles $\mathbf{y}$ were rendered by projecting the experimental PDB structure $\mathbf{X}^{\mathrm{expt}}$ using the image formation model in Section \ref{sec:particle_from_structure}. White Gaussian noise was added to the noiseless particles in the Fourier domain to match a specific signal to noise ratio (SNR).
\begin{equation*}
\hat{\mathbf{y}}^{\mathrm{noisy}} = \hat{\mathbf{y}}
+ \sigma_{\mathrm{noise}}\hat{\mathbf{N}}, \qquad \mathbf{N}\sim\mathcal{N}(\mathbf{0},\mathbf{I}),
\end{equation*}
The noise scale $\sigma_{\mathrm{noise}}$ was set using,
\begin{equation*}
\sigma_{\mathrm{noise}}
=
\frac{\operatorname{std}(\hat{\mathbf{y}})}
{\sqrt{\mathrm{SNR}}}.
\end{equation*}
Where $\mathrm{SNR}$ is the desired SNR for the dataset. Thus, the resulting particle satisfies
\begin{equation*}
\mathrm{SNR} = \frac{\mathrm{Var}[\hat{\mathbf{y}}]}
{\mathrm{Var}[\sigma_{\mathrm{noise}}\hat{\mathbf{N}}]}.
\end{equation*}
In all our synthetic experiments, we set $\mathrm{SNR}=0.01$ or $0.005$.
\paragraph{Real datasets.} All experimental datasets were downloaded from EMPIAR \citep{iudin2016empiar}. We preprocess the deposited particle ensemble generation. We only retain the particles listed in the depositors' final STAR files, and exclude particles identified as junk. The particles are then centered and Fourier-cropped to $128\times128$ pixels, followed by phase flipping. We then apply a soft circular mask with a radius equal to $0.4$ times the box size.

\subsection{Baselines}
\label{app:baselines}
\paragraph{Known poses.} We primarily compare against RELION \citep{zivanov2018new} and CryoDRGN \citep{zhong2021cryodrgn}. Both methods reconstruct 3D density from particle images. In the commands below, angle brackets denote dataset-specific inputs.
\begin{itemize}
    \item \textbf{CryoDRGN.} With known poses, CryoDRGN  \citep{zhong2021cryodrgn} volumes were trained using the following command:
\begin{shellcmd}
cryodrgn train_vae <particles.mrcs>
    --poses <poses.pkl> --ctf <ctf.pkl> [--ind <ind.pkl>]
    --zdim 8 -n 50 --no-amp
\end{shellcmd}
    \item \textbf{RELION.} With known poses, RELION \citep{zivanov2018new} volumes were generated using the following command:
\begin{shellcmd}
relion_refine --ctf --skip_align --iter 5
    --particle_diameter <diameter> --tau2_fudge 1
    --split_random_halves --flatten_solvent
    --solvent_mask <mask.mrc>
\end{shellcmd}
    Here, \texttt{<mask.mrc>} is a low-pass reference generated from the EMD map with a $60$\,\AA{} blob mask, and the particle diameter \texttt{<diameter>} depends on the protein.
\end{itemize}
\paragraph{\textit{Ab initio}.} We primarily compare against RELION (\textit{ab initio}) and CryoDRGN-AI \citep{levy2025cryodrgn}. These methods jointly estimate particle poses and 3D structure from the particle images without externally supplied orientations.
\begin{itemize}
    \item \textbf{CryoDRGN-AI.} CryoDRGN-AI volumes were trained using the following command:
\begin{shellcmd}
cryodrgn abinit <particles.mrcs> --ctf <ctf.pkl> --o <outdir>
    --zdim 8 --num-epochs 50 --no-amp --seed 0
    --num-workers 0 --lazy --checkpoint 50
\end{shellcmd}
    \item \textbf{RELION (\textit{ab initio}).} Without poses, RELION volumes were generated in two stages. First, an initial model was constructed using VDAM without externally provided poses:
\begin{shellcmd}
relion_refine --grad --denovo_3dref
    --i <particles.star> --o <init_dir>/run --iter 200 --K 1
    --sym <sym> --random_seed <seed> --particle_diameter <diameter>
    --flatten_solvent --zero_mask --ctf
    --ini_high 40 --oversampling 1 --healpix_order 1
    --offset_range 6 --offset_step 2 --pool 30 --pad 1
    --skip_gridding --dont_combine_weights_via_disc --gpu --j <threads>
\end{shellcmd}
    We then performed 3D auto-refinement from the resulting initial model:
\begin{shellcmd}
mpirun --oversubscribe --bind-to none -n 3 relion_refine_mpi
    --auto_refine --split_random_halves
    --i <particles.star> --ref <initial_model.mrc> --o <refine_dir>/run
    --ini_high 30 --random_seed <seed> --ctf --sym <sym>
    --particle_diameter <diameter> --flatten_solvent --zero_mask
    --oversampling 1 --healpix_order 2 --auto_local_healpix_order 4
    --offset_range 5 --offset_step 2 --low_resol_join_halves 40
    --norm --scale --pool 30 --pad 2
    --dont_combine_weights_via_disc --gpu --j <threads>
\end{shellcmd}
    In all Fold'EM \textit{ab initio} experiments, we use the poses and initial map generated in the first stage for the subsequent refinement.
\end{itemize}

\paragraph{Heterogeneous reconstruction.} For heterogeneous datasets, we first clustered the particles into structural groups and then reconstructed each density separately using the corresponding particle subset. This provides a density estimate for each discrete structural class without introducing additional heterogeneity modelling assumptions.

\subsubsection{AlphaFold3 and hardware resources}\label{sec:hardware}
For all our experiments, we used Protenix $\mathrm{v1.0.4}$ \citep{Zhang2026.02.05.703733} -- a PyTorch reimplementation of AlphaFold3 \citep{abramson2024accurate}. For map and pose estimation, we used RELION $5.0.0$ \citep{zivanov2018new}. For structural validation, we used Phenix \citep{liebschner2019macromolecular} $\mathrm{v}1.21.2-5419$. All experiments were carried out on NVIDIA H100 and L40S GPUs running Debian GNU/Linux 12.

\subsubsection{Collision and bond length loss} \label{app:losses}
To prevent the embedding optimization from generating physically invalid protein structures, we incorporate numerous stereochemical regularizers. Given an ensemble of conformations $\ens{X} = \{\mathbf{X}^{1}, \dots, \mathbf{X}^{n}\}$ corresponding to amino acid sequence $\mathbf{a}$, we minimize three loss functions that avoid elongated bonds and steric clashes.
\paragraph{Bond length loss.} The bond-length loss for a structure with $m$ atoms is given by,
\begin{equation*}
    \mathcal{L}_{\mathrm{bond}} (\mathbf{X}; \mathbf{a}) = \sum_{i=1}^{m} \sum_{j=i+1}^{m} B_{ij} \cdot ( \max(0, | d_{ij} - d^{\mathrm{ideal}}_{ij}| - \delta_{\mathrm{bond}}))^2
\end{equation*}
Here, $\mathbf{B} \in \mathbb{Z}^{m \times m}$ is a mask with every element $b_{ij} \in \{0, 1\}$. If atoms $i$ and $j$ are covalently bonded, $b_{ij} = 1$, otherwise $b_{ij} = 0$. $d^{\mathrm{ideal}}_{ij}$ is the sum of covalent radii of atoms $i$, $j$, and $d_{ij} = \|\mathbf{x}_{i} - \mathbf{x}_{j}\|_{2}$. $\delta_{\mathrm{bond}}=0.2\,\text{\AA}$ is a tolerance margin. 
\paragraph{Collision loss.} Collisions are caused when any two non-bonded atoms are close in space. To prevent this, we minimize:
\begin{equation}
\mathcal{L}_{\mathrm{collision}}  (\mathbf{X}; \mathbf{a}) = \sum_{i=1}^{m} \max_{j \neq i, B_{ij} = 0} (\max(0, (d^{\mathrm{ideal}}_{ij} + p^{\mathrm{collision}}) - d_{ij}))
\end{equation}
Here, $p^{\mathrm{collision}} = 0.4\,\text{\AA}$ is a padding distance to prevent over-penalization of near-contact atoms. Together, the resulting validity log-likelihood of an ensemble of conformations is given by
\begin{equation}\label{eq:validity}
     \log p (\mathbf{B} \mid \ens{X}, \mathbf{a}) = -  \sum_{k=1}^{n} (\lambda_{\mathrm{bond}} \mathcal{L}_{\mathrm{bond}} (\mathbf{X}^{k}; \mathbf{a}) + \lambda_{\mathrm{collision}} \mathcal{L}_{\mathrm{collision}} (\mathbf{X}^{k}; \mathbf{a}))
\end{equation}

\subsubsection{Relaxation} \label{app:relaxation}
Despite being regularized with loss functions that prevent collision and bond-violation, some minor clashes can still persist. Hence, all outputs of Fold'EM are relaxed using AMBER14 force field \citep{hornak2006comparison} in vacuum, in order to relax nucleic acids too. The energy of the structures is minimized for a maximum of $2000$ iterations with an energy tolerance threshold of $2.39 \,\mathrm{kcal}/\mathrm{mol}$ and stiffness of $10.0 \,\mathrm{kcal}/\mathrm{mol}/\AA^{2}$

\subsubsection{Metrics} \label{app:metrics}
\paragraph{$\mathrm{C}{\alpha}$ Root mean square deviation ($\mathrm{C}{\alpha}$ RMSD).} Given a structure $\mathbf{X}$ generated by Fold'EM, we measure its structural accuracy relative to the experimentally resolved structure $\mathbf{X}^{\mathrm{expt}}$ using the RMSD over matched $\mathrm{C\alpha}$ atoms. We first align $\mathbf{X}$ to $\mathbf{X}^{\mathrm{expt}}$ using the optimal rotation $\mathbf{R}^{\mathrm{kabsch}}$ and translation $\mathbf{t}^{\mathrm{kabsch}}$ obtained by Kabsch rigid-body alignment \citep{kabsch1976solution}. We then compute,
\begin{equation*}
    \mathrm{RMSD}_{\mathrm{C\alpha}} = \sqrt{\frac{1}{N_{\mathrm{C\alpha}}} \sum_{i \in \mathcal{I}_{\mathrm{C\alpha}}}^{m} \cdot |\mathbf{R}^{\mathrm{kabsch}} \mathbf{x}_{i} + \mathbf{t}^{\mathrm{kabsch}} - \mathbf{x}^{\mathrm{expt}}_{i}|_{2}^{2}}
\end{equation*}
where $\mathcal{I}_{\mathrm{C\alpha}}$ denotes the set of matched $\mathrm{C\alpha}$ atoms and $N_{\mathrm{C\alpha}} = |\mathcal{I}_{\mathrm{C\alpha}}|$.
\paragraph{Template Modeling score (TM-score).} For a Fold'EM generated structure $\mathbf{X}$ and experimental structure $\mathbf{X}^{\mathrm{expt}}$, the Template Modeling score (TM-score) measures the global structural similarity \citep{zhang2004scoring}. In contrast to RMSD, the TM-score is less sensitive to a small number of large local deviations. It is defined as 

\begin{equation}
    \mathrm{TMScore} = \max_{\mathbf{R},\mathbf{t}}
    \frac{1}{L_{\mathrm{ref}}}
    \sum_{i=1}^{L_{\mathrm{ali}}}
    \frac{1}{
        1+
        \left(
            \frac{d_i(\mathbf{R},\mathbf{t})}{d_0(L_{\mathrm{ref}})}
        \right)^2
    },
\end{equation}
where $L_{\mathrm{ref}}$ is the number of residues in the reference structure, $L_{\mathrm{ali}}$ is the number of aligned residues and
$d_i(\mathbf{R},\mathbf{t})$ is the distance between the $i$th pair of aligned $\mathrm{C}\alpha$ atoms after rigid-body superposition. The length-dependent normalization is given by
\begin{equation}
    d_0(L_{\mathrm{ref}}) =
    1.24(L_{\mathrm{ref}}-15)^{1/3}-1.8~\text{\AA}.
\end{equation}
TM-scores lie in $(0,1]$. Larger values indicate greater structural similarity.

\paragraph{MolProbity scores.} We assess stereochemical quality using \texttt{phenix.molprobity} \citep{williams2018molprobity}. We report the MolProbity score, clashscore and percentage of Ramachandran outliers. The clashscore is the number of serious all-atom steric overlaps per $1000$ atoms. The Ramachandran outlier percentage is the fraction of residues whose backbone $\phi$ and $\psi$ angles fall outside the allowed regions of the reference Ramachandran distributions. The MolProbity score is a composite measure combining all-atom clashes, side-chain rotamer outliers and Ramachandran statistics. Lower scores indicate better stereochemical
quality.

\paragraph{Fourier shell correlation$=0.143$.} For atomic models generated by Fold'EM, Fold'EM \textit{ab initio}, AlphaFold3, and obtained from the PDB, we report the resolution ($\AA$) at which the Fourier shell correlation (FSC) curve wrt the experimental map $\rho$ from EMD crosses the $0.143$ threshold, following the commonly used FSC criterion for cryo-EM resolution assessment \citep{rosenthal2003optimal}. We use \texttt{phenix.map\_box} to compute the FSC curve and determine the corresponding resolution at the $0.143$ threshold.

\subsection{Cryo-EM image formation model and likelihood function}
\label{sec:image-formation-model}

Given a particle image $\mathbf{y}$, its log-likelihood given a conformation $\mathbf{X}$ and pose $\boldsymbol{\phi}$ can be written as $\log p(\mathbf{y}\mid \mathbf{X}, \boldsymbol{\phi}) = \log p(\mathbf{y} \mid {\mathbf y}_\mathrm{c}(\mathbf{X}, \boldsymbol{\phi}))$, where ${\mathbf y}_\mathrm{c}$ is the simulated noiseless particle image and $\log p(\mathbf{y} \mid {\mathbf y}_\mathrm{c})$ is the noise log-likelihood.

\subsubsection{Image formation model}
\label{sec:image-formation}
Given an atomic structure $\mathbf X=\{(\mathbf x_i,e_i)\}_{i=1}^{m}$, where $\mathbf x_i\in\mathbb R^3$ is the position and $e_i$ the chemical element of atom $i$, we approximate its electrostatic potential as a superposition of atomic potentials, $\rho_{\mathbf X}(\mathbf r)=\sum_{i=1}^{m}\rho_{e_i}(\mathbf r-\mathbf x_i)$. Each atomic potential is specified by its electron scattering factor $f_e$, which we identify, up to the conventional proportionality constant, with its Fourier transform, $\widehat\rho_e(\boldsymbol\xi)=f_e(\|\boldsymbol\xi\|)$. The Fourier representation of the molecular potential is therefore
\begin{equation}
\widehat{\rho}_{\mathbf X}(\boldsymbol\xi)
=
\sum_{i=1}^{m}
f_{e_i}(\|\boldsymbol\xi\|)
\exp\!\left(-2\pi i\,\boldsymbol\xi^\top\mathbf x_i\right),
\label{eq:atomic_potential}
\end{equation}
with $\boldsymbol\xi$ denoting the 3D spatial frequency. For a pose $\boldsymbol{\phi}=(\mathbf R,\mathbf t)$, with $\mathbf R\in\mathrm{SO}(3)$ and in-plane translation $\mathbf t\in\mathbb R^2$, we define the projection operator $\mathcal P_{\boldsymbol{\phi}}$ projecting the density onto a plane oriented at $\boldsymbol{\phi}$. We express the projection in the Fourier domain through the slice theorem: projection of a rotated 3D potential corresponds to sampling a central plane of its 3D Fourier transform. Let $\boldsymbol\omega=(\omega_1,\omega_2)^\top$ denote a 2D spatial frequency. We define the pose-dependent slice operator
\begin{equation}\label{eq:fourier_slice}
(\mathcal S_{\boldsymbol{\phi}}\widehat\rho)(\boldsymbol\omega)
=
e^{-2\pi i\boldsymbol\omega^\top\mathbf t}
\widehat\rho\!\left(
\mathbf R^\top
\begin{bmatrix}
\boldsymbol{\omega}\\ 0
\end{bmatrix}
\right),
\end{equation}
where the rotation selects the corresponding central Fourier plane and the translation appears as a phase ramp. The noiseless particle image is formed by further applying the known contrast transfer function (CTF), $G(\boldsymbol\omega)$, in the Fourier domain
$\widehat{\mathbf y}_{\mathrm{c}}(\mathbf X,\boldsymbol{\phi})(\boldsymbol\omega)
=
G(\boldsymbol\omega)
\left(
\mathcal S_{\boldsymbol{\phi}}
\widehat\rho_{\mathbf X}
\right)(\boldsymbol\omega)$, or in the real space
$\mathbf y_{\mathrm{c}}(\mathbf X,\boldsymbol{\phi})
=
\mathcal F^{-1}
\left[
G\,\mathcal S_{\boldsymbol{\phi}}
\widehat\rho_{\mathbf X}
\right]$.
Known particle-specific imaging parameters, including the CTF, are suppressed from $\boldsymbol{\phi}$ throughout the paper; they can equivalently be included among the nuisance variables introduced above.

\subsubsection{Noise model and likelihood function}
The fundamental electron-detection noise is shot noise and is therefore Poisson at the level of individual electron counts. The particle images used for reconstruction, however, are processed rather than raw counting measurements: detector response, frame integration, motion correction, dose weighting, gain correction, background subtraction, normalization, and other sources of experimental uncertainty make a simple pixelwise Poisson model inappropriate for the resulting images. As is standard in likelihood-based single-particle cryo-EM reconstruction \citep{scheres2012relion}, we therefore model the residual image noise as additive Gaussian noise, allowing its variance to depend on spatial frequency.

Specifically, let $\mathbf y_c$ denote a candidate noiseless particle image and assume additive Gaussian noise in the Fourier domain, $\widehat{\mathbf y}=\widehat{\mathbf y}_c+\widehat{\mathbf N}$. We follow the community practice and model each Fourier coefficient of $\widehat{\mathbf N}$ as an independent zero-mean complex Gaussian with frequency-dependent variance $\sigma^2(\boldsymbol\omega)$. The corresponding noise log-likelihood is, up to an additive constant,
\begin{equation}
\log p(\mathbf y\mid\mathbf y_c)
=
-\frac{1}{2}
\sum_{\boldsymbol\omega\in\Omega}
\frac{
\left|
\widehat{\mathbf y}(\boldsymbol\omega)
-
\widehat{\mathbf y}_c(\boldsymbol\omega)
\right|^2
}{
\sigma^2(\boldsymbol\omega)
}.
\label{eq:image_likelihood}
\end{equation}
The Gaussian model should be understood as an effective likelihood for processed cryo-EM particle images rather than as a model of the underlying electron-arrival process.
In practice, we do not optimize this Gaussian log-likelihood directly. Instead, we use a band-limited normalized correlation between the observed and simulated particle images in Fourier space,
\[
\mathcal L_{\mathrm{corr}}(\mathbf y,\mathbf y_c)
=
\frac{
\operatorname{Re}\!\left[
\sum_{\boldsymbol\omega\in\Omega}
\widehat{\mathbf y}(\boldsymbol\omega)\,
{\widehat{\mathbf y}_c(\boldsymbol\omega)}^\ast
\right]
}{
\sqrt{\sum_{\boldsymbol\omega\in\Omega}
|\widehat{\mathbf y}(\boldsymbol\omega)|^2}
\sqrt{\sum_{\boldsymbol\omega\in\Omega}
|\widehat{\mathbf y}_c(\boldsymbol\omega)|^2}
+\epsilon
},
\]
where $\Omega=\{\boldsymbol\omega:k_{\min}\leq\|\boldsymbol\omega\|_2<k_{\max}+1\}$ is the frequency band used for guidance and $\epsilon$ is a small numerical stabilizer. Maximizing this objective is equivalent to minimizing squared error between the observed and simulated Fourier images after normalizing both to unit norm. It can therefore be viewed as a scale-invariant surrogate for the Gaussian image log-likelihood.

We use this surrogate because the overall amplitude and contrast of experimental cryo-EM particles can vary for reasons that are not informative about molecular structure. The normalization removes sensitivity to such global scale differences and makes the guidance depend primarily on agreement in the spatial-frequency content and phase structure of the particle. Restricting the comparison to $\Omega$ further excludes very low frequencies dominated by background and normalization effects, as well as high frequencies with little reliable signal. Compared with the full Gaussian likelihood, this objective discards explicit frequency-dependent noise weighting in favor of a simpler and empirically more robust similarity measure for inference-time optimization.

\subsubsection{Particle image rendering from atomic structure}\label{sec:particle_from_structure}

We implement the image-formation model of Section~\ref{sec:image-formation}
directly from the atomic structure
$\mathbf{X}=\{(\mathbf{x}_i,e_i)\}_{i=1}^m$, approximating the atomic
potential $\rho_{e_i}$ of atom $i$ by an isotropic Gaussian with amplitude
$Z_i$, the atomic number of element $e_i$. With $a$ the pixel size in
\text{\AA}/pixel, the Gaussian width is set by the Nyquist resolution $2a$,
$\sigma=2a/(\sqrt{2}\pi)$, that is,
$\sigma_{\mathrm{px}}=\sigma/a=\sqrt{2}/\pi\approx0.45$ pixels.

Because integrating an isotropic 3D Gaussian along the viewing axis yields a
2D Gaussian of the same width, the projection can be evaluated analytically
without constructing an intermediate 3D voxel grid. For a pose
$\phi=(\mathbf{R},\mathbf{t})$, with $\mathbf{R}\in SO(3)$ and in-plane
translation $\mathbf{t}\in\mathbb{R}^2$ measured in pixels, atom $i$ is
rotated into the camera frame, $\tilde{\mathbf{x}}_i=\mathbf{R}^\top\mathbf{x}_i$,
and its projected center, in pixel coordinates relative to the image center,
is $\mathbf{c}_i=a^{-1}(\tilde{x}_{i,1},\tilde{x}_{i,2})^\top+\mathbf{t}$. The
projected atomic centers define the continuous 2D projection
\[
    P_\phi(\mathbf{X})(\mathbf{r})
    =
    \sum_{i=1}^{m}
    Z_i
    \exp\left(
        -\frac{\lVert \mathbf{r}-\mathbf{c}_i\rVert_2^2}
        {2\sigma_{\mathrm{px}}^2}
    \right),
\]
where $\mathbf{r}=(u,v)^\top$ denotes image-plane coordinates in pixels.

Each Gaussian is integrated over the finite pixel area. Since the isotropic
2D Gaussian is separable, the integral factorizes into two one-dimensional
Gaussian integrals, and for a pixel centered at coordinate $p$ and a Gaussian
centered at $c$ the one-dimensional integral has the closed form
\[
    g(p;c)
    =
    \int_{p-\frac{1}{2}}^{p+\frac{1}{2}}
    \exp\left(
        -\frac{(u-c)^2}{2\sigma_{\mathrm{px}}^2}
    \right)\,\mathrm{d}u
    =
    \sigma_{\mathrm{px}}\sqrt{\frac{\pi}{2}}
    \left[
        \operatorname{erf}\left(
            \frac{p+\frac{1}{2}-c}
            {\sqrt{2}\sigma_{\mathrm{px}}}
        \right)
        -
        \operatorname{erf}\left(
            \frac{p-\frac{1}{2}-c}
            {\sqrt{2}\sigma_{\mathrm{px}}}
        \right)
    \right],
\]
with $\operatorname{erf}(x)=\frac{2}{\sqrt{\pi}}\int_0^x e^{-t^2}\,\mathrm{d}t$
the error function. The rendered value at pixel $(p,q)$ is therefore
\[
    I_\phi[p,q]
    =
    \sum_{i=1}^{m}
    Z_i\,
    g(p;c_{i,1})\,
    g(q;c_{i,2}).
\]

\subsubsection{Contrast transfer function}
\label{app:ctf}
We assume the CTF parameters of each particle to be known, model contrast
transfer under the weak-phase approximation \citep{wade1992brief}, and use the
CTF parameterization of \citet{rohou2015ctffind4}. For a two-dimensional
spatial frequency $\boldsymbol{\omega}=(\omega_x,\omega_y)^\top$, let
$s=\lVert\boldsymbol{\omega}\rVert_2$, measured in \text{\AA}$^{-1}$, and
$\theta=\operatorname{atan2}(\omega_y,\omega_x)$. To account for astigmatism,
the defocus is parameterized by two principal defocus values $d_u$ and $d_v$
and their orientation $\theta_d$, so that the effective defocus at
Fourier-space angle $\theta$ is
\[
    d_f(\theta)
    =
    \frac{1}{2}
    \left[
        d_u + d_v
        +
        (d_u-d_v)
        \cos\left(2(\theta-\theta_d)\right)
    \right],
\]
which reduces to a rotationally symmetric CTF with a single defocus value when
$d_u=d_v$.

The CTF phase is
\[
    \gamma(\boldsymbol{\omega})
    =
    2\pi
    \left[
        -\frac{1}{2}d_f(\theta)\lambda s^2
        +
        \frac{1}{4}C_s\lambda^3s^4
    \right]
    -
    \varphi,
\]
where $\lambda$ is the electron wavelength, $C_s$ the spherical aberration
coefficient and $\varphi$ the phase shift, and the CTF is
\[
    G(\boldsymbol{\omega}) = \sqrt{1-w^2}\,
    \sin\left(\gamma(\boldsymbol{\omega})\right)
    -
    w\,
    \cos\left(\gamma(\boldsymbol{\omega})\right),
\]
with $w$ the amplitude-contrast fraction. The electron wavelength is
determined by the accelerating voltage $V$ through the relativistic de Broglie
relation $\lambda(V)=hc/\sqrt{eV\left(eV+2m_ec^2\right)}$, where $h$ is
Planck's constant, $e$ the elementary charge, $m_e$ the electron rest mass and
$c$ the speed of light. With $V$ in volts and $\lambda$ in \text{\AA}, this is
$\lambda(V)=12.2642598/\sqrt{V\left(1+9.78475598\times10^{-7}\,V\right)}$.

Following the image-formation model of Section~\ref{sec:image-formation-model},
the CTF is applied in the Fourier domain. With
$\widehat{\mathbf{I}}_\phi(\boldsymbol{\omega})$ the Fourier transform of the
rendered projection of the previous section, the simulated noiseless particle
is
$\widehat{\mathbf{y}}_c(\mathbf{X},\phi)(\boldsymbol{\omega})
=
G(\boldsymbol{\omega})\,
\widehat{\mathbf{I}}_\phi(\boldsymbol{\omega})$,
and the real-space image is obtained by inverse Fourier transformation. For
phase-flipped experimental particle stacks, we apply the magnitude
$|G(\boldsymbol{\omega})|$ rather than the signed CTF $G(\boldsymbol{\omega})$,
to match the preprocessing of the observed images.

\subsection{B-RECOVAR: bootstrap-aggregated RECOVAR class assignment}
\label{app:class_assignments}

In the \textit{ab initio} heterogeneous setting, Algorithm~\ref{alg:FoldEM_alg} requires an initial class assignment \(c^{(0)}=(c_k^{(0)})_{k=1}^K\) for \(\boldsymbol{\eta}^{(0)}\), inferred directly from the particle images (Section~\ref{sec:initialization}). Standard low-dimensional embeddings, including cryoDRGN latent coordinates and principal components of the RECOVAR volume covariance, can be unreliable when few particles are available per class. Nuisance variation, such as continuous motion, defocus variation, and specimen-preparation effects, can have variance comparable to the structural signal of interest, and the leading covariance directions may therefore depend strongly on the sampled particles.

We therefore use \brecovar{}, a bootstrap-style, subsampled ensemble for class initialization. Each replicate estimates a low-dimensional residual subspace by pose-aware regularized back-projection, clusters the resulting particle coefficients, and refines the labels by cross-validated comparison with class reconstructions. Replicate labelings are then aggregated by spectral voting, and the normalized voting margin is reported as a particle-level confidence score.

\paragraph{Model and notation.}
Let \(\ens{Y}=\{\mathbf{y}_k\}_{k=1}^K\) denote \(K\) particle images drawn from a mixture of \(n\) discrete conformations with densities \(\rho_1,\ldots,\rho_n\), mixing proportions \(\pi_1,\ldots,\pi_n>0\), \(\sum_{j=1}^n\pi_j=1\), and class labels \(c_k\in\{1,\ldots,n\}\). Poses \(\boldsymbol{\phi}_k\) and CTFs \(G_k\) are obtained from a preceding consensus refinement and registered to a common coordinate system. The labels are unknown, and the procedure uses no reference structures or atomic models.

We restrict volumes to the Fourier shells \(\Omega_{\mathrm a}=\{\boldsymbol{\xi}: k_{\min}\leq\|\boldsymbol{\xi}\|_2<k_{\max}\}\) and images to the corresponding central-plane coefficients: the lowest-frequency shells are dominated by global contrast and normalization, whereas the highest-frequency shells are noise-dominated. On this band, the imaging operator for particle \(k\) is \(\mathbf A_k=G_k\mathcal S_{\boldsymbol{\phi}_k}\), where \(\mathcal S_{\boldsymbol{\phi}}\) is the central-slice operator of \eqref{eq:fourier_slice}, and we assume \(\mathbf y_k=\alpha_k^\star\mathbf A_k\rho_{c_k}+\boldsymbol{\varepsilon}_k\), with \(\alpha_k^\star>0\) a particle-specific intensity scale and \(\boldsymbol{\varepsilon}_k\) zero-mean noise, independent across particles and approximately white on \(\Omega_{\mathrm a}\).

\paragraph{Residual subspace.}
Define the mixture mean \(\bar\rho=\sum_{j=1}^n\pi_j\rho_j\) and the class deviations \(\delta_j=\rho_j-\bar\rho\). Since \(\sum_{j=1}^n\pi_j\delta_j=0\), the class deviations span a space of dimension \(q=\dim\operatorname{span}\{\delta_1,\ldots,\delta_n\}\leq n-1\). Let \(U_1,\ldots,U_q\) be a basis of this space and write \(\delta_j=\sum_{m=1}^q b_{jm}U_m\). After relative intensity normalization and consensus subtraction, the residual of particle \(k\) satisfies
\begin{equation}
  \mathbf r_k
  \approx
  \sum_{m=1}^q b_{c_km}\mathbf A_kU_m
  +
  \boldsymbol{\varepsilon}'_k ,
  \label{eq:lf_subspace}
\end{equation}
where any common scale is absorbed into the modes \(U_m\), and \(\boldsymbol{\varepsilon}'_k\) includes measurement noise and scale-estimation error. Thus, particles from class \(j\) have a common mean coefficient vector \(\mathbf b_j\in\mathbb R^q\), and if \(q=n-1\), these vectors form the vertices of an \((n-1)\)-simplex. For \(n=2\), the subspace is spanned by \(\Delta\rho=\rho_1-\rho_2\), with \(\delta_1=\pi_2\Delta\rho\) and \(\delta_2=-\pi_1\Delta\rho\), and the coefficient sign distinguishes the two classes up to a global label exchange. For a balanced two-state mixture, \(\|\delta_1\|=\|\delta_2\|\), so the residual energy alone carries no class information.

For a particle subset \(\mathcal K\), define the regularized back-projection
\[
  \mathcal{R}\big(\{(\nu_k,\mathbf d_k)\}_{k\in\mathcal K}\big)
  =
  \Big(
    \sum_{k\in\mathcal K}\nu_k\mathbf A_k^\ast\mathbf A_k
    +\lambda_{\mathrm w}\mathbf I
  \Big)^{-1}
  \sum_{k\in\mathcal K}\mathbf A_k^\ast\mathbf d_k,
\]
and let \(\mathcal M\) denote multiplication in real space by a smooth solvent mask derived from the consensus reconstruction. Using all particles, we form the consensus \(\tilde\rho=\mathcal R(\{(1,\mathbf y_k)\}_{k=1}^K)\), the scales \(\tilde\alpha_k=\operatorname{Re}\langle\mathbf y_k,\mathbf A_k\tilde\rho\rangle/\|\mathbf A_k\tilde\rho\|^2\), the normalized images \(\check{\mathbf y}_k=\tilde\alpha_k^{-1}\mathbf y_k\), and the residuals \(\mathbf r_k=\check{\mathbf y}_k-\mathbf A_k\tilde\rho\). The scale normalization makes the class coefficients comparable across particles.

We seek masked modes \(W_1,\ldots,W_q\) and coefficient vectors \(\tilde{\mathbf b}_k\in\mathbb R^q\) through the constrained factorization
\[
  \underset{
    \substack{
      \{W_m\},\{\tilde{\mathbf b}_k\}\\
      \|\tilde{\mathbf b}_{\cdot m}\|_2=1,\;m=1,\ldots,q
    }
  }{\operatorname{minimize}}
  \quad
  \sum_{k=1}^K
  \left\|
    \mathbf r_k-\sum_{m=1}^q\tilde b_{km}\mathbf A_kW_m
  \right\|^2
  +
  \lambda_{\mathrm w}\sum_{m=1}^q\|W_m\|^2 .
\]
The normalization removes the scale ambiguity between a mode and its coefficients. The modes are identifiable only up to a change of basis of the subspace, and the class labels only up to a permutation. For \(n=2\), omitting the mask yields a rank-one, one-dimensional RECOVAR fit without noise correction \citep{gilles2025cryo}. The smooth mask allows for variance reduction, as it suppresses solvent-dominated contributions, provided that it contains the class-specific density.

We estimate the factorization by greedy masked alternating least squares with deflation. Let \(\mathcal I\) be the particle subset used to fit one replicate; for a single full-data fit, \(\mathcal I=\{1,\ldots,K\}\). Set \(\mathbf r_k^{(0)}=\mathbf r_k\). For \(m=1,\ldots,q\), we initialize \(\mathbf a_{\mathcal I m}=(a_{km})_{k\in\mathcal I}\) as a random unit vector and iterate
\begin{align*}
  W_m
  &\leftarrow
  \mathcal M\,
  \mathcal R\big(
    \{(a_{km}^2,a_{km}\mathbf r_k^{(m-1)})\}_{k\in\mathcal I}
  \big),
  \\
  a_{km}
  &\leftarrow
  \frac{\langle\mathbf r_k^{(m-1)},\mathbf A_kW_m\rangle}
  {\|\mathbf A_kW_m\|^2},
  \qquad
  \mathbf a_{\mathcal I m}
  \leftarrow
  \frac{\mathbf a_{\mathcal I m}}{\|\mathbf a_{\mathcal I m}\|_2},
  \quad k\in\mathcal I .
\end{align*}
After the final update, we recompute the unnormalized coefficients \(\check a_{km}=\langle\mathbf r_k^{(m-1)},\mathbf A_kW_m\rangle/\|\mathbf A_kW_m\|^2\) and deflate, \(\mathbf r_k^{(m)}=\mathbf r_k^{(m-1)}-\check a_{km}\mathbf A_kW_m\), for \(k\in\mathcal I\). Once all modes are fitted, the coefficients of every particle are estimated jointly from the original residual, \(\tilde{\mathbf b}_k=\arg\min_{\mathbf b\in\mathbb R^q}\|\mathbf r_k^{(0)}-\sum_{m=1}^q b_m\mathbf A_kW_m\|^2\), \(k=1,\ldots,K\). Under \eqref{eq:lf_subspace}, the expected coefficient vector depends on the class but not on the pose, up to the fitted change of basis. We cluster \(\{\tilde{\mathbf b}_k\}_{k=1}^K\) by \(k\)-means with \(n\) clusters and multiple initializations. For \(n=2\), we cluster the standardized scalar \(\tilde b_{k1}\|\mathbf A_kW_1\|\), whose noise variance is approximately constant across particles.

\paragraph{Cross-validated refinement.}
At each refinement iteration, we partition the particles into two folds \(\mathcal J\) and \(\mathcal J'\) and, given the current labels \(\{\tilde c_k\}\), reconstruct one masked volume per class from each fold, \(\tilde\rho_j^{\mathcal F}=\mathcal M\,\mathcal R(\{(1,\check{\mathbf y}_k)\}_{k\in\mathcal F,\;\tilde c_k=j})\) for \(\mathcal F\in\{\mathcal J,\mathcal J'\}\). For \(k\in\mathcal F\), we score class \(j\) against the reconstruction from the opposite fold \(\mathcal F^{\mathrm c}\), \(s_{kj}=\mathcal L_{\mathrm{corr}}(\check{\mathbf y}_k,\mathbf A_k\tilde\rho_j^{\mathcal F^{\mathrm c}})\), where \(\mathcal L_{\mathrm{corr}}\) is the band-limited normalized correlation of Section~\ref{sec:image-formation-model} evaluated on \(\Omega_{\mathrm a}\), and center the scores of each class within the fold, \(\bar s_j^{\mathcal F}=|\mathcal F|^{-1}\sum_{k'\in\mathcal F}s_{k'j}\). We then update
\begin{equation}
  \tilde c_k
  \leftarrow
  \arg\max_{j\in\{1,\ldots,n\}}
  \big(s_{kj}-\bar s_j^{\mathcal F}\big),
  \qquad k\in\mathcal F .
  \label{eq:lf_centred}
\end{equation}
Cross-validation avoids direct self-correlation between a particle and a reconstruction containing its noise, and the fold-specific centering compensates for class-dependent score offsets caused by unequal reconstruction precision. The refinement requires sufficient particles from every current class in both folds.

\paragraph{Ensemble voting.}
With few particles per conformation, a single residual-subspace fit can be unstable even for \(n=2\): different initializations and particle samples may recover a nuisance direction with explained energy comparable to that of the class mode. The class subspace, however, should recur across particle subsamples and frequency bands containing structural signal, whereas sample-specific nuisance directions are less stable. We therefore fit \(B\) randomized replicates of Algorithm~\ref{alg:residual_mode}, each with a particle subset \(\mathcal I_b\), a selected frequency band, and an independent initialization. Modes are fitted on \(\mathcal I_b\), but coefficients, clustering, and refinement are evaluated for all particles. Each replicate produces labels \(\{\tilde c_k^{(b)}\}\), defined only up to permutation.

To align the labelings, we select as reference the replicate with the highest mean pairwise agreement and, for every replicate \(b\), choose the permutation \(P_b\) that maximizes the trace of its contingency matrix with the reference, using the Hungarian algorithm \citep{kuhn1955hungarian}; the aligned labels are \(\check c_k^{(b)}=P_b(\tilde c_k^{(b)})\).

For \(n=2\), let \(v_k^{(b)}\in\{-1,+1\}\) encode the aligned vote of replicate \(b\), and let \(\psi_b\) denote its balanced accuracy relative to the unknown labels. Under conditional independence of replicate errors given the class,
\begin{equation}
  \operatorname{Cov}\big(v^{(b)},v^{(b')}\big)
  =
  (1-\beta^2)(2\psi_b-1)(2\psi_{b'}-1),
  \qquad
  b\neq b',
  \label{eq:lf_sml}
\end{equation}
where \(\beta=\pi_1-\pi_2\). Thus, the off-diagonal covariance entries admit a rank-one factorization. Following the spectral meta-learner of \citet{parisi2014ranking}, we obtain \(\mathbf u\in\mathbb R^B\) from a spectral rank-one fit to these entries, orienting it relative to the reference replicate. The weighted vote \(\tilde v_k=\sum_{b=1}^B u_bv_k^{(b)}\) defines the binary assignment by its sign, and \(\kappa_k=|\tilde v_k|/\sum_{b=1}^B|u_b|\) is the confidence score. A reversed replicate receives a negative weight, and a chance-level replicate is downweighted. The relative gap between the two leading eigenvalues of the vote covariance indicates whether the replicates agree on a common split.

For \(n>2\), we apply the same meta-learner separately to the aligned one-versus-rest indicators \(v_{kj}^{(b)}=+1\) if \(\check c_k^{(b)}=j\) and \(-1\) otherwise. Let \(\mathbf u^{(j)}\) denote the resulting class-specific weights and define the class scores \(\bar v_{kj}=\sum_{b=1}^B u_b^{(j)}v_{kj}^{(b)}\big/\sum_{b=1}^B|u_b^{(j)}|\). We assign \(\tilde c_k=\arg\max_j\bar v_{kj}\) and report \(\kappa_k=\tfrac12(\bar v_{k,(1)}-\bar v_{k,(2)})\), where \(\bar v_{k,(1)}\geq\bar v_{k,(2)}\) are the two largest class scores. Class-specific weights allow a replicate that resolves only a subset of classes to contribute to those classes. Because replicates share particles, consensus maps, and masks, conditional independence is a working approximation; the margins and eigengaps are therefore diagnostic confidence measures rather than calibrated probabilities.

\paragraph{Comparison with RECOVAR.}
RECOVAR and \brecovar{} both use consensus poses and CTFs, require no learned network, and infer structural variability from pose-aware volume statistics. RECOVAR estimates a regularized, noise-whitened volume covariance and embeds particles in a high-dimensional PCA space; \brecovar{} instead estimates only the \(n-1\) masked, band-limited class-deviation modes from consensus-subtracted residuals. For \(n=2\), this is a rank-one, uncorrected analogue of one-dimensional RECOVAR. Restricting the fit to the class subspace avoids relying on the leading direction of the full covariance, which can be dominated by sample-dependent nuisance variation at low particle counts.

\brecovar{} further stabilizes the assignment by aggregating fits over particle subsamples, frequency bands, and initializations with the spectral meta-learner of \eqref{eq:lf_sml}; the normalized margin \(\kappa_k\) provides a particle-level confidence score. It then applies the cross-fitted refinement of \eqref{eq:lf_centred}, avoiding direct self-correlation between a particle and its comparison volume. For a matched comparison, both methods use the same label-free readout and spectral aggregation; only \brecovar{} uses cross-fitted refinement. RECOVAR randomizes its half-set split and downsampled box, whereas \brecovar{} randomizes the particle subset, frequency band, and initialization.

\paragraph{Implementation details.}
We use \(k_{\min}=3\) and \(k_{\max}\in\{24,32,40\}\) Fourier pixels, with \(q=n-1\). Although the shells below \(16\) pixels contribute little directly to class discrimination in our data, they are retained to stabilize the reconstructions. The regularization is \(\lambda_{\mathrm w}=0.1\) times the mean diagonal entry of \(\sum_k\mathbf A_k^\ast\mathbf A_k\). The solvent mask retains the \(8\%\) of voxels with the largest values of \(|\mathcal F^{-1}\hat\rho|\) and is smoothed with a Gaussian of standard deviation two voxels; mask fractions between \(4\%\) and \(15\%\) yield comparable results.
 
We use \(T=8\) alternating updates, \(B=24\) replicates, and independent \(80\%\) particle subsamples; the three frequency bands are allocated evenly among replicates. The consensus, normalized residuals, and mask are computed once per band using all particles. Each replicate uses \(L=4\) refinement iterations. In the experimental data, with fewer than approximately \(5{,}000\) particles per conformation, both the class direction and competing nuisance directions account for about \(1.5\%\) of the residual energy, which motivates the ensemble aggregation. The dominant computational cost is \(Bq\) single-mode fits, approximately \(10\) CPU-minutes for \(10^4\) particles on a \(D=128\) grid with \(n=2\). Replicates can be evaluated in parallel; modes within a replicate are fitted sequentially because of deflation.
 
\subsection{Hyperparameters and implementation details}\label{sec:impl_details}
We use the following hyperparameters for all experiments.
\begin{itemize}
    \item $k_{min} = 3$ and $k_{max}=48$.
    \item $D = 128$ pixels.
    \item Due to memory constraints, at each optimization step, we guide using $\mathrm{min}(|\ens{Y}|, 128)$ particles.
    \item Optimizer: Adam \citep{adam2014method}
    \item Learning rate: $0.06$
    \item Max gradient norm: $0.6$
    \item $\lambda_{\mathrm{bond}} = 10^{-3}$ and $\lambda_{\mathrm{collision}} = 0.09$
    \item $R$ (number outer iterations) was $3$ for real examples and $5$ for synthetic examples.
    \item \textbf{On the prior term.} In the formulation from \citet{maddipatla2026inference},  the prior $\log p(\ens{Z} \mid \mathbf{a})$ was essential: with $\ens{Z}$ treated as a free variable, nothing else prevented the optimization from drifting into regions of embedding space that the Pairformer never produces. In the present parameterization this role is partly assumed by the reparameterization itself, since displacements in $\ens{H}$ reach the conditioning variable only through $g$ and are therefore expressed in the basis of directions the trunk natively realizes.
\end{itemize}

\end{document}